\documentclass[twocolappendix,numberedappendix]{emulateapj}

\usepackage{mathtools}
\usepackage{graphicx}
\usepackage{epsfig}
\usepackage{tabularx}
\usepackage{multirow}
\usepackage{subfigure}
\usepackage{rotating}
\usepackage{booktabs}
\usepackage{array}
\usepackage{xfrac}
\usepackage{natbib}
\usepackage{hyperref}
\usepackage{xcolor}

\newcolumntype{R}{>{\raggedleft\arraybackslash}X}
\newcolumntype{C}{>{\centering\arraybackslash}X}

\def\iso#1#2{\mbox{${}^{#2}{\rm #1}$}}
\def\be1#1{\iso{Be}{1#1}}
\def\al2#1{\iso{Al}{2#1}}
\def\ca4#1{\iso{Ca}{4#1}}
\def\mn5#1{\iso{Mn}{5#1}}
\def\fe6#1{\iso{Fe}{6#1}}
\def\pu2#1#2{\iso{Pu}{2#1#2}}

\def\pfrac#1#2{\left( \frac{#1}{#2} \right)}

\def\msol{M_\odot}

\def\erf{{\rm erf~}}

\def\araa{ARA\&A}
\def\apj{ApJ}
\def\apjl{ApJ}
\def\apjs{ApJS}
\def\apss{Ap\&SS}
\def\aap{A\&A}
\def\baas{BAAS}

\def\mnras{MNRAS}
\def\na{New~Astron.} 
\def\prl{Phys.~Rev.~Lett.}
\def\sovast{Soviet~Ast.}
\def\nat{Nature}
\def\grl{Geophys.~Res.~Lett.}
\def\jgr{J.~Geophys.~Res.}

\def\beq{\begin{equation}}
\def\eeq{\end{equation}}
\def\beqar{\begin{eqnarray}}
\def\eeqar{\end{eqnarray}}

\newcommand*{\affaddr}[1]{#1} 
\newcommand*{\affmark}[1][*]{\textsuperscript{#1}}

\begin{document}
\raggedbottom

\title{Magnetic Imprisonment of Dusty Pinballs by a Supernova Remnant}

\author{
	Brian J. Fry\affmark[1,2], Brian D. Fields\affmark[2,3], and John R. Ellis\affmark[4,5,6]\\[6pt]
	\affaddr{\affmark[1]{\rm Department of Physics, United States Air Force Academy, Colorado Springs, CO 80840, USA}}\\
	\affaddr{\affmark[2]{\rm Department of Astronomy, University of Illinois, Urbana, IL 61801, USA}}\\
	\affaddr{\affmark[3]{\rm Department of Physics, University of Illinois, Urbana, IL 61801, USA}}\\
	\affaddr{\affmark[4]{\rm Theoretical Physics and Cosmology Group, Department of Physics, King's College London, London WC2R 2LS, UK}}\\
	\affaddr{\affmark[5]{\rm NICPB, R\"avala pst.~10, 10143 Tallinn, Estonia}}\\
	\affaddr{\affmark[6]{\rm Theory Division, Physics Department, CERN, CH-1211 Geneva 23, Switzerland}}\\
}

\begin{abstract} 
Motivated by recent measurements of deposits of \fe60 on the ocean floor and the lunar surface, we model the transport of dust grains containing \fe60 from a near-Earth (i.e., within 100 pc) supernova (SN).  We inject dust grains into the environment of an SN remnant (SNR) and trace their trajectories by applying a 1D hydrodynamic description assuming spherical symmetry to describe the plasma dynamics, and include a rudimentary, 3D magnetic field description to examine its influence on charged dust grains.  We assume the interstellar medium (ISM) magnetic fields are turbulent, and are amplified by the SNR shock, while the SN wind and ejecta fields are negligible.  We examine the various influences on the dust grains within the SNR to determine when/if the dust decouples from the plasma, how much it is sputtered, and where within the SNR the dust grains are located.  We find that Rayleigh-Taylor instabilities are important for dust survival, as they influence the location of the SN's reverse shock.  We find that the presence of a magnetic field within the shocked ISM material limits the passage of SN dust grains, with the field either reflecting or trapping the grains within the heart of the SNR. These results have important implications for {\it in situ} \fe60 measurements, and for dust evolution in SNRs generally.

\end{abstract}

\section{Introduction}
\label{sect:intro}

Supernovae (SNe) are some of both the most destructive and creative events in the universe.  An SN explosion blasts apart a massive star, and its outward propagating shock wave shatters dust grains floating in the interstellar medium (ISM).  However, the explosion also leads to the formation of a new compact object, creates heavy elements beyond iron and nickel, and, as the SN remnant (SNR) expands, new dust grains condense from within the ejecta.  This clash of simultaneously destroying and creating dust raises the question whether SN are net producers or demolishers of dust.

Looking at another facet of SNe:  they are estimated to occur at a rate of $1-3$ per century within the Milky Way \citep[e.g.,][and references therein]{adams13} and, given the size of the Milky Way, this suggests that one (probably more) has occurred close enough to have produced detectable effects on Earth.  These effects could range from delivery of SN material onto the Earth's surface to biological effects.  Studies of the possible biological effects of a near-Earth SN have a long history in the literature \citep[e.g.,][]{shk69,alva80,es95,mel17,fields19}, but the delivery of SN material onto Earth has only more recently been examined, first by \citet{efs96} \citep[see also][]{kor96}, who suggested looking for long-lived radioactive isotopes ($\tau_{1/2} \sim$ Myr) such as \fe60 and \pu244, whose presence would constitute direct evidence of such an event, since these isotopes are not manufactured within the Solar System and any pre-solar isotopes would have decayed by today \citep[see also][]{kor19}.

The first evidence for such extra-solar radioisotopes was found by \citet{knie99} in a sample of ferro-manganese (Fe-Mn) crust from Mona Pihoa in the South Pacific.  \citeauthor{knie99} used accelerator mass spectrometry (AMS) to find an anomaly in \fe60 concentration that suggested that an SN occurred near Earth sometime within the last 5 Myr (the time within this range could not be specified).  This study was later confirmed by \citet{knie04} using a different Fe-Mn crust sample from the equatorial Pacific Ocean floor, which found a distinct signal in \fe60 abundance $\sim2.2$ Myr ago.  Later \citet{fit08} also found an \fe60 signal in the Fe-Mn crust, but analysis of sea sediment samples from the northern Atlantic Ocean found no clear signal.  Fitoussi et al. noted several reasons for the discrepancy, including variations in the background and differences in the uptake efficiencies between the Fe-Mn crust and sediment, or a signal duration much longer (and hence diluted) than the then-expected timescale of $\sim few \ kyr$.  Subsequently, results from Eltanin sediment samples from the southern Indian Ocean were reported in \citet{feige14}, confirming the \citet{knie04} Fe-Mn crust detection in these sea sediment samples.  Most recently, \citet{wall16}, using AMS but in a different laboratory and with different samples, also found \fe60 in Fe-Mn crusts and nodules, confirming the $2-3$ Myr signal but also reporting evidence for a \emph{second} signal at $\sim 6-8$ Myr.  \citet{wall16} also detected \fe60 in several deep-ocean sediments, again finding the earlier signal, while measuring a deposition timescale of $\approx 1$ Myr, consistent with the \citet{fit08} result.  \citet{ludw16} detected \fe60 in sediments by isolating the isotope from iron-bearing microfossils; they confirmed the deposition timescale $\sim 1$ Myr.

We note also that cosmic-ray studies by \citet{kach15} and \citet{sav15} found a signature in the proton cosmic-ray spectrum suggesting an injection of cosmic-rays associated with an SN occurring $\sim 2$ Myr ago, and the discovery of \fe60 in cosmic-rays by \citet{binns16}, suggest an SN origin within the last $\sim 2.6$ Myr located $\lesssim 1$ kpc of Earth, based on the \fe60 lifetime and cosmic-ray diffusion. Additionally, lunar regolith samples \citep{cook09,fimi12,fimi14,fimi16} have also shown an excess of \fe60, although only the presence of an excess is detectable, not the precise arrival time or fluence \citep{feige13}, because of the nature of the regolith.  However, \citet{fry16} suggested the use of lunar regolith radioisotope distributions with lunar latitude as an ``antenna'' to find the direction to the source of the \fe60 material.

Many studies have thus confirmed the $2-3$ Myr \fe60 signal, using a variety of sampling techniques from around the Earth and on the Moon.  All sediment studies are consistent with a $\sim 1$ Myr deposition time.  It is thus well-established that (at least one) recent near-Earth SN deposited its ejecta on the Earth and Moon.  On the other hand, different papers have reported fluences that have varied by an order of magnitude.  The study by \citet{fry16} found that terrestrial atmospheric and oceanic processes could explain such differences in the fluence values between these studies, including the weak signal in the \citeauthor{fit08} sediment sample.  

The possible nucleosynthesis site of the \fe60 material was most thoroughly examined by \citet{fry15}, who considered all known astrophysical sources of \fe60 (including core-collapse SNe, neutron-star mergers and thermonuclear/Type Ia SNe).  They found that core-collapse SNe are excellent candidates; in particular they found that an Electron-Capture SN (ECSN) arising from a Zero-Age Main Sequence (ZAMS) mass $\approx 8-10 \ \msol$ (``$\odot$'' refers to the Sun) to be the most likely progenitor. However, they were not able to rule out completely a Super Asymptotic Giant Branch (SAGB) star with a ZAMS mass $\approx 6.5-9 \ \msol$ as a possible source, based on nucleosynthesis criteria alone.  For this work we will assume that the \fe60 source is a core-collapse SN, and we will highlight the ECSN case.

With regards to the location of the \fe60-producing SN, \citet{bene02} suggested that the source event occurred in the Sco-Cen OB association.  This association was $\sim 130$ pc away at the time of the \fe60-producing event, and its members were described in detail by \citet{fuchs06}.  \citet{breit12,breit16,feige16,feige17,schul17,schul18} have modeled the formation of the Local Bubble and used hydrodynamic simulations to model SNe occurring within the Sco-Cen association and track the \fe60 dust entrained within the blast.  Comparably, \citet{sore17} examined SN activity in open clusters within 1000 pc of Earth over the past 35 Myr and found several passing within 200 pc of Earth.  \citet{mam16} suggested that the Tuc-Hor group, which was within $\sim 60$ pc of Earth at the time of the \fe60-producing event, could have provided an ECSN, based on the masses of the current group members.   \citet{hyde18} considered both the Sco-Cen and Tuc-Hor sites and found that Tuc-Hor could be the site of either the $2-3$ or $6-8$ Myr ago events, but argued that the $2-3$ Myr ago event arose from the Upper Centaurus Loop component of Sco-Cen, assigning the earlier event to Tuc-Hor.  Additionally, \citet{neuh19} sought to link the arrival of \fe60 with a runaway star, finding one runaway-pulsar-pair met their criteria, although there was a discrepancy between the time of the \fe60 signal peak ($\sim 2.2$ Myr ago) and the time of common distance ($1.78 \pm 0.21$ Myr ago).

With regard to the deposition of SN \fe60 on Earth, \citet{faj08} used hydrodynamic models to show that for plausible SN distances, the SNR plasma cannot penetrate the Solar Wind to 1 AU.  However, SN radioisotopes including \fe60 are generally in refractory elements that readily form dust \citep{bene02}. \citet{af2011} and \citet{fry16} showed that SN material in the form of dust can have sufficient mass and velocity to overcome the Solar Wind and reach Earth.  The \fe60 detections thus imply that at least some SN iron was condensed into dust grains, and survived passage to the Solar System while still retaining high enough mass and velocity to pass within 1 AU of the Sun.  In the broader context of the nature of SNe, this raises the following question:  \emph{How can an SN, which is quite proficient at destroying dust, transport dust material effectively across light-years of interstellar space to the Solar System without destroying it?}

Many studies have examined general dust processing \citep[e.g.,][and references therein]{dwek92,drain03}, and within an SNR in particular \citet{noz06,noz07,koz09}.  Several studies consider only one type of action such as formation \citep{cher09,cher10,cher11,cher13,dwek16}, or examine only one process such as charging \citep{lafon81,drain87,bark94} or sputtering \citep{shull77,scalo77,tiel87,dwek96,jones96,janev01}.  Other studies have focused on a specific event within the grain's journey in the SNR, such as the passage of the reverse shock (RS) \citep{silv10,silv12,bisc16}.  More comprehensive studies such as \citet{noz07,nath08,mice16,bocc16} follow the grains through the entire SNR, but do not include magnetic fields, which could potentially affect the trajectory of the grains within the SNR.

To date, studies of near-Earth SNe, \citep[e.g.,][]{af2011,fry15,fry16,breit16,feige16,feige17,schul17} have assumed that the \fe60 material would be coupled to the SNR plasma, and most likely confined to the leading edge of the SNR.  This paper relaxes this assumption, allowing the grains to decouple from the SNR plasma earlier in the SNR evolution, potentially escaping the SNR.  We utilize a similar approach to that conducted by \citet{noz07,nath08,mice16,bocc16}, namely applying a hydrodynamic simulation of the SN evolution, then examining dust processes from those conditions, but we include grain charging, a rudimentary treatment of the SNR's magnetic field (specifically the ISM magnetic field), and examine the specific case of \fe60-containing dust grains from a near-Earth SN.  Our work assumes the magnetic fields in the SN ejecta and stellar wind are negligible for the purposed of dust propagation; including those fields would only reinforce our main conclusions.

We find that:
\begin{itemize}
	\item Magnetic fields (specifically, the shocked ISM field) dominate the fate of larger Fe grains ($a_{\rm gr} \gtrsim 0.05$ $\mu$m), effectively confining them within the inner SNR, whilst drag dominates the fate of smaller Fe grains, eventually slowing them to the SNR plasma velocity.  
	\item Dust grains exhibit ``\emph{pinball}'' behavior due to magnetic reflections, and ricochet in the interior of the SNR.
	\item The inclusion of Rayleigh-Taylor instabilities is important, since the earlier passage of the RS exposes the dust grains to erosion and drag for a longer time period.
	\item Grain propagation studies omitting the effects of grain-grain collisions and shock encounters may be missing important grain influences.
	\item Our results indicate that the Sco-Cen association could not have hosted the SN progenitor, because the SN would not have been able to push the ISM magnetic field, $\boldsymbol{B}_{\rm ISM}$, beyond the Solar System so as to allow the \fe60-containing dust grains to penetrate the Solar System and reach Earth.  
	\item The Tuc-Hor association is still a possible source for the \fe60.
\end{itemize}
Our findings also suggest that earlier assumptions of dust grains being entrained in the leading edge of the SNR \citep[e.g.,][]{fry15,breit16,feige16,feige17,schul17} are not appropriate.

\section{SNR Evolution}
\label{sect:SNR}

A SNR will transition through four main phases as it evolves \citep{ost88,padmV2,drain11,jank16}.  The first phase is Free Expansion, and it is characterized by a constant velocity after the explosion.  The ejected material moves outward supersonically, and produces a shock wave in the surrounding, ambient material (hereafter referred to as the ``forward shock'' or FS).  The presence of an ambient medium causes the ejected material to slow down, but in the early phases of expansion, when the mass of the ejected material far exceeds that of the swept up ambient material, this deceleration is negligible when examining the expansion of the FS.  However, this slight deceleration creates a second shock wave (hereafter referred to as the ``reverse shock'' or RS), which communicates the presence of the ambient medium to the ejected material.  Analytic solutions to this phase were found by \citet{chev82,chev92} and \citet{nad85} (referred to as Chevalier-Nadezhin); these self-similar solutions yield good descriptions for the position of the FS and incorporate the presence of a RS.  In the Free Expansion phase, both the forward and reverse shocks are moving outward.

There are, however, some difficulties with the Chevalier-Nadezhin solutions. First, at the interface (also known as the contact discontinuity) between the ejecta and ambient medium, the Chevalier-Nadezhin density profile solutions produce singularities (resulting in either an infinite or zero density depending on the density profile of the ambient medium).  Both cases are unphysical and, if one attempted to examine dust dynamics in such a state, the grain would encounter an imaginary wall or vacuum.  Additionally, the Chevalier-Nadezhin solutions ignore the presence of Rayleigh-Taylor instabilities along the RS.  Rayleigh-Taylor instabilities drive the RS further inward \citep{hera94,blell01,blond01} and, when considering dust, the location of the RS is of extreme importance since its passage shatters grains and exposes them to the hot SNR material.

As the FS sweeps up more material, the SNR transitions into the second, Energy-/Adiabatic-Conserving, phase.  This phase is often called the Sedov-Taylor phase after \citet{sed59} and \citet{tay50} who found self-similar descriptions of the  expansion.  When the swept-up material is approximately equal in mass to the ejecta material, the RS will cease moving outward and be driven inward, deeper into the ejecta, eventually proceeding all the way to the center of the SNR.  Studying dust dynamics during the Sedov-Taylor phase would be fairly straightforward, since the plasma density, velocity, and pressure within the SNR are described smoothly.  However, because the dust grains are initially formed during the Free Expansion phase, a description that includes a transition between phases is required.  \citet{tm99} found analytic solutions for this transition, describing the positions of the FS and RS through both phases. However, these do not include descriptions of the plasma properties needed to describe the grain dynamics (except as initial conditions) and ignore the effects of Rayleigh-Taylor instabilities along the RS.

As the SNR expands and cools, ions within the SNR shell will combine with electrons and radiate photons.  As the SNR outer shell becomes radiative, the FS will lose thermal pressure support and the expansion will be determined by the momentum within the shell \citep{blond98, drain11}.  This third phase is the Momentum-Conserving phase, with the SNR shell slowing as it collects more surrounding material (this phase is also referred to as the ``snowplow'' phase because of this accumulating action).  The snowplow phase typically begins around 50,000 yr after the SN explosion but, since the typical grain lifetime is around 100,000 yr \citep{drain11}, a detailed examination of the grain properties at the end of the Sedov-Taylor phase will be needed first, since the dust could be severely (possibly completely) ablated before reaching the snowplow phase.

Eventually the FS will slow to the sound speed of the ambient medium.  At this point the SNR effectively stops expanding, and the SNR enters the Fade-Away phase as the shock transitions to a sound wave in the ambient medium.  The SNR will eventually be dispersed through random processes in the ISM.  Any dust grains that survive to this stage will behave the same as they would in the general ISM.

Before specifying the SNR environment we assume, we first mention the quantities of the SNR environment we require in order to describe accurately the dynamics of our dust grains.  \S \ref{sect:grainproc} will describe in much greater detail why these quantities are important.  The density, velocity, and temperature (and how they evolve with time) within the SNR are required, as these determine the drag experienced by the grains as well as the degree of erosion by the plasma.  The composition of the plasma should also be detailed, since larger ions such as O/Si/S in sufficient concentrations can enhance erosion beyond that due to H and He.  The grain's charge and the direction and strength of the SNR magnetic field are important since dust grains spiraling around magnetic field lines could potentially become trapped within the SNR, as we discuss in \S \ref{subsubsect:mag} below.  The charge is dependent on the material of the dust grain, which is itself dependent on where the dust is formed within the ejecta.  Finally, the location of the dust grain's birthplace in the ejecta is significant, since it also affects when it will encounter the RS and its initial position and velocity.

\section{Grain Processes}
\label{sect:grainproc}

\begin{figure*}[t]
	\centering
	\subfigure[]
	{\includegraphics[width=\textwidth]{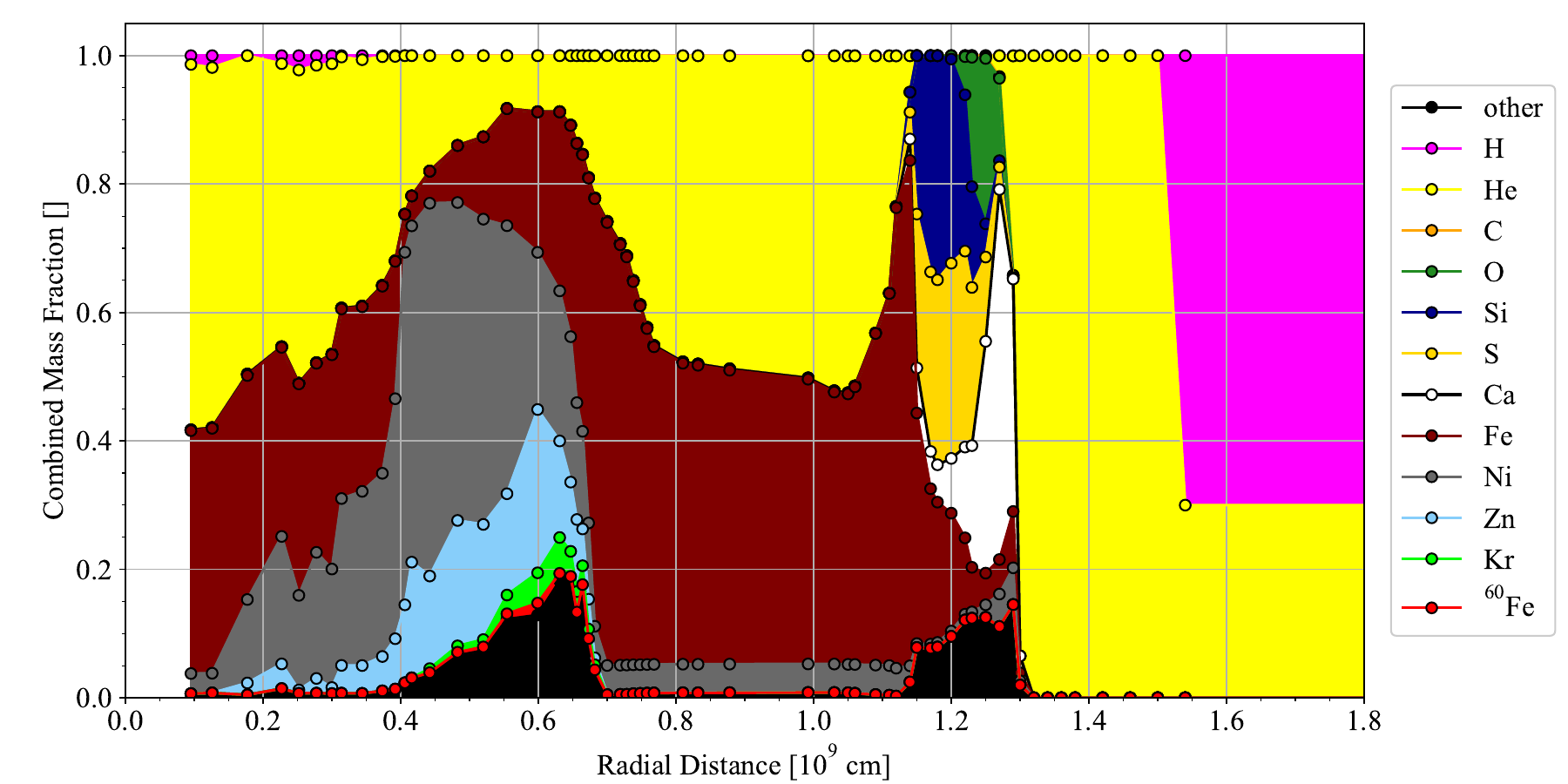}
		\label{fig:nucleoprod}}
	\subfigure[]
	{\includegraphics[width=\textwidth]{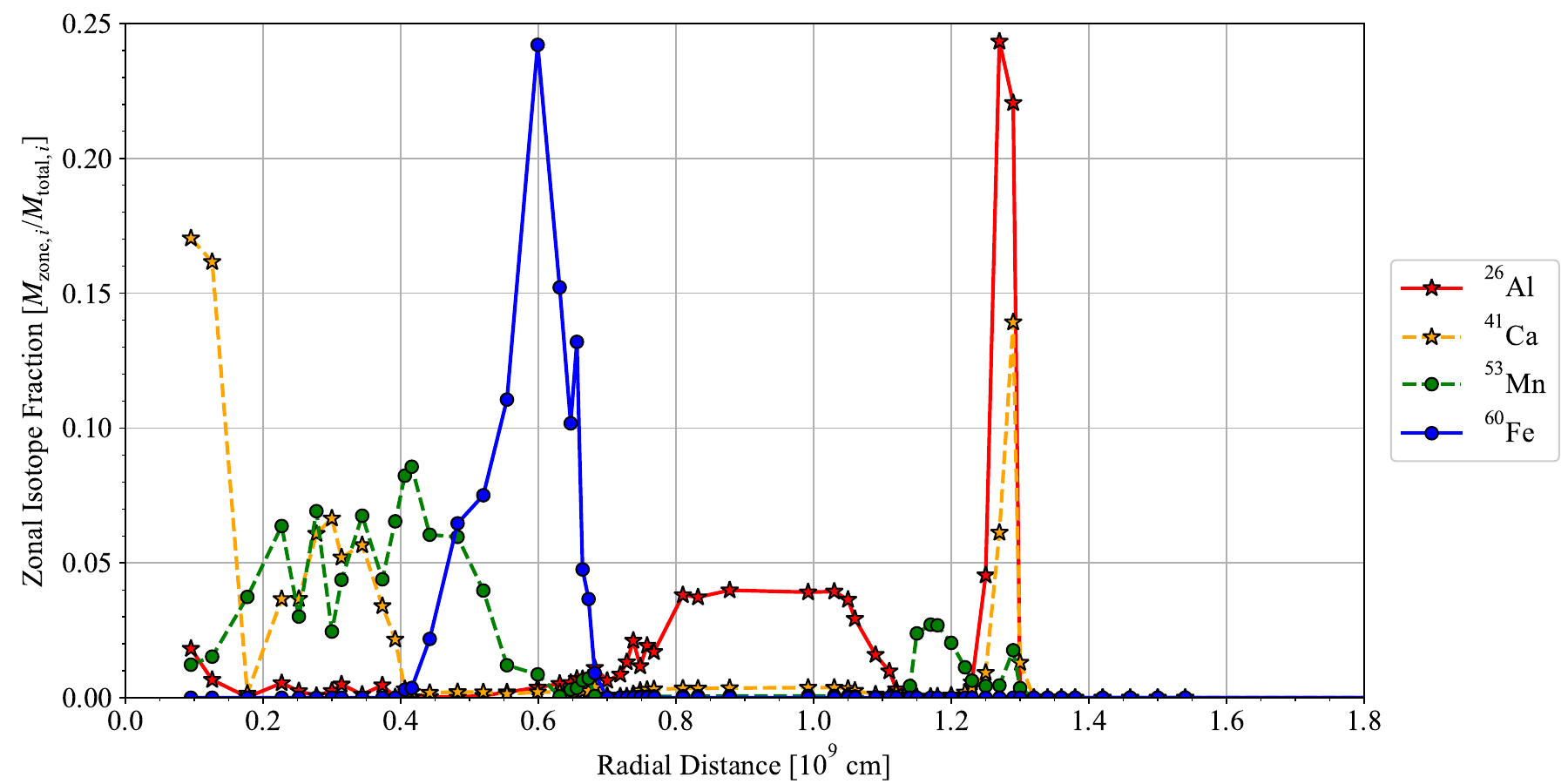}
		\label{fig:radioprod}} \\
	\caption[Nucleosynthesis products within each zone.]{Nucleosynthesis products within each zone.  The upper panel shows the mass fractions for the main stable nucleosynthesis products, and the lower panel shows the relative distribution of the main radioisotopes.  Note that the mass fractions for each element are stacked, not absolute. To find the absolute mass fraction for an element, subtract the value of the element plotted just below it.  By comparing the concentrations of the radioisotopes in panel (a) to the stable products in panel (b), we can estimate the dust molecules into which the radioisotopes are most likely to condense.
		\label{fig:products}
	}
\end{figure*}

We first examine the influences acting on the dust grain.  The radioisotopes will be formed deep within the ejecta; as the SNR expands, the ejecta will cool and overdensities in the ejecta will form clouds \citep[also referred to as clumps,][]{silv10,silv12}.  This will begin $\sim 1-5$ yr after the SN.  The radioisotopes will chemically bond with the surrounding elements forming molecules first, then combining to form larger and larger grains.  We assume our primary radioisotopes (\fe60, \al26, \mn53, \ca41) will form compounds like their stable isotopes.  In the case of \fe60, the bulk of which forms in a primarily Fe-Ni region (see Figure~\ref{fig:products}), we assume it will condense and form into metallic Fe grains rather than silicates, oxides, or sulfides since the associated elements are not present in that region of the ejecta.  Conversely, \al26, which is created in an O-rich region, will likely form into AlO and Al$_{2}$O$_{3}$ molecules, and some of the \mn53 will likely form MnS since it is created in an S-rich region \citep{field75}.  Although mixing due to Rayleigh-Taylor instabilities is present, this mixing is macroscopic not microscopic, meaning the composition choice of our grains will not be affected.  This is supported observationally by examinations of Cassiopeia A \citep{douv99} and discussed in detail in \citet{cher10,cher11}.  Knowing the type of compound the radioisotope resides in is important since different compounds have different densities, are more/less resistant to erosion, and absorb/emit electrons and photons differently.

The grains will continue to grow until the elemental products run out, the density drops too low, or the RS arrives.  Up to this point, the ejecta gas, overdense clouds, and the dust grains within have been traveling together with negligible relative velocities.  The RS will then slow and heat the gas and send a shock wave through the cloud, crushing it and shattering some of the dust grains.  Several studies have examined this process \citep{silv10,silv12,bisc16}, and our examination of the dust grains will begin just after this processing.  The cloud containing the dust grains will dissipate, and the dust grains will be exposed to the hot SNR plasma.  Because of their high mass compared to the surrounding gas, the dust grains will have decoupled from the plasma and will be moving with a large velocity relative  to the plasma.  A number of influences will now act on the dust grain \citep[see ][and references therein]{dwek92}, and we will now examine the most important processes in greater detail.  We assume the grains to be spherical in shape with radius, $a_{\rm gr}$, and uniform in composition.

To track the trajectory of a dust grain within the SNR, we solve a system of 7 ordinary differential equations:
\begin{align}
\frac{d\boldsymbol{r}_{\rm gr}}{dt} &= \boldsymbol{v}_{\rm gr} \nonumber \\[6pt]
\frac{d\boldsymbol{p}_{\rm gr}}{dt} &= \sum_{i} \boldsymbol{F}_{i}(a_{\rm gr}, q_{\rm gr}, \rho, T, ...) \nonumber \\[6pt]
\frac{da_{\rm gr}}{dt} &= \sum_{i} {\cal N}_{i}(a_{\rm gr}, q_{\rm gr}, \rho, T, ...)
\end{align}
where the summed processes, $i$, are dependent on the grain properties (size, charge, etc.) and the SNR environment (density, temperature, etc.).  In addition, we solve for the grain charge analytically:
\beq
q_{\rm gr} = q_{\rm gr}(a_{\rm gr}, v_{\rm gr}, \rho, T, ...) \, ,
\eeq
Our specific initial grain conditions will be given in \S \ref{sect:results} but, in qualitative terms, we follow our dust grains from time $t_{0}$, which corresponds to the time the RS passes the location of the grain.  While we expect dust grains to reside within overdense cloud, we assumed our dust grains to be at the leading edge of the cloud at the arrival of the RS.  The dust grains are immediately exposed to the hot, SNR environment rather than being relatively protected until the cloud dissipates.\footnote{It is expected the cloud dissipates in approximately 3 cloud crushing times, $\sim 3 \tau_{\rm cc}$ where $\tau_{\rm cc} = (\rho_{\rm cloud}/\rho_{\rm ICM})^{1/2} (a_{\rm cloud}/v_{\rm RS})$ \citep{klein94}.  Typically, the ratio of the cloud to the intercloud medium (ICM) densities is $(\rho_{\rm cloud}/\rho_{\rm ICM}) \sim 100-1000$ with a cloud size of $a_{\rm cloud} = 10^{16}$ cm \citep{silv10,silv12}.  In our simulation, the speed of the RS is $\sim 175$ km s$^{-1}$, giving a cloud dissipation time of $3\tau_{\rm cc} \sim 540-1700$ yr.  The dust grain would largely retain its initial velocity through the passage of the RS until the cloud has dissipated}  This represents the worst case scenario for dust grain survival.  The initial velocity, $v_{0}$, is determined by the velocity of the surrounding gas at the time of condensation.  We assume that dust condensation took place over 100-1000 days after the explosion \citep{sara15,slud18,cher17}.\footnote{Because of Free Expansion and the fact that we are not modeling grain growth, our results are not sensitive to the condensation time.}  The initial position of the grain, $r_{0}$, is the location of the grain at the time the RS passes, and the initial grain size, $a_{0}$, will be the post-RS/post-shattering size (in case the RS caused any grain shattering).  Since we do not model grain growth, we examine a range of sizes.  Also, we do not include grain-grain interactions \citep[e.g., impacts and charge influences;][found that collisions were rare, with a collisional timescale of $8 \times 10^{5}$ yr, roughly the length of our simulation timescale]{bocc16} or grain heating \citep[][found grain sublimation to be negligible]{bian07,noz07,bocc16}.  Lastly, the initial charge will effectively be zero, $q_{0} = 0$, since the grains are formed in a cool, dense cloud, and they will quickly charge/discharge to equilibrium (see \S \ref{subsect:graincharging}).

\subsection{Grain Dynamics}
\label{subsect:graindynamics}

\subsubsection{Drag Force}
\label{subsubsect:drag}

Because the dust grains have a relative velocity compared to the surrounding plasma, they will experience drag.  Drag will be due to collisions with plasma particles and, since the plasma is ionized and the grains charged (see \S \ref{subsect:graincharging}), Coulomb drag (also called plasma drag) may also be relevant.  The combined drag of both sources is given by \citep{drain79,drain11}:
\begin{align}
F_{\rm drag} =&~2 \pi a_{\rm gr}^{2} k_{\rm B}T \nonumber \\[6pt] 
&\times \left\lbrace \sum_{j} n_{j} \left[ G_{0}\left( s_{j} \right) + Z_{j}^{2} \Phi^{2} \ln \left[ \Lambda / Z_{j} \right] G_{2}\left( s_{j} \right) \right] \right\rbrace \, ,
\end{align}
(See Appendix \ref{app:graindynamics} for descriptions of the variables.)

\subsubsection{Motion of a charged dust grain in a magnetic field}
\label{subsubsect:mag}

Because there will be magnetic fields present within the SNR and the grain will be charged, we include the Lorentz force on the grain:
\beq
\boldsymbol{F}_{\rm mag} = \frac{q_{\rm gr}}{c} \boldsymbol{v}_{\rm rel} \times \boldsymbol{B} \, .
\eeq
Because of flux-freezing, the magnetic field, $\boldsymbol{B}$, will be moving with the plasma, so we use the grain's velocity relative
to the plasma, $\boldsymbol{v}_{\rm rel} = \boldsymbol{v}_{\rm gr} - \boldsymbol{v}_{\rm plasma}$.

Under the influence of the Lorentz force, charged dust grains in a magnetic field will spiral around magnetic field lines \citep{north84}.  The radius of this spiraling, $R_{\rm gyro}$, is given by \citep{murr04}:
\begin{align}
R_{\rm gyro} =&~9.75 \times 10^{-4} \ {\rm pc} \pfrac{\rho_{\rm gr}}{7.87 \ {\rm g \ cm^{-3}}} \pfrac{172.5 \ {\rm V}}{|U_{\rm gr}|} \nonumber \\[6pt]
&\times \pfrac{1 \ \mu{\rm G}}{B} \pfrac{v_{\perp, \rm gr}}{175 \ {\rm km \ s^{-1}}} \pfrac{a_{\rm gr}}{0.1 \ \mu{\rm m}}^{2} \, ,
\label{eq:gyrodist}
\end{align}
where $\boldsymbol{B}$ is the magnetic field, $\rho_{\rm gr}$ is the mass density, $\boldsymbol{v}_{\perp, \rm gr}$ is the velocity perpendicular to the magnetic field, $a_{\rm gr}$ is the radius, and $U_{\rm gr}$ is the potential of the grain.  Additionally, the period of this spiraling can be determined:
\begin{align}
\tau_{\rm gyro} =&~34 \ {\rm yr} \pfrac{\rho_{\rm gr}}{7.87 \ {\rm g \ cm^{-3}}} \pfrac{172.5 \ {\rm V}}{|U_{\rm gr}|} \nonumber \\[6pt]
&\times \pfrac{1 \ \mu{\rm G}}{B} \pfrac{a_{\rm gr}}{0.1 \ \mu{\rm m}}^{2} \, ,
\label{eq:gyrotime}
\end{align}

In the case of a magnetic field with varying magnitude, the spiraling dust grain will conserve adiabatic quantities \citep[see e.g., \S 12.5,][]{jack98}.  Of particular interest for our examination is the adiabatic invariant $p_{\perp, \rm gr}^{2}/B$, where $p_{\perp, \rm gr}$ is the momentum of the grain perpendicular to the magnetic field and the parallel component $v_\parallel = \sqrt{v^2 - v_{\perp}^{2}}$, where $v$ is the total velocity.  Since magnetic fields do not perform work on the grain, we know the speed of the grain at later times will be the same as when it entered the field, $v = v_{\rm ini}$.  If the magnetic field increases with position, $B(r)$, then by the adiabatic invariance:
\beq
\frac{v_{\perp}^{2}}{B(r)} = \frac{v_{\perp,\rm ini}^{2}}{B_{\rm ini}} \, , \nonumber
\eeq
\beq
\Rightarrow v_{\parallel}^{2} = v_{\rm ini}^{2} - v_{\perp,\rm ini}^{2} \frac{B(r)}{B_{\rm ini}} \, .
\label{eq:magrelation}
\eeq
As the magnetic field increases, $v_{\perp}$ will increase, which means $v_{\parallel}$ will decrease in order to maintain the original speed of the grain.  There will be a position, $R_{\rm bounce}$, where the right side of Equation~\ref{eq:magrelation} vanishes, and the grain's movement along the magnetic field lines will reverse direction.  Essentially, the grain will ``bounce'' off the stronger magnetic field. This is referred to as a magnetic mirror in \cite{jack98}, and leads to the `pinball' behavior we discuss in \S \ref{sect:results}-\ref{sect:conclusions}.

We can find an expression for the strength of the magnetic field able to bounce a dust grain assuming $\tau_{\rm gyro} \ll \tau_{\rm sputtering}$ (see Equation~\ref{eq:sputtertime}).  At bounce, $B_{\rm bounce} \equiv B(R_{\rm bounce})$:
\beq
v_{\rm ini}^{2} = v_{\perp,\rm ini}^{2} \frac{B_{\rm bounce}}{B_{\rm ini}} \, ,
\eeq
and if we consider an average case ($v_{\perp} \sim v_{\parallel}$):
\beq
\left\langle v_{\perp,\rm ini} \right\rangle \sim \frac{1}{\sqrt{2}} v_{\rm ini} \, ,
\eeq
then the magnetic field at bounce is:
\beq
v_{\rm ini}^{2} \approx \frac{1}{2} v_{\rm ini}^{2} \frac{B_{\rm bounce}}{B_{\rm ini}}
\eeq
\beq
\Rightarrow B_{\rm bounce} \approx 2 B_{\rm ini} \, .
\eeq
If the magnetic field varies with some characteristic length scale, $\lambda_{\rm mag}$, then when $R_{\rm gyro} \lesssim \lambda_{\rm mag}$, the grain will be ``captured'' by the magnetic field (i.e., the grain spirals around the magnetic field lines).  After capture, if the magnetic field strength doubles, the dust grain will be reflected.  We will see this effect is relevant when considering dust grains encountering shocked ISM material within an SNR.

\subsection{Grain Sputtering}
\label{subsect:grainsputtering}

In addition to drag from the grains' high velocity relative to the plasma, the grains will also be eroded/sputtered by impacts with plasma particles.  In addition to kinetic sputtering from bulk motion of plasma onto the grains, at high temperatures the thermal velocities of plasma particles will also erode the grains.  Because of high relative velocities and high temperatures within the SNR, we include both kinetic and thermal sputtering.  The erosion rate due to sputtering (both kinetic and thermal) is given by \citep{dwek92}, and we use the approach by \citet{noz06,bisc16}:
\begin{align}
\frac{da_{\rm gr}}{dt} =&~-\frac{m_{\rm sp}}{4 \rho_{\rm gr}} \sum_{j} \frac{n_{j}}{s_{j}} \pfrac{8k_{\rm B}T}{\pi m_{j}}^{1/2} \exp\left[ -s_{j}^{2}\right] \nonumber \\[6pt]
&\times \int d\epsilon_{j}~\sqrt{\epsilon_{j}} \exp\left[-\epsilon_{j}\right] \sinh \left( 2 s_{j} \sqrt{\epsilon_{j}} \right) Y_{j}^{0}(\epsilon_{j}) \, ,
\end{align}
(See Appendix \ref{app:grainsputtering} for descriptions of the variables.)

For our grain parameters, the sputtering time for dust grains is \citep{drain11}:
\begin{align}
\tau_{\rm sputtering} =&~10^{6}~{\rm yr} \left[ 1 + \pfrac{10^{6}~{\rm K}}{T}^{3} \right] \pfrac{a_{\rm gr}}{0.1~{\rm \mu m}} \nonumber \\[6pt]
&\times \pfrac{0.1~{\rm cm}^{-1}}{n_{\rm ISM}}
\label{eq:sputtertime}
\end{align}

\subsection{Grain Charging}
\label{subsect:graincharging}

As grains move within the SNR, they will acquire/lose electrons and ions due to impacts with the plasma and photons.  Several processes can influence the total charge of the grain, so the total charging rate, $dq_{\rm gr}/dt$ is:
\beq
\frac{dq_{\rm gr}}{dt} = \sum_{j} {\cal I}_{j} \, ,
\eeq
which is summed over $j$ processes of currents, ${\cal I}_{j}$.  These currents are due to impinging plasma particles, ${\cal I}_{\rm imp}$, and the associated secondary electrons emitted, ${\cal I}_{\rm see}$, transmitted plasma particles, ${\cal I}_{\rm trans}$, and photoelectron emission, ${\cal I}_{\gamma}$.  The derivations are the same as used by \citet{kim98}. However, these derivations are very computationally intensive (see Appendix \ref{app:graincharging} for an in-depth discussion).  In order simplify calculations, we employ an analytic description of the charging processes.

If we compare the gyro period given in Equation~\ref{eq:gyrotime} with a basic approximation for the charging time from accumulating/dispersing electrons/ions \citep{shuk02}:
\begin{align}
\tau_{\rm charge, {\it e/i}} =&~0.06 \ {\rm yr} \sqrt{\frac{T}{10^{6} \ {\rm K}}} \pfrac{0.1 \ {\rm cm^{-3}}}{n_{\rm ISM}} \pfrac{0.1 \ \mu{\rm m}}{a_{\rm gr}} \nonumber \\[6pt]
&\times \pfrac{1}{1+\sqrt{\frac{m_{p}}{m_{e}}}\exp\left[-\pfrac{\Phi}{2} \right]} \, ,
\end{align}
we can see that the charging time is much less than the gyro period ($\tau_{{\rm charge, }e} \ll \tau_{\rm gyro}$), allowing us to use an analytic approximation of the grain charge when solving for the grain's gyroscopic motions (this means that we do not solve for $dq_{\rm gr}/dt$ in our system of ordinary differential equations).  In order to employ a faster description of grain charging processes, we apply an analytic approach introduced by \citet{shull78} and extended by \citet{mckee87}; in this approach we solve numerically for the steady-state value of the grain potential at various plasma temperatures ($T$), relative velocities ($v_{\rm rel}$), and grain radii ($a_{\rm gr}$), then fit a function to the results.  It should be noted that this approach inherently ignores the cooling/heating history of the grains and the grain potential will be single-valued at a given temperature \citep[for more information, see][]{meyer82,hora90}.  However, because of Fe's low secondary electron emission yield ($\delta_{\rm max, Fe} = 1.3 < 6$) our Fe grains should have single-valued potentials across all temperature values.\footnote{From \citet{hora90}, the presence of multi-valued potentials occurs for substances with $\delta_{\rm max} \gtrsim 6$.}

For a more detailed description of our analytic approximation, see Appendix \ref{app:chargingapprox}.

\section{SNR Model Description}
\label{sect:model}

\begin{figure*}[t]
	\centering
	\subfigure[]
	{\includegraphics[width=\textwidth]{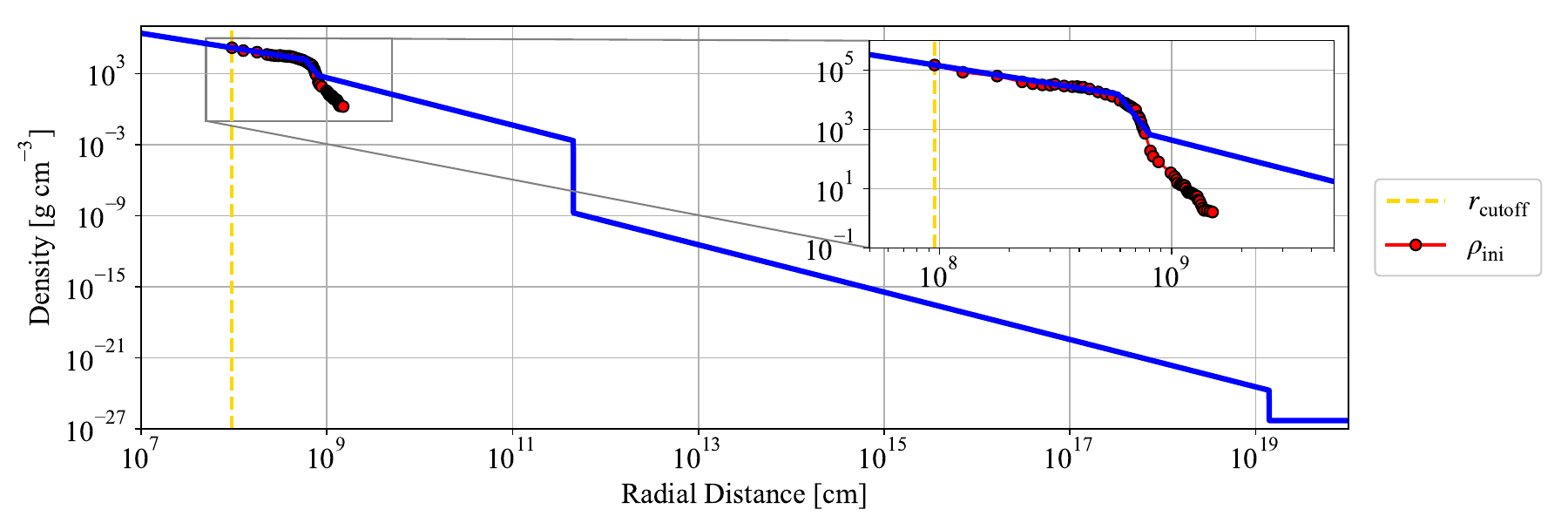}
		\label{fig:initdensity}} 
	\subfigure[]
	{\includegraphics[width=\textwidth]{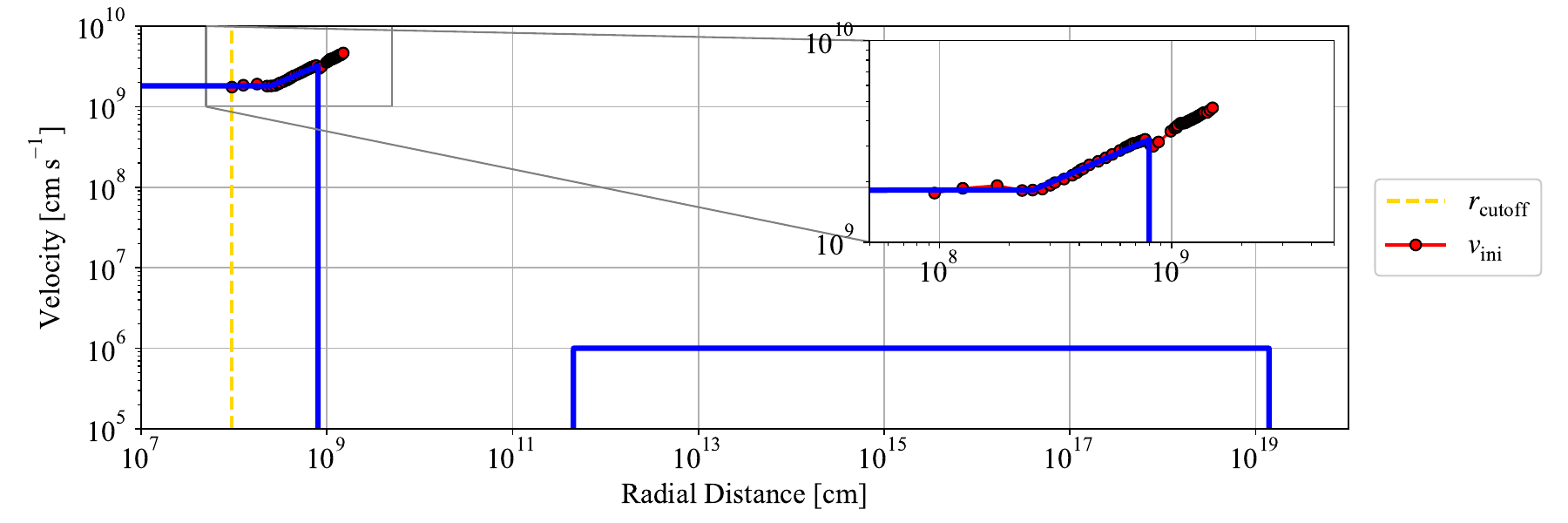}
		\label{fig:initvelocity}} 
	\subfigure[]
	{\includegraphics[width=\textwidth]{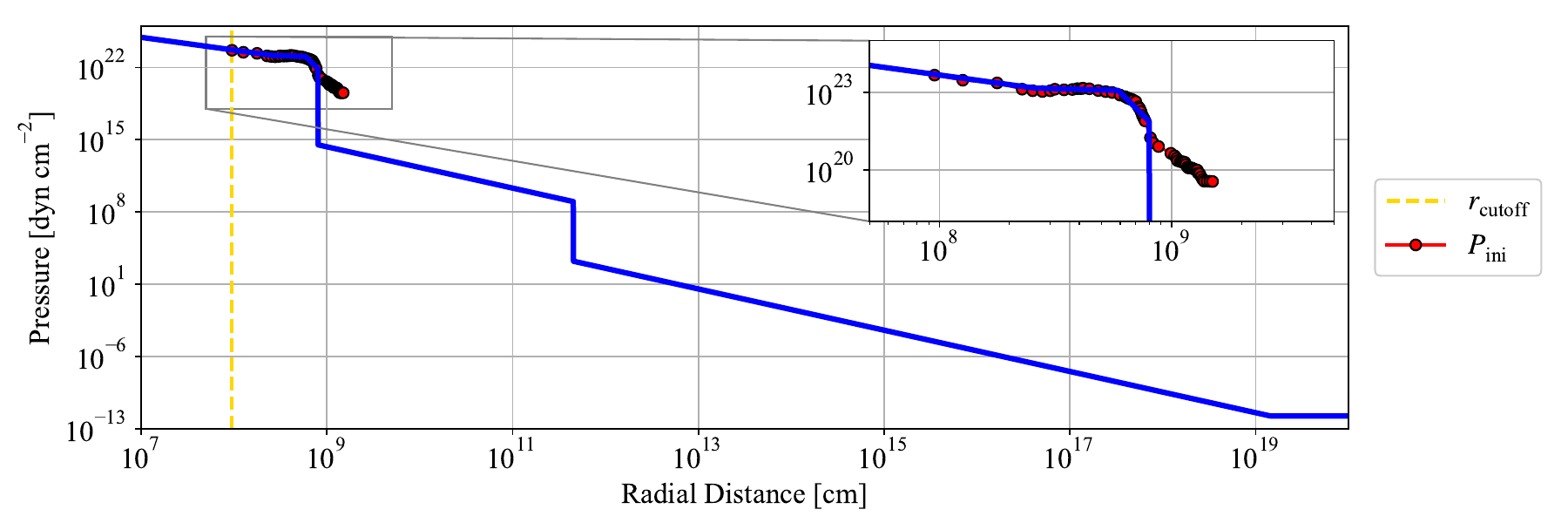}
		\label{fig:initpressure}}
	\caption[Initial density, velocity and pressure profiles.]{Initial density, velocity and pressure profiles.  Our initial profiles are shown in blue with the azimuthally-averaged \citet{wana13b} ECSN results plotted in red.  Because our model includes an outer envelope \citep[similar to][]{jank08}, the envelope profile contains some of the nucleosynthesis products.  The composition of the envelope was adjusted to include the nucleosynthesis products' mass, but these products were given the initial density, velocity, and pressure of the envelope.
		\label{fig:inithydro}
	}
\end{figure*}

We need to model the SNR's density, $\rho$, velocity, $\boldsymbol{v}$, pressure, $P$, temperature, $T$, and magnetic field, $\boldsymbol{B}$, and use the magnetohydrodynamics (MHD) equations as the framework.  Since we are dealing with an SN explosion, the relative thermal velocities between particles will be small compared to the bulk velocity (meaning thermal conduction can be ignored), the plasma is collisionless (meaning resistivity and Ohmic heating due to electron-ion collisions can be ignored), and the ejecta velocity is radial and much greater than the escape velocity ($v_{\rm ej} \gg v_{\rm esc}$) for the central compact object (meaning Coriolis and gravitational effects can be ignored).  Additionally, if we limit our examination to the early phases of the SNR expansion (the Free Expansion and Sedov-Taylor phases), we can ignore radiative effects.  Thus we use the ideal MHD equations (in Lagrangian form and cgs/esu units):
\begin{align}
\frac{D\rho}{Dt} &= -\rho \boldsymbol{\nabla} \cdot \boldsymbol{v}  & {\rm (Mass)} \label{eq:MHDmass} \, , \\[6pt]
\frac{D\boldsymbol{v}}{Dt} &= -\frac{1}{\rho} \boldsymbol{\nabla} P + \frac{\boldsymbol{J}}{\rho c} \times \boldsymbol{B} & {\rm (Momentum)} \label{eq:MHDmomentum} \, , \\[6pt]
\frac{DP}{Dt} &= -\Gamma P \boldsymbol{\nabla} \cdot \boldsymbol{v} & {\rm (Energy)} \label{eq:MHDenergy} \, , \\[6pt]
\frac{D\boldsymbol{B}}{Dt} &= \left( \boldsymbol{B} \cdot \boldsymbol{\nabla} \right) \boldsymbol{v} - \boldsymbol{B} \left( \boldsymbol{\nabla} \cdot \boldsymbol{v} \right) & {\rm (Induction)} \, ,
\label{eq:MHDflux}
\end{align}
where $\frac{D}{Dt} \equiv \frac{\partial}{\partial t} + \boldsymbol{v} \cdot \boldsymbol{\nabla}$, $\boldsymbol{J} = \frac{c}{4 \pi} \boldsymbol{\nabla} \times \boldsymbol{B}$, and $\Gamma = 5/3$ for the medium we study.  Additionally, we assume a polytropic, Newtonian fluid with $P = (\Gamma - 1) \mathbb{E}$ where $\mathbb{E}$ is the energy density.  Expanding the momentum equation (Equation~\ref{eq:MHDmomentum}), we find:
\begin{align}
\frac{D\boldsymbol{v}}{Dt} &= -\frac{1}{\rho} \boldsymbol{\nabla} P + \frac{1}{\rho c} \pfrac{c}{4 \pi} \left( \boldsymbol{\nabla} \times \boldsymbol{B} \right) \times \boldsymbol{B} \\[6pt]
&= -\frac{1}{\rho} \boldsymbol{\nabla} P + \frac{1}{4 \pi \rho} \left( \left( \boldsymbol{B} \cdot \boldsymbol{\nabla} \right) \boldsymbol{B} - \frac{1}{2} \boldsymbol{\nabla} B^{2} \right) \, .
\end{align}
The first term in the parentheses, $\left( \boldsymbol{B} \cdot \boldsymbol{\nabla} \right) \boldsymbol{B}$, represents the magnetic tension \citep[$\sim B^{2}/4\pi$,][]{galt16}\footnote{This term is sometimes referred to as ``magnetic stress'' \citep[see e.g.,][]{david01}.} and the second term, $\frac{1}{2} \boldsymbol{\nabla} B^{2}$, represents magnetic pressure $\sim B^{2}/8\pi$.  In a typical SNR during the Free Expansion and Sedov-Taylor phases, expanding at $\sim 200$ km s$^{-1}$ with a peak density of $4 m_{H}$ cm$^{-3}$, the ram pressure is $P_{\rm ram} = \rho v^{2} \sim 10^{-9} \ {\rm dyne~cm}^{-2}$.  In contrast, for a typical ISM magnetic field ($B_{\rm ISM} \sim 1~\mu$G) the magnetic tension and pressure are much weaker, $P_{\rm mag} \sim 10^{-13} \ {\rm dyne~cm}^{-2}$ (see \S \ref{subsect:magfield}).  Because of this, they can be ignored, and Eqs.~(\ref{eq:MHDmass}) - (\ref{eq:MHDenergy}) simplify to:
\beq
{\rm Basic~Fluid~Equations} 
\begin{cases}
	\displaystyle \frac{D\rho}{Dt} = -\rho \boldsymbol{\nabla} \cdot \boldsymbol{v} \, , \\[12pt]
	\displaystyle \frac{D\boldsymbol{v}}{Dt} = -\frac{1}{\rho} \boldsymbol{\nabla} P \, , \\[12pt]
	\displaystyle \frac{DP}{Dt} = -\Gamma P \boldsymbol{\nabla} \cdot \boldsymbol{v} \, .
\end{cases}
\eeq
These are the basic fluid equations we use, where we see that during the early stages of SNR evolution, the expansion can be determined without including the magnetic field influence.

\subsection{Nucleosynthesis Products}
\label{subsect:nucleo}

\citet{fry15} found that an ECSN was the most likely source for the \fe60 signal.  This determination was based on the nucleosynthesis results in \citet{wana09, wana2013a, wana13b}, so we use the results of these nucleosynthesis simulations for several aspects of our model.  First, the results of the nucleosynthesis simulations are the initial conditions for our hydrodynamics simulations.  The nucleosynthesis results also describe the composition of the ejecta, allowing us to determine the types and concentrations of elements interacting with the grains as they transit the SNR.  Lastly, the nucleosynthesis results give the initial positions of the radioisotopes within the ejecta.  This allows us to determine the initial velocities and types of grains that will mostly likely be formed containing specific radioisotopes \citep{sara13,sara15,bisc14,slud18}.  The major nucleosynthesis products and radioisotopes are shown with their initial positions in Figure~\ref{fig:products}.

We have assumed for definiteness an ECSN progenitor with a ZAMS mass of 8.8 $\msol$, leaving a 1.363 $\msol$ neutron star, and ejecting a 0.014 $\msol$ inner core that contains the SN synthesized products and a 1.249 $\msol$ outer envelope composed of 70\% Hydrogen and 30\% Helium (we assume large-scale convection that thoroughly mixes the envelope) giving a total ejecta mass of $M_{\rm ej} = 1.263~\msol$.  This is similar to the treatment by \citet{jank08} and \citet{hoff08}.  The envelope is assumed to be in hydrostatic equilibrium and a single isothermal sphere with a temperature of 3500 K and completely ionized ($\mu_{\rm envel} = 0.61$) (for more detail see \S \ref{subsect:hydroinit}).  The energy delivered as kinetic energy into the ejecta is $E_{\rm SN} = 1.5 \times 10^{50}$ erg \citep{wana09}.

\subsection{Hydrodynamic Initial Conditions}
\label{subsect:hydroinit}

The density, velocity, and pressure profiles are based on the \citet{wana13b} nucleosynthesis results, the expected configuration of an ECSN progenitor, and the properties of the Local Bubble at the time of the SN.  The ejecta is divided into an inner core region and an outer shell region based on the \citet{wana13b} results.  The two-dimensional \citet{wana13b} results contained values at different azimuths and radii, so we averaged the values across azimuths, and fit a power law profile to the density and velocity averages.  A comparison of the averaged results and fits are shown in Figures~\ref{fig:initdensity} and \ref{fig:initvelocity}; note that the core cutoff positions for density and velocity ($r_{\rm core, 1}$ and $r_{\rm core, 2}$, respectively) are slightly different in order to provide a better fit.

The progenitor of an ECSN is assumed to be a super-AGB star \citep{smar09,woos15} with an envelope that has been completely mixed due to its thermal-pulsing (TP-SAGB) phase \citep{herw05,poel08,pumo09,pumo10,jones16}.  This implies that the progenitor will be a red supergiant and have an extended, isothermal envelope.  We chose an envelope temperature of 3500 K \citep[the approximate surface temperature of Betelgeuse;][]{frey02} with $\rho \propto r^{-2}$ and $v = 0$.  The edge of the envelope, $r_{\rm envel}$ marks the edge of the progenitor, and we assume the presence of a pre-SN wind.  Several studies have examined the wind and mass loss during the TP-SAGB phase \citep[see e.g.,][and references therein]{doh13}, and we assumed a mass loss of $\dot{M}_{\rm wind} = 7 \times 10^{-5}~\msol~{\rm yr}^{-1}$ and a wind velocity of $v_{\rm wind} = 10$ km s$^{-1}$.

The pre-SN stellar wind will extend until the ram pressure of the wind, $P_{\rm wind} = \rho v_{\rm wind}^{2}$, equals the pressure of the ISM, $P_{\rm ISM}$ \citep{cast75}.  Because the Local Bubble shows evidence of multiple SNe \citep[e.g.,][and references therein]{breit16}, we assume the source of the \fe60 signal to be the most recent SN and that this SN would have occurred in a region similar to that currently observed in the Local Bubble.  Thus we assumed ISM values of:  $n_{\rm ISM} = 0.005 \ {\rm cm}^{-3}$, $\mu_{\rm ISM} = 0.61$, $P_{\rm ISM} = 1.8 \times 10^{-12} \ {\rm dyn~cm}^{-2}$ \citep{faj08} and $T_{\rm ISM} = 2.6 \times 10^{6}$ K.

Combining these parameters, we assume the following initial conditions for our hydrodynamic simulations:
\beq
\rho(r) = 
\begin{cases}
	\rho_{\rm core}(r) & \quad r_{\rm cutoff} \leq r < r_{\rm core,2} \, , \\[6pt]
	\rho_{\rm shell}(r) & \quad r_{\rm core,2} \leq r < r_{\rm shell} \, , \\[6pt]
	\rho_{\rm envel}(r) & \quad r_{\rm shell} \leq r < r_{\rm envel} \, , \\[6pt]
	\rho_{\rm wind}(r) & \quad r_{\rm envel} \leq r < r_{\rm wind} \, , \\[6pt]
	\rho_{\rm ISM} & \quad r_{\rm wind} \leq r  \, ;\\
\end{cases}
\eeq
where
\begin{align}
\rho_{\rm core}(r) &= 1.51 \times 10^{5}~{\rm g~cm^{-3}} \pfrac{r}{9.52 \times 10^{7}~{\rm cm}}^{-5/4} \, , \nonumber \\[6pt]
\rho_{\rm shell}(r) &= 1.75~{\rm g~cm^{-3}} \pfrac{r}{1.46 \times 10^{9}~{\rm cm}}^{-10}  \, , \nonumber \\[6pt]
\rho_{\rm envel}(r) &= 743.67~{\rm g~cm^{-3}} \pfrac{r}{7.68 \times 10^{8}~{\rm cm}}^{-2} \, , \nonumber \\[6pt]
\rho_{\rm wind}(r) &= \frac{\dot{M}_{\rm wind}}{4 \pi v_{\rm wind} r^{2}} \, , \nonumber \\[6pt]
\rho_{\rm ISM} &= \mu_{\rm ISM} m_{H} n_{\rm ISM} \, ;
\end{align}
and
\begin{align}
r_{\rm cutoff} &= 9.52 \times 10^{7} \ {\rm cm} \nonumber \\[6pt]
r_{\rm core, 2} &= 5.88 \times 10^{8} \ {\rm cm} \nonumber \\[6pt]
r_{\rm shell} &= 8.01 \times 10^{8} \ {\rm cm} \nonumber \\[6pt]
r_{\rm envel} &= 4.51 \times 10^{11} \ {\rm cm} \nonumber \\[6pt]
r_{\rm wind} &= 1.40 \times 10^{19} \ {\rm cm} \, ;
\end{align}
\beq
v(r) = 
\begin{cases}
	v_{\rm core}(r) & \quad r_{\rm cutoff} \leq r < r_{\rm core,1} \, , \\[6pt]
	v_{\rm shell}(r) & \quad r_{\rm core,1} \leq r < r_{\rm shell} \, , \\[6pt]
	v_{\rm envel}(r) & \quad r_{\rm shell} \leq r < r_{\rm envel} \, , \\[6pt]
	v_{\rm wind} & \quad r_{\rm envel} \leq r < r_{\rm wind} \, , \\[6pt]
	v_{\rm ISM} & \quad r_{\rm wind} \leq r \, ; \\[6pt]
\end{cases}
\eeq
where 
\begin{align}
v_{\rm core}(r) &= 1.81 \times 10^{9}~{\rm cm~s^{-1}} \, , \nonumber \\[6pt]
v_{\rm shell}(r) &= 1.81 \times 10^{9}~{\rm cm~s^{-1}} \pfrac{r}{r_{\rm core, 1}}^{1/2}  \, , \nonumber \\[6pt]
v_{\rm envel}(r) &= 0~{\rm cm~s^{-1}} \, , \nonumber \\[6pt]
v_{\rm wind}(r) &= 10~{\rm km~s^{-1}} \, , \nonumber \\[6pt]
v_{\rm ISM} &= 0~{\rm cm~s^{-1}} \, ;
\end{align}
and $r_{\rm core, 1} = 2.52 \times 10^{8} \ {\rm cm}$;
\beq
P(r) = 
\begin{cases}
	P_{\rm core}(r) & \quad r_{\rm cutoff} \leq r < r_{\rm core,2} \, , \\[6pt]
	P_{\rm shell}(r) & \quad r_{\rm core,2} \leq r < r_{\rm shell} \, , \\[6pt]
	P_{\rm envel}(r) & \quad r_{\rm shell} \leq r < r_{\rm envel} \, , \\[6pt]
	P_{\rm wind}(r) & \quad r_{\rm envel} \leq r < r_{\rm wind} \, , \\[6pt]
	P_{\rm ISM} & \quad r_{\rm wind} \leq r \, ; \\[6pt]
\end{cases}
\eeq
\begin{align}
P_{\rm core}(r) &= \rho(r) v(r)^{2} \, , \nonumber \\[6pt]
P_{\rm shell}(r) &= \rho(r) v(r)^{2} \, , \nonumber \\[6pt]
P_{\rm envel}(r) &= \frac{\rho(r)k_{\rm B}T_{\rm envel}}{\mu_{\rm envel} m_{H}} \, , \nonumber \\[6pt]
P_{\rm wind}(r) &= \rho(r) v(r)^{2} \, , \nonumber \\[6pt]
P_{\rm ISM} &= P_{\rm ISM} \, .
\end{align}

\subsection{Hydrodynamic Simulations}
\label{subsect:hydrosim}

Our hydrodynamic simulations use the RT1D code written by \citet{duff16}.  This is a 1-D+, adaptive, moving mesh code that includes Rayleigh-Taylor instabilities in SNe.  Although the code is one-dimensional, it includes the multi-dimensional effects of Rayleigh-Taylor instabilities as source terms in the fluid equations, chosen to represent turbulent perturbations averaged over solid angle.  This enables simulations of non-radiative SNR expansion based on the basic fluid equations, that run from the Free Expansion through the Sedov-Taylor phases.  As noted by \citeauthor{duff16}, the incorporation of Rayleigh-Taylor instabilities eliminates the singularities inherent in the Chevalier-Nadezhin solutions and provides a more accurate position of the RS than that found by \citet{tm99}.  The simulations are run in characteristic units; these are dimensionless units of the hydrodynamic quantities (i.e., in characteristic units, density is given by $\rho^{*} = \rho/\rho_{\rm ch}$, where $\rho_{\rm ch}$ is the characteristic density for the SN environment).  The characteristic values' definitions and our adopted values are:
\begin{align}
r_{\rm ch} &\equiv \pfrac{M_{\rm ej}}{\rho_{\rm ISM}}^{1/3} &= &~25.6 \ {\rm pc} \, , \nonumber \\[6pt]
t_{\rm ch} &\equiv \frac{M_{\rm ej}^{5/6}}{\rho_{\rm ISM}^{1/3} E_{\rm SN}^{1/2}} &= &~10,200 \ {\rm yr} \, , \nonumber \\[6pt]
M_{\rm ch} &\equiv M_{\rm ej} &= &~1.263 \ \msol \, , \nonumber \\[6pt]
\rho_{\rm ch} &\equiv \rho_{\rm ISM} &= &~0.00305 m_{p} \ {\rm cm^{-3}} \, , \nonumber \\[6pt]
v_{\rm ch} &\equiv \frac{r_{\rm ch}}{t_{\rm ch}} &= &~2440 \ {\rm km~s^{-1}} \, , \nonumber \\[6pt]
P_{\rm ch} &\equiv \rho_{\rm ch} v_{\rm ch}^{2} &= &~3.05 \times 10^{-10} \ {\rm dyn~cm^{-2}} \, , \nonumber \\[6pt]
T_{\rm ch} &\equiv \frac{\mu_{\rm ej} m_{\rm u} P_{\rm ch}}{k_{\rm B} \rho_{\rm ch}} & & \nonumber \\[6pt]
		   &= \frac{\mu_{\rm ej} m_{\rm u}}{k_{\rm B}} v_{\rm ch}^{2} &= &~4.41 \times 10^{8} \ {\rm K} \, .
\end{align}

These definitions are based on \citet{tm99} and assume a uniform ambient medium \citep[for power-law, i.e., stellar wind, mediums see also:][]{tm99,lam03,haid16}.\footnote{The characteristic temperature, $T_{\rm ch}$, is more commonly defined as  $\displaystyle T_{\rm ch}=\pfrac{3}{16} \pfrac{\mu m_{p} v_{\rm ch}^{2}}{k_{\rm B}}$\citep{tm99,mckee91}.  This is a result of characteristic values defined based on jump conditions:  $\displaystyle P_{\rm ch} = \frac{\rho_{\rm ISM} v_{\rm ch}^{2}}{\Gamma+1} = \frac{\rho_{\rm ch} k_{\rm B}T_{\rm ch}}{\mu m_{p}}$, $\displaystyle \rho_{\rm ch} = \pfrac{\Gamma+1}{\Gamma-1} \rho_{\rm ISM}$ with $\Gamma = 5/3$ \citep[][pg 402, \S 36.2.5]{drain11}.  This description is a more accurate gauge of the SN, but since our use of characteristic units are confined to unit conversions, our conclusions will not be affected by the distinction.}  Since we are examining the SNR expansion both inside and outside the pre-SN stellar wind, we applied the uniform case since the forward shock is in a uniform medium for the vast majority of our simulation duration.  As noted in \S \ref{subsect:hydroinit}, our choices for $E_{\rm SN}$ and $M_{\rm ej}$ correspond to our adopted ECSN model.

The hydrodynamics simulations were begun at $t = 300$ ms ($t^{*} = 9.285 \times 10^{-13}$) after the core bounce, which corresponds to the end of the \citet{wana13b} results, and run through $t = 512$ kyr ($t^{*} = 50$).  This encompasses the entire Free Expansion and Sedov-Taylor phases.  The position values began at the cutoff radial position, $r_{\rm cutoff} = 9.52 \times 10^{7}$ cm ($r_{\rm cutoff}^{*} = 1.206 \times 10^{-12}$) and extended through the outermost radial position, $R = 25,600$ pc ($R^{*} = 1000$), and were initially divided into 1024 zones.  In the simulation run, results were generated at 1000 logarithmically spaced time intervals.  Each output includes the radial position of the zone's midpoint, zone radial width, density, velocity, pressure, mixture fraction (the fraction of the zone comprised of ejecta material), and the turbulent factor (which was a measure of the Rayleigh-Taylor fluctuations, for our purposes this was not used).  Because many of the results had nearly power-law profiles, 2-D linear spline interpolation functions were generated for the common logarithms of the SNR quantities (i.e., $\log{\rho^{*}}$, $\log{\left|v^{*}\right|}$, etc) done across $\log{t^{*}} \times \log{r^{*}}$.  The temperature interpolation was done with $\log{T^{*}} = \log{(P^{*}/\rho^{*})}$.

Additionally, with the assumption of spherical symmetry, the mass enclosed, $M_{\rm enclosed}$, by a sphere at a given radial position, $r$, is described by:
\beq
M_{\rm enclosed} = \int_{r_{\rm min}}^{r} 4 \pi \rho(r') r'^{2}~dr' \, ,
\eeq
where $r_{\min}$ is the innermost zone position in the RT1D simulation.  Given the enclosed mass, the average elemental composition of SNR can be determined at any point.  With the interpolation functions, we have the means to determine the density, velocity, temperature, and composition at all locations within the SNR.

The RT1D code allows simulations to be run with or without Rayleigh-Taylor instabilities included.  Using the initial conditions defined in \S \ref{subsect:hydroinit}, without Rayleigh-Taylor instabilities our dust grains encounter the RS after $\sim$8600 yr, but with Rayleigh-Taylor instabilities our dust grains encounter the RS much sooner, at $\sim$5000 yr.

\subsection{Magnetic Field}
\label{subsect:magfield}

\begin{figure*}[t]
	\begin{center}
		\includegraphics[width=\textwidth]{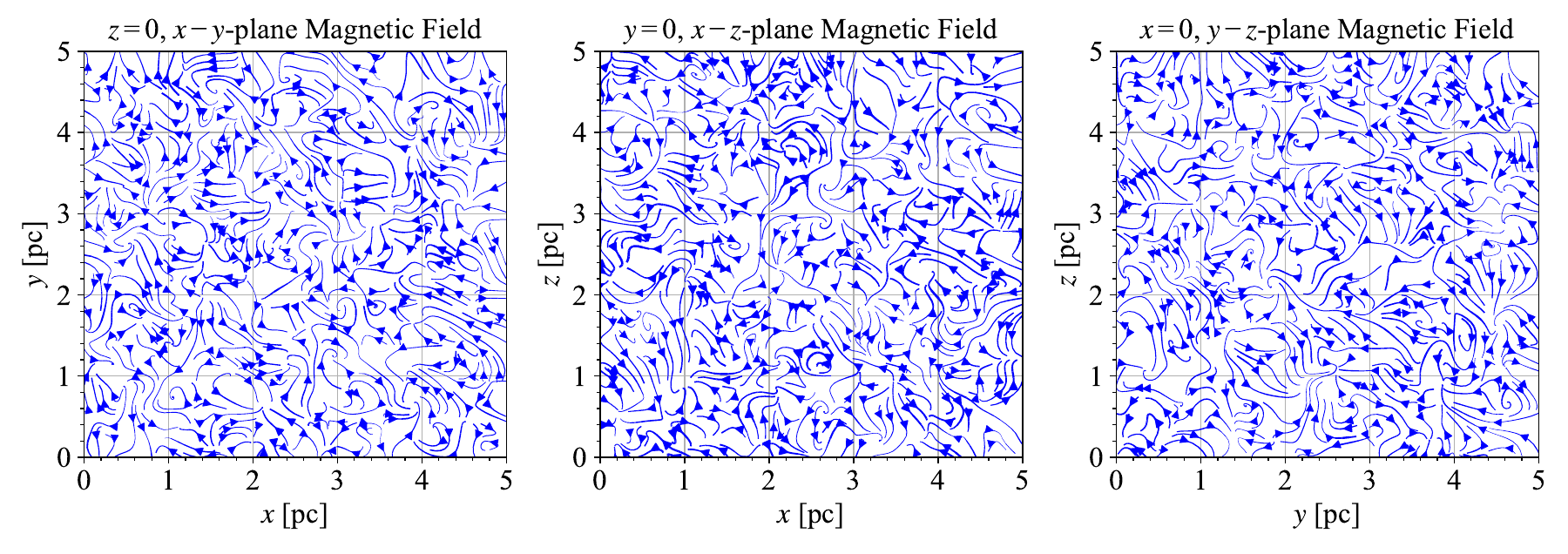}
		\caption{Sample turbulent field.  The energy spectrum uses a Kolmogorov profile (${\cal E}_{k} \sim k^{-5/3}$) with $\mathbb{C} = 1.144 \times 10^{-13}$ G$^{2}$ pc$^{17/3}$, $\lambda_{\rm outer} = 5$ pc, and $\lambda_{\rm inner} = 0.05$ pc.
			\label{fig:magdiv}
		}
	\end{center}
\end{figure*}

In our simulation we split the magnetic field into three regions:  the star/ejecta field, the stellar wind field, and the ISM field.  In the star/ejecta region, we expect a surface field $B \sim 1$ G \citep[this is the average surface field for Betelgeuse, which is similar in mass to our expected progenitor,][]{petit13}.  In addition, massive stars generally seem to exhibit weak surface magnetic fields \citep[$B \lesssim 1$ G,][]{aug19}.  From magnetic flux conservation \citep[e.g.,][]{padmV2}, $B \approx B_{\rm ini} \left( r_{\rm ini}/r \right)^{2}$, and if we estimate that the dust will decouple from the plasma at $\sim 1$ pc (a \emph{very} rough estimate) then the stellar field will have weakened to: $B \sim 10^{-8} \ \mu$G ($\Rightarrow R_{\rm gyro} \sim 100$ kpc $> \lambda_{\rm Milky~Way}$, see \S \ref{subsubsect:mag}).  Furthermore, because the stellar wind magnetic field is the surface field stretched even further due to flux-freezing within the pre-SN wind, we expect the stellar wind magnetic field to be weak as well.  With this reasoning, we set $B_{\rm star} = B_{\rm wind} = 0$.

For the ISM magnetic field, because we focus our examination within the Local Bubble, the site of multiple SNe, we assume the initial magnetic field to be nonuniform and weakened compared to the average ISM \citep[$\left\langle \left| \boldsymbol{B}_{\rm ISM} \right| \right\rangle \sim 3.57 \ \mu{\rm G}$ and $\sqrt{\left\langle B_{\rm ISM}^2 \right\rangle} \approx 8.94 \ \mu{\rm G}$,][]{bals01}.  Using flux conservation, from \citet{padmV2} again, and the fact that $\rho \propto r^{-3}$:
\beq
\Rightarrow B \approx B_{\rm ini} \pfrac{\rho}{\rho_{\rm ini}}^{2/3} \, ,
\eeq
so our initial magnetic field will have an average magnitude of $\left\langle \left| \boldsymbol{B}_{\rm ini} \right| \right\rangle \approx 0.06 \ \mu{\rm G}$ and dispersion of $\sqrt{\left\langle B_{\rm ini}^2 \right\rangle} \approx 0.15 \ \mu{\rm G}$.

As our model SN expands into the ISM, we expect it to encounter an ISM magnetic field that has been twisted by turbulence from the passages of previous SNe ejecta.  This assumption is based on evidence that the Local Bubble has experienced multiple SNe, which means the ISM will be ionized. The presence of density perturbations within the ISM and ejecta lead to instabilities that will drive turbulence in the ISM plasma that will, in turn, drag the magnetic field with it.  In order to generate a model for the magnetic field encountered by our dust grains, we will build our magnetic field in three parts:
\begin{enumerate}
	\item Generate a grid of initial magnetic field values with an energy spectrum appropriate to turbulent media.
	\item Interpolate between the grid values in order to describe the initial magnetic field at all points while remaining divergence-free.
	\item Transform the initial magnetic field to account for the passage of the FS and varying density in order to determine the final magnetic field encountered by our dust grains.
\end{enumerate}
We assume that the turbulence in the ISM is homogeneous and isotropic and is fully-developed and stationary (time-translation invariant).  In order to generate a vector field with these properties, we use a random realization that generates a 3-D grid of values with the desired specific energy spectrum, ${\cal E}_{k} \sim k^{\eta}$ (for more detail see Appendix \ref{app:turbmag}).  This grid has discrete values from $k_{\rm outer} = 2\pi / \lambda_{\rm outer}$ to $k_{\rm inner} = 2\pi / \lambda_{\rm inner}$, where $\lambda_{\rm outer}$ is the outer turbulent scale where is energy is injected and $\lambda_{\rm inner}$ is the inner turbulent scale where energy is dissipated by viscous forces.  For this work, we chose $\lambda_{\rm outer} = 5$ pc \citep[this is within the range of radio polarization variations in several SNRs, i.e., $3-13$ pc,][]{furst04, uyan04, han14, ma16}, and due to limitations with high resolution discrete Fourier transforms, we chose $\lambda_{\rm inner} \lesssim 0.05$ pc.  This $256^3$ grid is then Fourier transformed from $k$ space to real space, giving a cube of dimensions $\lambda_{\rm outer}^3$.  In order to minimize memory requirements, this volume is rotated to random orientations and stacked together in order to completely fill in the total simulation volume.\footnote{It should be noted that we do not include a description of magnetic field amplification along the FS \citep[$B_{\rm ISM} \sim 1$ mG,][]{inoue09,xu17}.  As will be seen in \S \ref{sect:results}, since our grains do not reach the FS, this should not affect our conclusions.}

In order to determine the value of the initial magnetic field everywhere and ensure it has zero divergence, we use a radial basis function to interpolate between the turbulent grid vector values \citep{mcnal11} (see Appendix \ref{app:interpmag}).  An inherent property of this type of interpolation ensures $\boldsymbol{\nabla} \cdot \boldsymbol{B} = 0$ for our initial magnetic field even if the random grid values alone were not necessarily divergence free \citep[for comparison, an alternate SNR turbulent magnetic field is given by][]{west17}, see Figure~\ref{fig:magdiv}.

Although magnetic fields are dynamically unimportant in the early evolution of an SNR, it is still possible to determine the evolution of the magnetic fields in terms of the other fluid quantities \citep{chev74} via the flux-freezing assumption.  Namely, in order to determine the magnetic field, $\boldsymbol{B}$, we combine Equation~\ref{eq:MHDflux} with Equation~\ref{eq:MHDmass}, $\frac{D}{Dt}\pfrac{1}{\rho} = \frac{\boldsymbol{\nabla} \cdot \boldsymbol{v}}{\rho}$, yielding:
\beqar
\frac{D\boldsymbol{B}}{Dt}
& = & \left( \boldsymbol{B} \cdot \boldsymbol{\nabla} \right) \boldsymbol{v} - \boldsymbol{B} \left( \rho~\frac{D}{Dt}\pfrac{1}{\rho} \right) \nonumber \\[6pt]
\Rightarrow \frac{1}{\rho} \frac{D\boldsymbol{B}}{Dt} + \boldsymbol{B}~\frac{D}{Dt}\pfrac{1}{\rho}
& =  & \frac{1}{\rho} \left( \boldsymbol{B} \cdot \boldsymbol{\nabla} \right) \boldsymbol{v} \nonumber \\[6pt]
\Rightarrow \frac{D}{Dt}\pfrac{\boldsymbol{B}}{\rho}
& = & \pfrac{\boldsymbol{B}}{\rho} \cdot \boldsymbol{\nabla} \boldsymbol{v} \, .
\label{eq:MHDfluxfreeze}
\eeqar
When compared to the flux conservation Lagrangian derivative, $\frac{D}{Dt}({d\boldsymbol{l}}) = d\boldsymbol{l} \cdot \boldsymbol{\nabla} \boldsymbol{v}$, this means that the magnetic flux is ``frozen in'' the fluid.  Because Equation~\ref{eq:MHDfluxfreeze} relates the evolution of the magnetic field to the evolution of only the density (which can be determined using the fluid equations), we can solve for the evolution of the magnetic field using that of the density.

For an infinitesimally small fluid element, the magnetic field will be uniform through the entire fluid element, and we can decompose the vector $\boldsymbol{B}$ into a component parallel to the direction of expansion, $\boldsymbol{B}_{\parallel}$, and a component orthogonal to the direction of expansion, $\boldsymbol{B}_{\perp}$:
\beqar
\boldsymbol{B}_{\parallel} & \equiv & \left( \boldsymbol{B} \cdot \boldsymbol{\hat{r}} \right) \boldsymbol{\hat{r}} \, , \quad \boldsymbol{B}_{\perp} \equiv \boldsymbol{B} - \left( \boldsymbol{B} \cdot \boldsymbol{\hat{r}} \right) \boldsymbol{\hat{r}} \, , \\[6pt]
\boldsymbol{B}
& = & \boldsymbol{B}_{\perp} + \boldsymbol{B}_{\parallel} \, , \quad B^{2} = B_{\perp}^{2} + B_{\parallel}^{2} \, ,
\eeqar
where $\boldsymbol{\hat{r}}$ is the radial unit vector.  Using the flux-freezing condition and spherical symmetry \citep{deav05}, we find the following relations for the initial and final magnetic fields (for further detail, see Appendix \ref{app:fluxfreeze}):
\beqar
B_{\rm \parallel, fin} & =  & B_{\rm \parallel, ini} \pfrac{r_{\rm ini}}{ r_{\rm fin}}^{2} \, , \\[6pt]
B_{\rm \perp, fin} &  =  & B_{\rm \perp, ini} \pfrac{\rho_{\rm fin} r_{\rm fin}}{\rho_{\rm ini} r_{\rm ini}} \, .
\eeqar
With this relation between the initial and final magnetic fields, we can relate their divergences as well:
\begin{align}
\boldsymbol{\nabla} \cdot \boldsymbol{B}_{\rm fin} &= \boldsymbol{\nabla} \cdot \boldsymbol{B}_{\rm \perp, fin} + \boldsymbol{\nabla} \cdot \boldsymbol{B}_{\rm \parallel, fin} \nonumber \\[6pt]
&= \boldsymbol{\nabla} \cdot \boldsymbol{B}_{\rm \perp, ini} \pfrac{\rho_{\rm fin} r_{\rm fin}}{\rho_{\rm ini} r_{\rm ini}} + \boldsymbol{\nabla} \cdot \boldsymbol{B}_{\rm \parallel, ini} \pfrac{r_{\rm ini}}{ r_{\rm fin}}^{2} \nonumber \\[6pt]
&= \pfrac{\rho_{\rm fin} r_{\rm fin}}{\rho_{\rm ini} r_{\rm ini}} \boldsymbol{\nabla} \cdot \boldsymbol{B}_{\rm \perp, ini} + \pfrac{r_{\rm ini}}{ r_{\rm fin}}^{2} \boldsymbol{\nabla} \cdot \boldsymbol{B}_{\rm \parallel, ini} \, .
\end{align}
Because our initial magnetic field has been interpolated to be divergence free, the magnetic field will remain divergence free at all times:
\begin{align}
\boldsymbol{\nabla} \cdot \boldsymbol{B}_{\rm ini} = 0 &\Rightarrow \boldsymbol{\nabla} \cdot \boldsymbol{B}_{\rm \perp, ini} = \boldsymbol{\nabla} \cdot \boldsymbol{B}_{\rm \parallel, ini} = 0 \nonumber \\[6pt]
&\Rightarrow \boldsymbol{\nabla} \cdot \boldsymbol{B}_{\rm fin} = 0 \, .
\end{align}

\begin{figure*}[ht]
	\begin{center}
		\includegraphics[width=\textwidth]{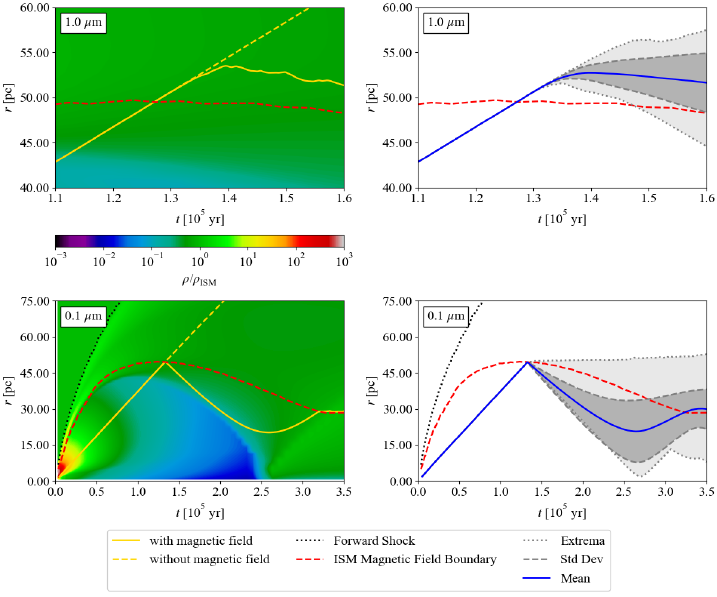} \\
		(a)
		\caption[Trajectories of metallic Fe-grains of varying initial sizes on a density contour.]{Sample trajectories of metallic Fe-grains of varying initial sizes on a density contour (left) and means/standard deviations/extrema of trajectories (right).  The $\gtrsim 0.05$-$\mu$m grains showed little influence by drag, traveling on similar trajectories when no magnetic field is present, but experiencing strong interactions in the presence of a magnetized ISM.  Panel (a):  Most $1$-$\mu$m grains became trapped in the ISM magnetic field; their greater mass contributing to greater penetration before capture, but preventing later escape.  In contrast, the $0.1$-$\mu$m grains experience mostly reflections at the first contact with the ISM magnetic field, while most grains become trapped at the second contact (although some grains do show additional reflections).  Panel (b):  The $\lesssim 0.05$-$\mu$m grains showed increasing influence by drag, delaying interaction with the magnetized ISM.  Most $0.05$-$\mu$m grains became trapped in the ISM magnetic field, although occasional reflections did occur.  The $0.005$-$\mu$m grains demonstrated no reflections and were completely trapped by the ISM magnetic field.  Below $\leq 0.004$-$\mu$m, drag and sputtering on dust grains becomes significant, with the smallest grains never reaching ISM material before being completely sputtered.  The standard deviation assumes a normal distribution.
			\label{fig:sizesolution1}
		}
	\end{center}
\end{figure*}

\setcounter{figure}{3}

\begin{figure*}[ht]
	\begin{center}
		\includegraphics[width=\textwidth]{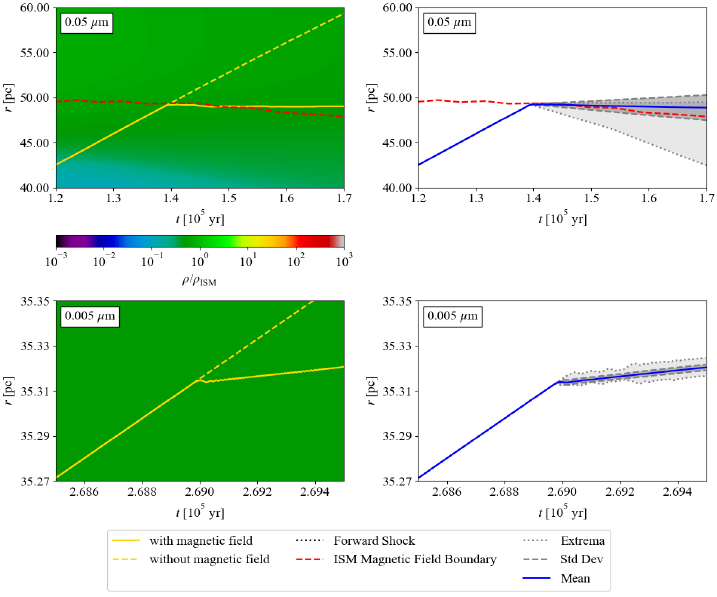} \\
		(b)
		\caption[]{(Continued.)
			\label{fig:sizesolution2}
		}
	\end{center}
\end{figure*}

\section{Results}
\label{sect:results}

\begin{figure*}[t]
	\centering
	\includegraphics[width=\textwidth]{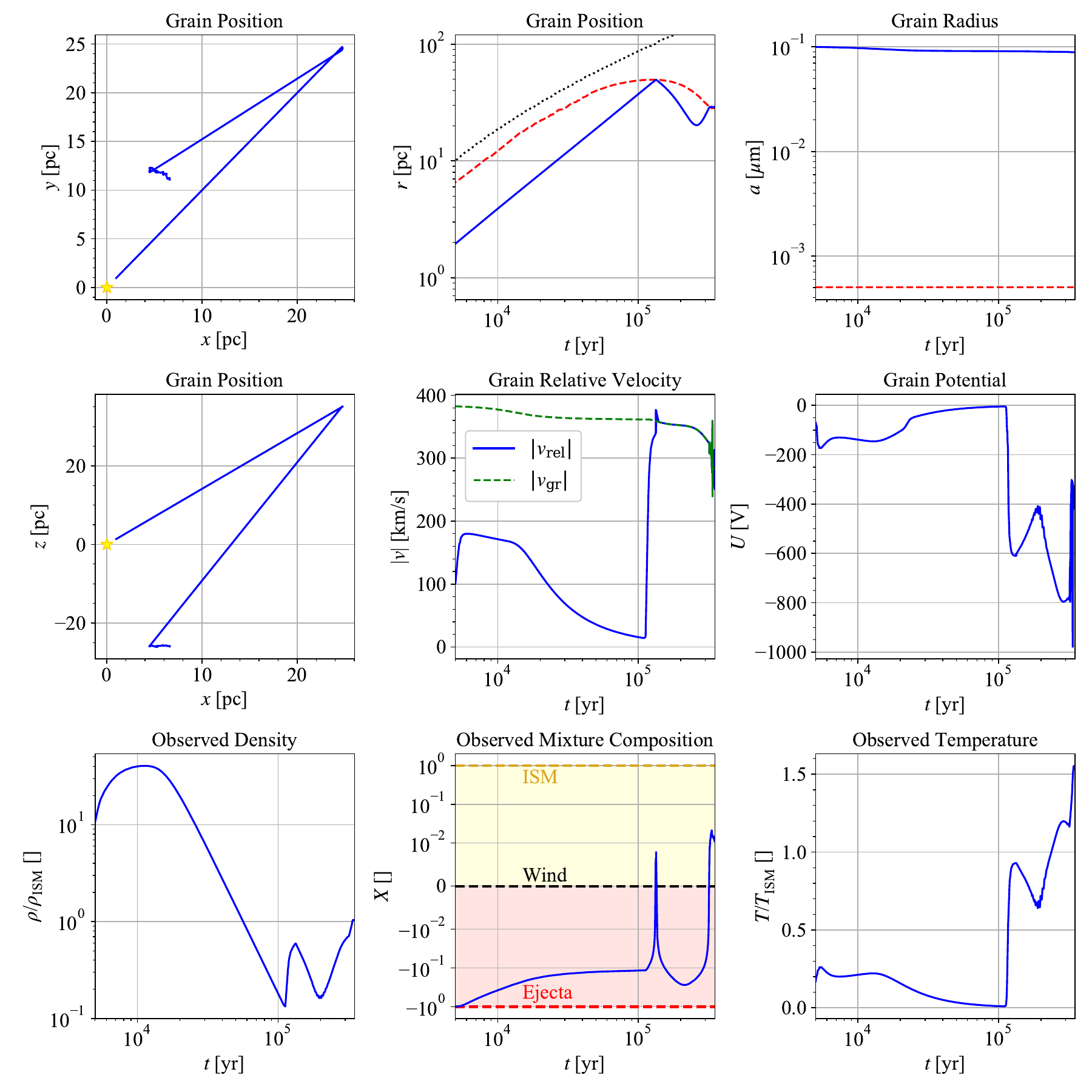}
	\caption[Summary of grain parameters for a sample 0.1-$\mu$m metallic Fe-grain.]{Summary of grain parameters for a sample 0.1-$\mu$m metallic Fe-grain.  The top, left panels are projections of the grain's positions onto the $x-y$- (upper left) and $x-z$-planes (center left) with the yellow star representing the site of the SN.  The upper center plot shows the grain's radial position with the dotted, black line indicating the position of the forward shock and the dashed, red line indicating the boundary between the pre-SN stellar wind and the ISM material and ISM magnetic field.  The first reflection occurs at $t \approx 1.3 \times 10^{5}$ yr, and the grain becomes trapped in the magnetic field at $t \approx 3 \times 10^{5}$ yr.  The rate of deceleration due to drag changes as the grain moves through different densities (bottom left) and is most pronounced following reflections, when the relative velocity, $v_{\rm rel}$, between the grain and plasma is greatest (center).  The rate of grain erosion due to sputtering remains fairly constant throughout the entire simulation (upper right), but the grain potential makes sharp fluctuations while generally staying negative (center right) and strongly mirrors the surrounding plasma temperature fluctuations (bottom right).  The dashed red line on the Grain Radius plot represents the sputtering limit; below this, the grain is assumed to have been destroyed.  The bottom center plot shows the pre-SN material the grain passes through; -1 is pure ejecta, 0 is pure wind, and 1 is pure ISM material with fractional values representing mixtures (there is no ISM/ejecta mixing).  The grain's reflection/trapping directly correlates to the grain's encounters with ISM material.
		\label{fig:combinedsolution}
	}
\end{figure*}

We have examined the trajectories for dust grains containing \fe60 in an SNR expanding into an ISM containing a turbulent magnetic field with a Kolmogorov spectrum ($\eta = -5/3$).  We assumed the grains contained material located originally at $r = 0.6 \times 10^9$ cm at the beginning of the hydrodynamic simulation; this corresponds to the highest concentration of \fe60 within the ejecta (see Figure~\ref{fig:products}).  The \fe60 was assumed to condense into metallic-Fe grains between 100-1000 days after the SN explosion, which corresponds to $r_{\rm gr} = 111$ AU and $v_{\rm gr} = 382$ km s$^{-1}$.  We assume that the grain is entrained within its surrounding dust cloud from the initial time-step until encountering the RS at $t = 5000$ yr (with Rayleigh-Taylor instabilities present; by contrast without those instabilities, the RS would not arrive until $t = 8600$ yr).  As the RS passes, we assume the grain is immediately exposed to the shocked SNR environment, the simulation begins at $t_{0} = 5000$ yr, $r_{\rm gr,0} = 1.96$ pc, $v_{\rm gr,0} = 382$ km s$^{-1}$ (because we focused on the most stressing case, we performed runs only with Rayleigh-Taylor instabilities included).

To begin with, we examined the unmagnetized case.  This serves as a basis of comparison for our magnetized examinations, as well as a comparison with previous works \citep[see e.g.,][]{noz07,nath08,mice16,bocc16}.  For various grain sizes ($0.005-1~\mu$m), the grains demonstrate purely radial motion, gradually slowing as they approach the FS, see Figure~\ref{fig:sizesolution1}, left panels.  As should be expected, the larger grains maintain their velocities relative to smaller grains due to the former's greater mass.  Qualitatively, we were able to reproduce the previous cited works' results.

Next, we examined a variety of post-RS grain sizes ranging from $0.005-1~\mu$m with a magnetized ISM, and ran the simulation 100 times to examine different configurations of the magnetic field, see Figure~\ref{fig:sizesolution1}, right panels.  The $a_{\rm gr,0} = 0.1$-$\mu$m grains showed the most noteworthy trajectories, with some being reflected nearly radially backwards into the SNR, and others being deflected along the edge of the SNR or becoming trapped.  The $a_{\rm gr,0} = 1$-$\mu$m grains also showed strong reflections into the inner SNR as well as some trapping.\footnote{It should be noted that $1$-$\mu$m is an \emph{exceptionally} large dust grain, and that we do not expect many, if any, such grains to form.  We include them here for completeness.}  As the grains decrease in size, drag becomes more important delaying contact with the ISM magnetic field; $a_{\rm gr,0} = 0.05$-$\mu$m grains will become mostly trapped by the ISM magnetic field although some are reflected.  The influence of drag on $a_{\rm gr,0} = 0.005$-$\mu$m grains show complete trapping and no reflections (see Figure~\ref{fig:sizesolution1}).  The $a_{\rm gr,0} \lesssim 0.002$-$\mu$m grains completely sputter before reaching ISM material.

\begin{figure}[t]
	\centering
	\includegraphics[width=0.48\textwidth, trim=1.1in 0.3in 0.1in 0.95in, clip=True]{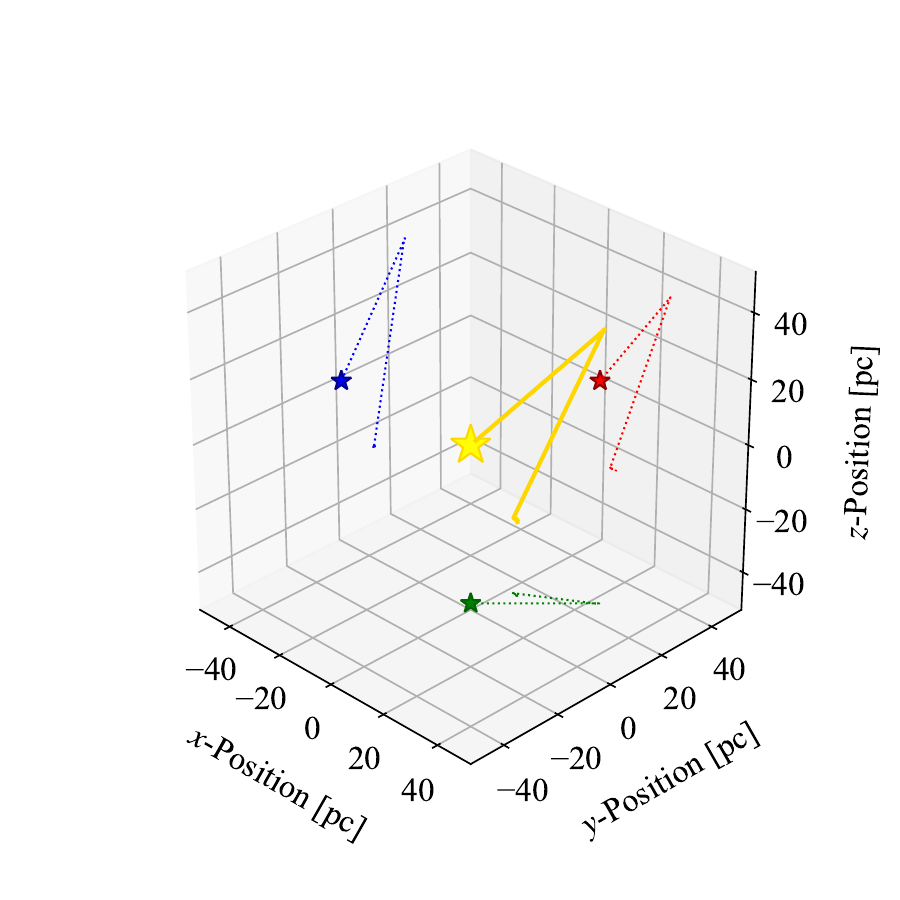}
	\caption[Three-dimensional plot of a sample 0.1-$\mu$m metallic Fe-grain.]{Three-dimensional plot of a sample 0.1-$\mu$m metallic Fe-grain.  The yellow lines are the 3-D plot of the grain trajectory with the green, red, and blue lines showing the $x-y$-, $x-z$-, and $y-z$-planes respectively.  The stars represent the location of the SN.  Note:  this is the same sample shown in Figure~\ref{fig:combinedsolution}.
		\label{fig:3Dtrajectory}
	}
\end{figure}

\begin{figure}[t]
	\centering
	\includegraphics[width=0.48\textwidth]{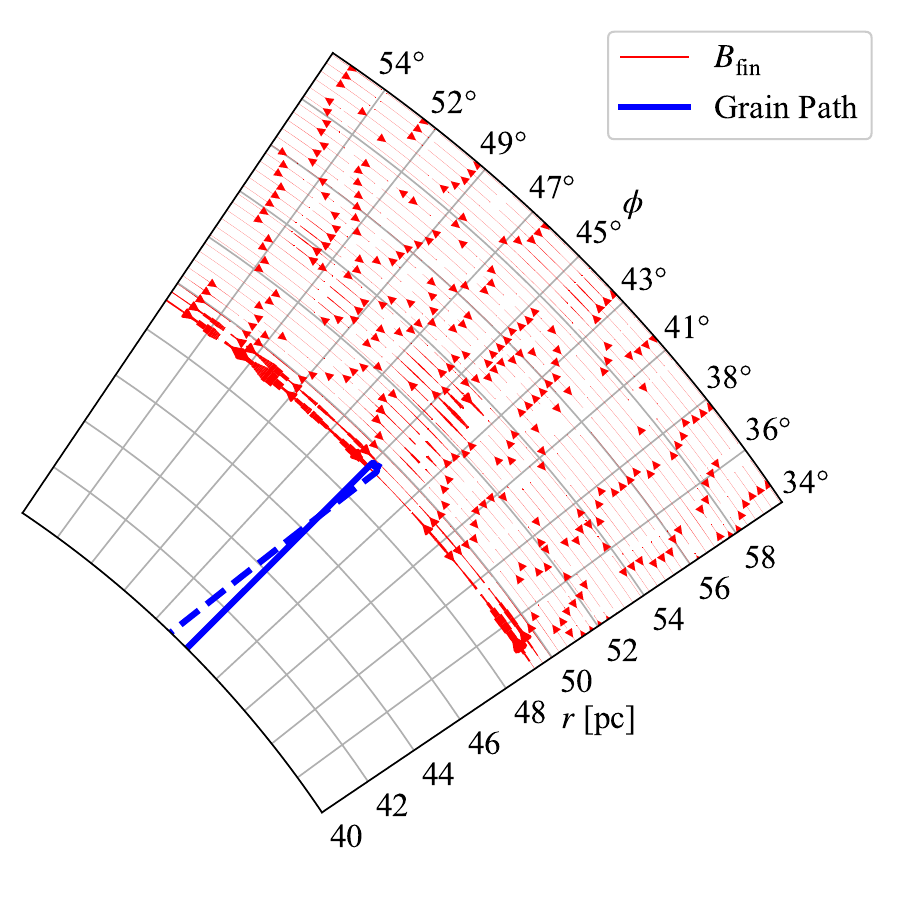}
	\caption[Dust Grain Trajectory and Magnetic Field Lines]{The azimuthal trajectory of a sample 0.1-$\mu$m metallic Fe-grain along its initial radial direction.  The grain path is shown as a solid blue until reflection, and a dashed, blue line afterwards.  The red lines are the magnetic field lines at the moment of reflection ($t \approx 130$ kyr, $r \approx 50$ pc).  Note the stretching of the magnetic field lines parallel to the FS and the lack of a magnetic field interior to 49 pc representing pre-SN stellar and wind material.  Because of mixing of the wind and ISM material, the magnetic field gradually increases in strength, until strong enough to reflect the dust grain.
		\label{fig:magbounce}
	}
\end{figure}

The $a_{\rm gr,0} = 0.1$-$\mu$m grains' `pinball' behavior is particularly dramatic.  They experience some drag and sputtering, but this effect is relatively minor since the relative velocity $v_{\rm rel} \lesssim 175$ km s$^{-1}$ for much of their transit, see Figure~\ref{fig:combinedsolution}.  There is no deflection (i.e., non-radial motion) of the grain's trajectory while the grain is traveling through pre-SN circumstellar material, see Figure~\ref{fig:3Dtrajectory}.  This is because there is (effectively) no magnetic field in this material.  In contrast, shortly after encountering shocked ISM material, the frozen-in ISM magnetic field reflects the grain back into the SNR.  This action is repeated as the grain transits the SNR and again encounters shocked ISM material.  The charged dust grains ricochet inside the magnetized ISM material like pinballs \citep[this boundary is roughly equivalent to the contact discontinuity, see e.g.,][]{wang02}.  Figure~\ref{fig:magbounce} shows the grain penetrates the ISM material to some extent and allows us to verify that the grain is being reflected due to the magnetic field rather than a discontinuity inherent to our model.

\begin{figure*}[t]
	\centering
	\includegraphics[width=\textwidth]{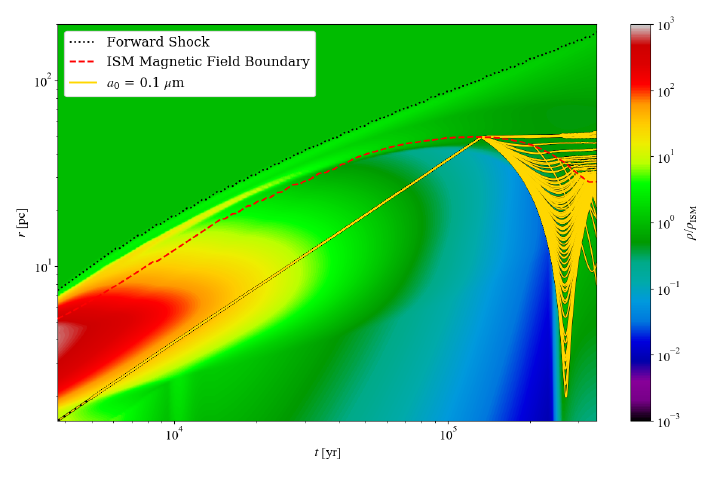}
	\caption[Multiple trajectories of metallic Fe-grains on a density contour.]{Multiple trajectories of metallic Fe-grains on a density contour.  Trajectories for 100 grains (plotted in yellow) are shown with an initial grain radius of $a_{\rm gr} = 0.1$-$\mu$m and encounter the ISM magnetic field.  The grain trajectory prior to cloud crushing is shown with dashed, yellow lines and solid, yellow lines when the cloud has dissipated and the grain is exposed the SNR plasma.  Note that most grains remain in the $r \lesssim 50$ pc region while bouncing within the shell.  The FS is shown with the dotted, black curve, and the contact discontinuity between ejecta and ISM material are shown with a dashed, red line.
		\label{fig:multiplesolution}
	}
\end{figure*}

Figure~\ref{fig:multiplesolution} shows results from 100 runs with an initial $a_{\rm gr,0} = 0.1$-$\mu$m grain.  Almost every grain is reflected nearly opposite to its initial radial direction, with some grains experiencing multiple reflections.  These features appear as `U'- or `W'-shaped trajectories in Figure~\ref{fig:multiplesolution}.

\section{Predictions and Implications} 
\label{sect:implications}

\subsection{Predictions:  SN Dust Confinement and Evolution}
\label{subsect:predictions}

The dust grain trajectories in our simulations are largely confined by the ISM magnetic field, and although our selection of SN parameters are motivated by the near-Earth scenario, this dust confinement effect should not be sensitive to that particular scenario.  The grain reflections and trapping occurred in regions with magnetic field strengths $\sim$ tens of nanogauss.  Even if our assumption that the ISM magnetic field scales as $B \sim \rho^{2/3}$ is not strictly adhered to, it is still expected the magnetic field would scale in the range of $B \sim \rho^{[0,1]}$ \citep{deav05}.  This would mean that regardless of the scaling used, dust grains would eventually encounter magnetic field strengths sufficient to cause reflections/trapping as they propagate radially outwards in our simulations (albeit at slightly different location depending on the scaling value).  Further, our chosen magnetic field is weaker than the average ISM magnetic field suggesting dust confinement is a property of SNe in general rather than this specific scenario.

Given that propagation in the ISM magnetic field is the limiting mechanism for the larger SN dust grains, it is possible to estimate the maximum radial position for the dust grains within the SNR.  Since the ISM material must be pushed away by the ejecta, we can assume the dust's maximum radial position is where the total ejected and stellar wind material equals the swept up ISM material:
\beq
M_{\rm ej} + M_{\rm wind} = M_{\rm swept-up~ISM} = \frac{4}{3} \pi \rho_{\rm ISM} R_{\rm dust}^{3} \, ,
\eeq
\begin{align}
\Rightarrow R_{\rm dust} =&~24 \ {\rm pc} \pfrac{M_{\rm ej} + M_{\rm wind}}{90~\msol}^{1/3} \pfrac{0.61}{\mu_{\rm ISM}}^{1/3} \nonumber \\[6pt]
&\times \pfrac{0.1~{\rm cm}^{-3}}{n_{\rm ISM}}^{1/3} \, .
\end{align}
For an observational comparison, the core-collapse SNR Sagittarius (Sgr) A East is $\sim 10$~kyr old and recent SOFIA observations have confirmed the presence of dust within the center of the SNR \citep{lau15}.  The dust is confined to the inner $\sim 2$~pc diameter \citep{lau15}, while the outer radio shell (that corresponds to the FS) is has a diameter of $\sim 7$ pc \citep{maeda02,lau15}.  Using the estimated conditions of the progenitor:  $M_{\rm ej} = 2~\msol$ \citep{maeda02}, $M_{\rm wind} = 10~\msol$ \citep{mez89}, and $n_{\rm ISM} = 1000~{\rm cm}^{-3}$ \citep{maeda02,lau15}, this gives $R_{\rm dust} = 0.57$~pc, which is in good agreement with the observed dust region of Sgr A East SNR.

\subsection{Dust Delivery of Radioisotopes from Near-Earth SNe}
\label{subsect:delivery}

The Sco-Cen OB association has frequently been expected as the source of the \fe60 given the large number of SN-capable progenitors within that association.  However, based on these results, the possibility of a Sco-Cen progenitor ($D \sim 130$ pc) as the source of the \fe60 seems extremely unlikely, given that magnetic fields restrict the movement of the larger (i.e., $a_{\rm gr} \gtrsim 0.002~\mu$m) grains and drag halts the movement of smaller grains.  Although Sco-Cen may have yielded a larger, more powerful progenitor \citep[e.g., a 15-$\msol$ CCSN,][]{hyde18}, the additional explosive energy ($E_{\rm SN}$) is not expected to be able to push back the ISM magnetic field over 100 pc.  ISM magnetic fields would severely restrict the passage of dust grains.  In order for a Sco-Cen source of \fe60, \emph{some mechanism(s) would be needed to either drive the ISM magnetic field back or allow charged dust grains to pass more efficiently}.  However, Tuc-Hor is still a likely source, as our simulation showed consistent dust propagation out to $\sim 50$ pc, which (considering the uncertainty in the initial ISM density) is consistent with the distance to Tuc-Hor ($\sim 45-60$ pc).

The implications of magnetic reflections are that the grains are not confined to the shell region as assumed by \citet{fry15,fry16,breit16,feige16,feige17,schul17}, but confined to the interior of the SNR.  These reflections also alter the assumption of a plane-wave arrival of SN dust grains into the Solar System.  It even appears likely that, after the SNR envelops the Solar System, some dust grains will bounce off the contact discontinuity and approach Earth from a direction nearly opposite of the SN!  Further characterization of this passage is needed to determine the viability of using lunar samples to determine the direction to the SN as proposed by \citet{fry16}.  This potentially explains the surprisingly extended $\sim 1$ Myr duration of the signal \citep{fit08,wall16,ludw16,fields19}, but negates the proposal by \citet{fry15} of using time-resolved samples as an alternate gauge of the SN's distance.  On the other hand, the time-resolved samples will yield a measure of the SNR's propagation and internal dust distribution.

Finally, we note that our approach and findings are complementary with the work of \citet{elli97}.  These authors showed that supernova dust should be dynamically decoupled from the plasma, and that some grains can be accelerated.  They argued that sputtering of these fast grains provides the nonthermal ions that are injected into and accelerated by the supernova shocks.  \citet{elli97} concluded that this mechanism could be the origin of the cosmic-ray enhancement in refractory elements.  Our work extends and complements this picture:  the \fe60-bearing dust that survives sputtering leads to the terrestrial and lunar signals, while the sputtered ions supernova may be responsible for the \fe60 seen in cosmic-rays by \citet{binns16}.

\subsection{Magnetic Field Discussion}
\label{subsect:discussion}
In this exploratory calculation we have treated the magnetic field in an idealized manner.  We have neglected entirely any field in the supernova wind and ejecta, and we have treated the ISM field as a random Gaussian field with an entirely turbulent, power-law (i.e., ${\cal E}_{k} \sim k^{\eta}$) spectrum.  These choices can influence our results.

The absence of magnetic fields in the supernova wind and ejecta leads to the undeflected, radial motion of the dust particles within this material.  Then the encounter with the ISM field results ultimately in magnetic mirroring and/or trapping.  Mirroring occurs for grains moving from field strengths $B_{\rm min}$ to $B_{\rm max}$ with velocities that satisfy the `loss cone' condition (from Equation~\ref{eq:magrelation}):
\beq
\frac{v_\parallel}{v_\perp} < \sqrt{\frac{B_{\rm max}}{B_{\rm min}}-1}
\label{eq:mirror}
\eeq
where $v_\parallel$ and $v_\perp$ are the velocity components parallel and perpendicular to the field.  In the limit where $B_{\rm min} \rightarrow 0$, all are reflected.  Under our expected magnetic field initial conditions, (see \S \ref{subsect:magfield}), our expected values for $v_\parallel/v_\perp \lesssim 1000$, suggesting that the vast majority of (but not all) encounters should experience magnetic reflections.  Because our simulations allow for mixing between non-ISM and ISM material, the dust grain's encounter with a magnetic field is not abrupt, and so magnetic reflections are not guaranteed in our simulation environment, and, in fact, not all encounters in our results do exhibit reflections.  However, a more complete magnetic field description would better ensure that any grain behavior is not the result of a model's intrinsic construction.

Another choice was to use a completely turbulent (i.e., \emph{disordered}) field with no uniform (i.e., {\emph{ordered}}) component.  If the field has a nonzero \emph{ordered} component, this can direct the grains along field lines and impede motion perpendicular to the magnetic field.  Examining a strictly uniform case and a combination of uniform-turbulent configurations would further characterize the magnetic mirroring effect in future follow-up studies.

Lastly, the selection of a random Gaussian field leads to turbulent field behavior that has its dominant component on the largest scales ($\lambda_{\rm outer} = 5~{\rm pc}$, see also Figure~\ref{fig:magdiv}).  This can create an overly smooth field at small scales as suggested by turbulence simulations \citep[e.g.,][]{wils98, burk09} and as seen in solar wind data \citep[e.g.,][]{sorr99}.  Examining the influence of this difference at small scales is beyond the scope of this work, but should be included in future work to examine its influence on trapping and mirroring effects.

\section{Conclusions}
\label{sect:conclusions}

We have studied the motion and evolution of dust grains created in an unmagnetized SNR exploding into a magnetized ISM.  The SNR evolution is described via a 1-D+ model assuming spherical symmetry that includes angle-averaged effects of Rayleigh-Taylor instabilities and resultant mixing.  The ISM magnetic fields are initialized with a turbulent magnetic field with a Kolmogorov spectrum, and evolve kinematically via flux-freezing, and thus are altered by the shock.  We included the effects of drag, sputtering, and charging on the dust grains, and simulate the motions of grains of different sizes.

In the absence of magnetic fields, or equivalently for uncharged dust, our results are similar to those of other groups \citep[see e.g.,][]{noz07,nath08,mice16,bocc16}.  We assume the grains are initially entrained with the gas from which they are born, and thus have radial trajectories.  After the ejecta encounter the RS, the dust grains decouple from the decelerated gas, and move towards the FS.  For large grains, the effects of drag and sputtering are small enough that the grains survive to pass close to or across and beyond the FS into the ISM.  The result would be a ``halo'' of the largest dust grains that precedes the FS.  

However, we find that magnetic fields have a dramatic effect, leading to qualitatively new dust trajectories and fate compared to the unmagnetized case.  In particular, we find that the dust grains typically suffer large deflections when encountering the shocked ISM, in which the pre-existing turbulent magnetic fields have been altered.  The main effect we observe is magnetic trapping and mirroring, occurring at the interface between the SN ejecta and the shocked ISM.  The reflected particle moves back into the SN ejecta, traversing the SNR until it encounters  the ejecta/ISM interface again; effectively the dust grain has become a pinball within the SNR.  The resulting motion is thus a series of ricochets inside the SN ejecta region.  The presence of ISM magnetic fields means dust is not distributed throughout the entire SNR, but confined much deeper within the SNR.  

Our results show that the inclusion of Rayleigh-Taylor instabilities is important. Figure~\ref{fig:multiplesolution} shows that the grains enter the shocked medium as early as 5000 yr after the explosion, rather than later at $\sim 10^5$ yr when the RS proceeds inward to the center of the SNR.  By entering the shocked plasma earlier, grains are subject to erosion and drag longer.

Lastly, the presence of magnetic reflections also suggest that grain-grain interactions and shattering due to shock crossings may not be entirely negligible.  The grain reflections into the SNR greatly increase the likelihood of collisions compared to the purely radial trajectories assumed to date \citep[an example of a grain-grain collision approach is used by][]{kirsch19}.  Additionally, as seen in Figure~\ref{fig:multiplesolution}, grains crossing the RS multiple times will be subject to repeated shattering, reducing the likelihood for long-term survival \citep[see also][]{will16}.

The overarching message of our study is that magnetic fields have a dramatic effect on the evolution and survival of SN dust.  This has important implications not only for the terrestrial and lunar deposition of \fe60 and other radioisotopes, but also for the evolution of dust in SNRs generally, and possibly for the the role of supernova dust in cosmic-ray acceleration.  We will explore these implications more in future work.

Further simulations using other radioisotopes are planned.  Based on these \fe60 results, it appears that \al26 and \ca41 (which form in the front portion of the ejecta, making them more likely to encounter the RS before \fe60) will be exposed to the hot SNR plasma earlier.  Simulations examining \al26 could readily be compared to results from \citet{feige18}.  Since their density is less than metallic Fe, they will be more sensitive to drag and the magnetic fields due to their lower mass.  A portion of \mn53 is synthesized slightly deeper in the ejecta and may form MnS, but the bulk of \ca41 and \mn53 are deeper in the ejecta than \fe60.  The question remains into what type of dust, if any, they will be incorporated.  Additional simulations are also planned to characterize fully the dust grains' movements within the SNR.  These include varying the ISM density and magnetic field, varying the grain composition/sizes further, and examining the case in which there is no H/He envelope around the pre-SN star \citep{frem16}.  Additionally, the use of a Nonuniform Fast Fourier Transform (NUFFT) for the turbulent magnetic field description would allow characterization down to the dissipation scale \citep[see e.g.,][]{haines88,ham00,green04,ham15}.

\acknowledgments
B.J.F. would like to thank Paul Duffell for his assistance in using the RT1D code, and Shinya Wanajo for kindly sharing his ECSN nucleosynthesis data; this work would not have been possible without either.  B.J.F. would also like to thank Jenny Feige for her comments and suggestions regarding this work.  We are grateful for encouragement and illuminating discussions with Bruce Draine, Paul Ricker, Charles Gammie, Ada Ertel, and Jesse Miller.  The work of J.E. is supported in part by STFC (UK) via the research grant ST/L000326/1, and in part by the Estonian Research Council via a Mobilitas Pluss grant.

\appendix
\label{app:appendix}

\section{List of Variables}
\label{app:variables}

\begin{flushleft}
	\underline{Variable - Description [common value or unit of measure]} \\
	$*$ - (as superscript) parameter in characteristic units [dimensionless] \\
	$\parallel$ - (as subscript) parallel component \\
	$\perp$ - (as subscript) perpendicular component \\
	$0$ - (as subscript) `initial value for simulation' \\
	$a_{\rm gr}$ - radius of dust grain [$\mu$m] \\
	$a_{\rm cloud}$ - radius of cloud [km] \\
	$a_{\rm sc}$ - screening length [$\mu$m] \\
	$A$ - magnetic vector potential [G cm] \\
	$\mathbb{A}$ - local vector potential [G cm$^{2}$] \\
	$\alpha$ - angle [radians] \\
	$\boldsymbol{b}$ - perturbed magnetic field [G] \\
	$\boldsymbol{B}$ - magnetic field [G] \\
	$c$ - speed of light [$\sim 3 \times 10^5$ km s$^{-1}$] \\
	$C_{\rm abs}$ - absorption cross-section [cm$^{2}$] \\
	$C_{\rm coll}$ - collisional cross-section [cm$^{2}$] \\
	${\rm ch}$ - (as subscript) `characteristic scale' \\
	$\mathbb{C}$ - scaling constant [G$^{2}$ pc$^{2}$] \\
	$\chi$ - generic/dummy variable [dimensionless] \\
	$D$ - distance to Earth [pc] \\
	$\mathbb{D}$ - dilution factor [dimensionless] \\
	$\delta_{j}$ - secondary electron yield [dimensionless] \\
	$\delta_{\rm max}$ - maximum yield from a bulk solid [dimensionless] \\
	$\Delta_{j}$ - penetration threshold energy [eV] \\
	$e$ - elementary charge [$4.803 \times 10^{-10}$ Fr] \\
	$E$ - energy [erg] \\
	$E_{1}$ - energy constant [eV] \\
	$E_{2}$ - energy constant [keV] \\
	$E_{\rm bind}$ - binding energy [eV] \\
	$E_{e}$ - most probable energy from electrons [eV] \\
	$E_{\gamma}$ - most probable energy from photons [eV] \\
	$E_{H}$ - ionization energy of hydrogen [13.6 eV] \\
	$E_{\rm ion}$ - most probable energy from ions[eV] \\
	$E_{\rm low}$ - minimum photoelectric emission energy [eV] \\
	$E_{\rm max}$ - energy at maximum yield from a bulk solid [eV] \\
	$E_{\rm min}$ - minimum emission energy [eV] \\
	$E_{\rm SN}$ - energy of SN [erg] \\
	$E_{\rm th}$ - threshold energy [eV] \\
	${\cal E}$ - specific energy [erg g$^{-1}$] \\
	$\mathbb{E}$ - energy density [erg cm$^{-3}$] \\
	$\epsilon$ - reduced energy [dimensionless] \\
	$\eta$ - spectral index [dimensionless] \\
	$f_{1}$ - fitting function [dimensionless] \\
	$f_{2}$ - fitting function [dimensionless] \\
	$f_{j}$ - Maxwellian velocity distribution function [dimensionless] \\
	${\rm fin}$ - (as subscript) `final value' \\
	$\boldsymbol{F}$ - force [dyn] \\
	${\cal F}$ - fluence [atoms cm$^{-2}$] \\
	$\mathbb{F}_{\rm mag}$ - magnetic flux [G cm$^{2}$] \\
	$\mathbb{F}_{\gamma, h\nu}$ - spectral photon flux [photons cm$^{-2}$ s$^{-1}$ eV$^{-1}$] \\
	$g$ - maximum fraction energy transfer [dimensionless] \\
	$G_{0}$ - collisional drag function [dimensionless] \\
	$G_{2}$ - Coulomb drag function [dimensionless] \\
	$\Gamma$ - adiabatic index [dimensionless] \\
	$h$ - Planck's constant [$6.626 \times 10^{-27}$ erg s$^{-1}$] \\
	${\rm ini}$ - (as subscript) `initial value' \\
	${\cal I}$ - charge current [Fr s$^{-1}$] \\
	$\boldsymbol{J}$ - charge current density [Fr cm$^{-2}$ s$^{-1}$] \\
	$k$ - (as subscript) `Fourier counterpart' \\
	$\boldsymbol{k}$ - wavenumber [cm$^{-1}$] \\
	$k_{\rm B}$ - Boltzmann constant [$1.38 \times 10^{-16}$ erg K$^{-1}$] \\
	$\kappa$ - sputtering free parameter [dimensionless] \\
	$l$ - length [cm] \\
	$\lambda$ - length scale [km] \\
	$\Lambda$ - charge parameter [dimensionless] \\
	$m$ - mass (e.g., of $p^{+}$, $e^{-}$, $H$, dust grain, etc.) [g] \\
	$m_{\rm u}$ - atomic mass unit [$1.66 \times 10^{-24}$ g] \\
	$m_{\rm sp}$ - sputtered mass [g] \\
	$M_{\rm ej}$ - mass of the ejecta [$\msol$] \\
	$M_{\rm enclosed}$ - mass enclosed in spherical shell [$\msol$] \\
	$\dot{M}_{\rm wind}$ - stellar wind mass loss [$\msol$ yr$^{-1}$] \\
	$\mu_{\rm ej}$ - mean mass of the ejecta [dimensionless] \\
	$\mu_{\rm envel}$ - mean mass of the stellar envelope [dimensionless] \\
	$\mu_{\rm ISM}$ - mean mass of the ISM [dimensionless] \\
	$n$ - number density [cm$^{-3}$] \\
	$N$ - number of values/points [dimensionless] \\
	${\cal N}$ - material current [cm s$^{-1}$] \\
	$\nu$ - frequency [Hz] \\
	$\omega$ - grain penetration factor [dimensionless] \\
	$\Omega$ - solid angle [sr] \\
	$p$ - momentum [g cm s$^{-1}$] \\
	$P$ - pressure [dyn cm$^{-2}$] \\
	${\cal P}$ - specific power [erg cm$^{2}$ g$^{-1}$] \\
	$\phi$ - angle [radians] \\
	$\Phi$ - potential parameter [dimensionless] \\
	$\psi$ - radial basis function [dimensionless] \\
	$\pounds_{e}$ - electron range power index [dimensionless] \\
	$q_{\rm gr}$ - charge of grain [Fr] \\
	$\boldsymbol{r}$ - radial position [cm] \\
	$R_{\rm bounce}$ - position at bounce [cm] \\
	$R_{e}$ - electron range [nm] \\
	$R_{\rm gyro}$ - gyro radius [pc] \\
	$R_{m}$ - reduced range [dimensionless] \\
	$R_{\rm SN}$ - position of forward shock [pc] \\
	$\mathbb{R}$ - range constant [nm] \\
	$\rho$ - mass density (e.g., of ISM, etc.) [g cm$^{-3}$] \\
	$\rho_{\rm gr}$ - mass density of grain [g cm$^{-3}$] \\
	$\varrho_{e}$ - energy distribution for secondary electrons emitted by electrons [eV$^{-1}$] \\
	$\varrho_{\gamma}$ - energy distribution for secondary electrons emitted by photons [eV$^{-1}$] \\
	$\varrho_{\rm ion}$ - energy distribution for secondary electrons emitted by ions [eV$^{-1}$] \\
	$s$ - velocity parameter [dimensionless] \\
	$S$ - stopping cross section [cm$^{2}$ erg] \\
	$\mathbb{S}$ - scaling factor [cm$^{-2}$] \\
	$\sigma_{A}$ - standard deviation of the magnetic vector potential [G pc$^{-1}$] \\
	$\varsigma$ - elastic reduced stopping cross section [1] \\
	$t$ - time [s] \\
	$T$ - temperature [K] \\
	$T_{5} \equiv T~(10^{5}~{\rm K})^{-1}$ [dimensionless] \\
	$\tau_{1/2}$ - half-life [Myr] \\
	$\tau_{\rm cc}$ - cloud crushing time [yr] \\
	$\tau_{\rm charge}$ - charging time [yr] \\
	$\tau_{\rm gyro}$ - gyro period [yr] \\
	$\theta$ - angle [radians] \\
	$\boldsymbol{u}$ - perturbed velocity [cm s$^{-1}$] \\
	$U_{\rm gr}$ - potential of dust grain [V] \\
	$\boldsymbol{v}$ - velocity [cm s$^{-1}$] \\
	$v_{7} \equiv v~(10^{7}~{\rm cm~s^{-1}})^{-1}$ [dimensionless] \\
	$\boldsymbol{v}_{\rm gr}$ - velocity of grain relative to center of explosion [cm s$^{-1}$] \\
	$v_{\rm ej}$ - velocity of the ejecta [km s$^{-1}$] \\
	$v_{\rm esc}$ - escape velocity [km s$^{-1}$] \\
	$v_{\rm rel}$ - relative velocity [km s$^{-1}$] \\
	$v_{\rm RS}$ - relative velocity of reverse shock [km s$^{-1}$] \\
	$v_{T}$ - thermal velocity [km s$^{-1}$] \\
	$v_{\rm wind}$ - stellar wind velocity [km s$^{-1}$] \\
	$W$ - work function [eV] \\
	$\xi$ - sputtering function [dimensionless] \\
	$X$ - composition fraction [dimensionless] \\
	$Y^{0}$ - backward sputtering yield at normal incidence [atoms ion$^{-1}$] \\
	$Z$ - charge number [dimensionless] \\
\end{flushleft}

\section{Turbulent Magnetic Field}
\label{app:turbmag}

\setcounter{figure}{0}
\makeatletter 
\renewcommand{\thefigure}{B\@arabic\c@figure}
\makeatother

\begin{figure*}[t]
	\centering
	\subfigure[]
	{\includegraphics[width=0.48\textwidth]{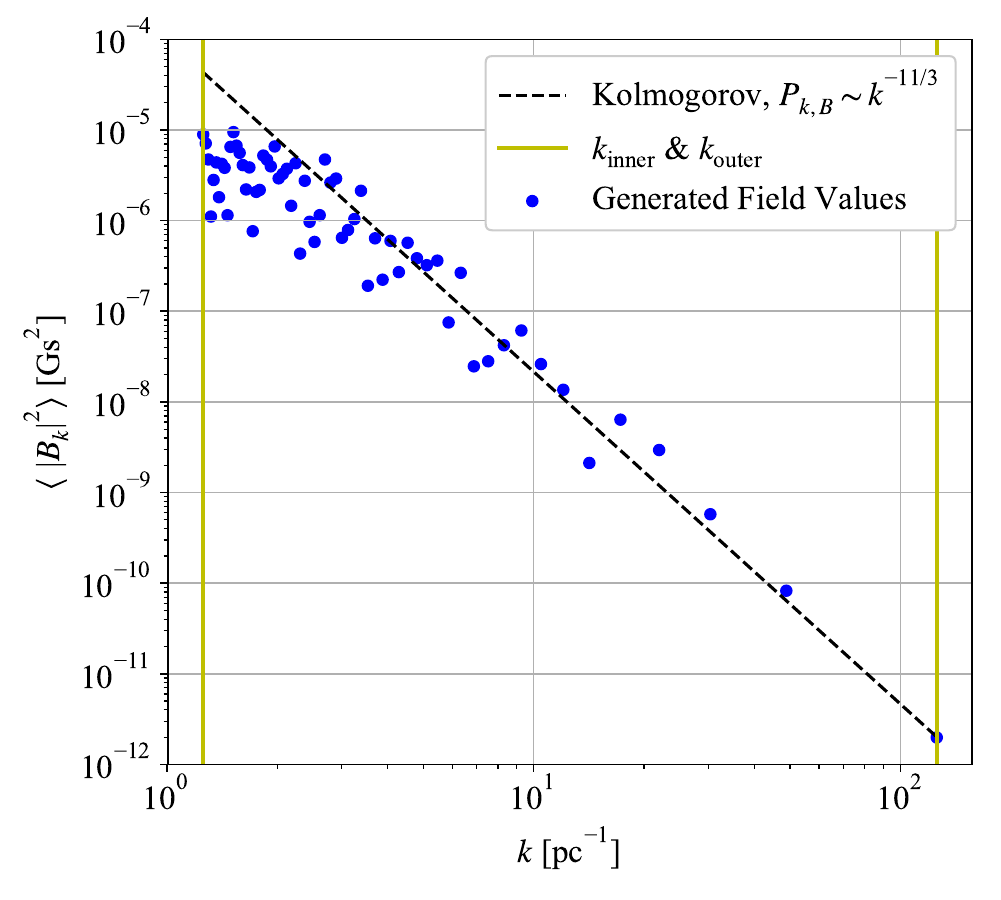}
		\label{fig:kolmogen}} ~	
	\subfigure[]
	{\includegraphics[width=0.48\textwidth]{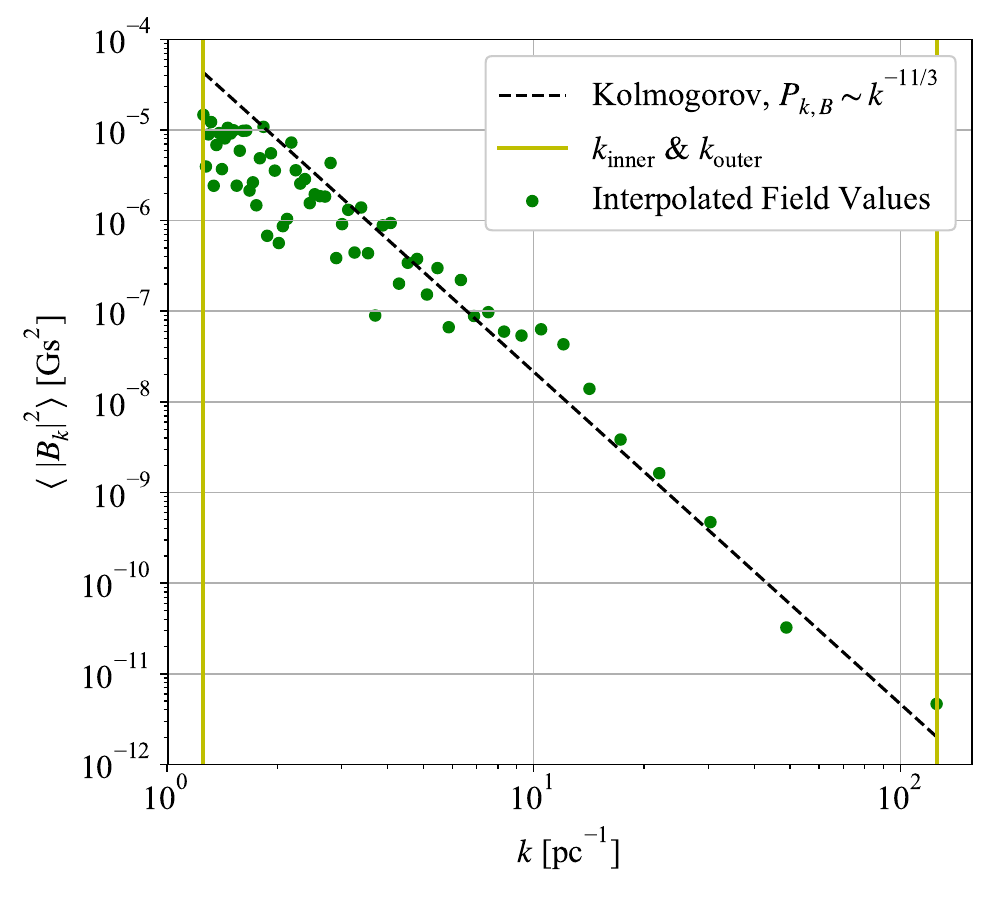}
		\label{fig:kolmoint}} \\
	\caption{Sample power spectra for MHD turbulence.  The left panel (a) shows the generated grid power spectrum for a Kolmogorov profile (${\cal E}_{k} \sim k^{-5/3}$) with $\mathbb{C} = 1$ G$^{2}$  pc$^{17/3}$, $\lambda_{\rm outer} = 5$~pc, and $\lambda_{\rm inner} = 0.05$~pc.  The right panel (b) shows the interpolated field power spectrum for the same profile.  The dashed line is shown for reference.
		\label{fig:kolmo}
	}
\end{figure*}

We begin by assuming the ISM has fully-developed and stationary (time-translation invariant) MHD turbulence that is homogeneous and isotropic.  We define the turbulent field's total velocity, $\boldsymbol{v}$, and magnetic field, $\boldsymbol{B}$, at a point, $\boldsymbol{x}$:
\begin{align}
\boldsymbol{v}(\boldsymbol{x}) = \langle \boldsymbol{v} \rangle + \boldsymbol{u}(\boldsymbol{x}) &\rightleftharpoons \boldsymbol{v}_{k} = \langle \boldsymbol{v}_{k} \rangle + \boldsymbol{u}_{k} \, , \nonumber \\[6pt]
\boldsymbol{B}(\boldsymbol{x}) = \langle \boldsymbol{B} \rangle + \boldsymbol{b}(\boldsymbol{x}) &\rightleftharpoons \boldsymbol{B}_{k} = \langle \boldsymbol{B}_{k} \rangle + \boldsymbol{b}_{k} \, ,
\end{align}
where $\langle \boldsymbol{v} \rangle$ and $\langle \boldsymbol{B} \rangle$ are the average velocity and magnetic fields respectively, $\boldsymbol{u}$ and $\boldsymbol{b}$ are the perturbed velocity and magnetic fields respectively, and the subscript, $k$, denotes the Fourier counterpart (i.e., $\boldsymbol{B}(\boldsymbol{x}) \rightleftharpoons \boldsymbol{B}_{k}$).  Because the turbulence is homogeneous and isotropic, we assume $\langle \boldsymbol{B} \rangle \approx 0 \Rightarrow \boldsymbol{B}(\boldsymbol{x}) \approx \boldsymbol{b}(\boldsymbol{x})$.  Additionally, by Fourier analysis, the specific energy spectrum, ${\cal E}_{k}$, of the turbulence is:
\beq
{\cal E}_{k} = \frac{1}{2} \langle u_{k}^{2} \rangle + \frac{1}{8 \pi \rho} \langle b_{k}^{2} \rangle \, ,
\eeq
where $\rho$ is the average mass density of the ISM.  For fully-developed turbulence, we assume equipartition between the kinetic and magnetic energies $\left( \frac{1}{2} \langle u_{k}^{2} \rangle \approx \frac{1}{8 \pi \rho} \langle b_{k}^{2} \rangle \right) \Rightarrow {\cal E}_{k} \propto \langle b_{k}^{2} \rangle$.  In the typical description of turbulence, energy is injected into the system at some outer scale ($\lambda_{\rm outer} = 2 \pi/k_{\rm outer}$), and cascades from larger to smaller scales according to a power-law relation (${\cal E}_{k} \sim k^{\eta}$), until it is dissipated at some inner scale ($\lambda_{\rm inner} = 2 \pi/k_{\rm inner}$) by viscous forces.

In order to create a vector field with this behavior, we use the technique outlined by \citet{zel70} and \citet{efst85} and applied to turbulence in several papers \citep[e.g.,][]{dub95,wall98,wieb98,wats01}.  Most of these examples used this technique to generate incompressible velocity fields (i.e., $\boldsymbol{\nabla} \cdot \boldsymbol{v} = 0$) but, since we require our magnetic field to be divergence-free (i.e., $\boldsymbol{\nabla} \cdot \boldsymbol{B} = 0$), this technique is appropriate here as well.

First, we calculate the specific power spectrum, ${\cal P}$, from the energy spectrum of the desired field:
\beq
{\cal E}_{k}~dk = {\cal P}_{k}~d^{3}k \Rightarrow {\cal P}_{k, \boldsymbol{B}} \equiv \langle |B_{k}|^{2} \rangle \sim k^{\eta-2} \, .
\eeq
The magnetic field is divergence-free, $\boldsymbol{\nabla} \cdot \boldsymbol{B} = 0$, so:
\beq
\boldsymbol{B} = \boldsymbol{\nabla} \times \boldsymbol{A} \rightleftharpoons \boldsymbol{B}_{k} = i \boldsymbol{k} \times \boldsymbol{A}_{k} \, ,
\label{eq:fourierdiv}
\eeq
where $\boldsymbol{A}$ is the vector potential of the magnetic field.  This implies that the power spectrum of the potential is:
\beq
{\cal P}_{k, \boldsymbol{A}} \equiv \langle |A_{k}|^{2} \rangle \sim k^{\eta-4} \, ,
\eeq
for the Fourier components that are described by a Gaussian distribution \citep{dub95}.  In order to prevent an increase in the spectrum beyond scales where the energy is injected, a cutoff wavenumber is introduced so that \citep{dub95}:
\beq
{\cal P}_{k, \boldsymbol{A}} \equiv \langle |A_{k}|^{2} \rangle \sim \left( k^{2}+k_{\rm outer}^{2} \right)^{(\eta-4)/2} \, .
\eeq
The vector components of $\boldsymbol{A}_{k}$ (both real and imaginary) are generated via a Gaussian distribution \citep{wall98}:
\beq
f\left( A_{k} ; \sigma_{A} \right) = \frac{1}{\sigma_{A} \sqrt{2 \pi}} \exp \left[ - \pfrac{A_{k}^{2}}{2 \sigma_{A}^{2}} \right] \, ,
\eeq
\beq
\langle |A_{k, \{x, y, z\}}^{\rm \{Re, Im\}}|^{2} \rangle = \sigma_{A}^{2} = \mathbb{C} \left( k^{2}+k_{\rm outer}^{2} \right)^{(\eta-4)/2} \, ,
\eeq
\begin{align}
&A_{k, x} = A_{k, x}^{\rm Re} + i A_{k, x}^{\rm Im} \, , \nonumber \\[6pt]
&A_{k, y} = A_{k, y}^{\rm Re} + i A_{k, y}^{\rm Im} \, , \nonumber \\[6pt]
&A_{k, z} = A_{k, z}^{\rm Re} + i A_{k, z}^{\rm Im} \, , \nonumber \\[6pt]
&\boldsymbol{A}_{k} = \left\langle A_{k, x}, A_{k, y}, A_{k, z} \right\rangle \, ,
\end{align}
where $\sigma_{A}$ is the standard deviation and $\mathbb{C}$ is a constant that is the same for all components and is adjusted to scale to the desired value of $\left\langle B \right\rangle $.  The value of $\boldsymbol{B}_{k}$ is then found by Equation~\ref{eq:fourierdiv}, then the inverse Fourier transform of $\boldsymbol{B}_{k}$ is taken to find $\boldsymbol{B}(\boldsymbol{x})$.  The corresponding positions for $\boldsymbol{B}(\boldsymbol{x})$ are found by $\{x, y, z\} = 2 \pi/\{k_{x}, k_{y}, k_{z}\}$, see Figure~\ref{fig:kolmo}.

The spectral index, $\eta$, is chosen based on the desired phenomenology; for example, the Kolmogorov spectrum \citep{kol41,gold95}:  $\eta = -5/3$, the Iroshnikov-Kraichnan/strong spectrum \citep{iro64,krai65}:  $\eta = -3/2$, and the universal/weak spectrum:  $\eta = -2$.  The Kolmogorov spectrum assumes an incompressible fluid which is not the case in most astrophysical environments. Nevertheless, as noted by \citet{dub95,gold95}, this spectrum appears in many contexts including solar wind turbulence \citep{matt82}.

This procedure will create a turbulent field within a $\lambda_{\rm outer}^{3}$ grid, see Figure~\ref{fig:magdiv}.  In order to fill in more volume while minimizing computation time and memory, the entire simulation volume (200$^{3}$ pc$^{3}$) is divided into $\lambda_{\rm outer}^{3}$ boxes and each box is filled with 1 of 24 randomly-chosen, possible orientations of the generated turbulent field.  These grid values of $\boldsymbol{B}(\boldsymbol{x})$ can now be interpolated in order to find the initial  $\boldsymbol{B}(\boldsymbol{x})$ field at all points.

\section{Interpolating the Magnetic Field}
\label{app:interpmag}

In creating a scheme for interpolating a magnetic field, $\boldsymbol{B}$, from $N$ data points/values, it is first helpful to stipulate the properties of the final scheme.  First, the interpolated field must satisfy Gauss' Law for Magnetic Fields, i.e., the magnetic field should be divergence-free everywhere ($\boldsymbol{\nabla} \cdot \boldsymbol{B} = 0$).  Secondly, the scheme should yield the value of the data input magnetic field, $\boldsymbol{B}_i$, at each data point, $\boldsymbol{x}_i$, i.e., $\boldsymbol{B}(\boldsymbol{x}_i) = \boldsymbol{B}_i(\boldsymbol{x}_i)$, see also \citet{mcnal11}.

To begin, let an individual magnetic field, $\boldsymbol{B}_{i}$, be the curl of a vector potential, $\boldsymbol{A}_{i}$, such that $\boldsymbol{B}_{i} = \boldsymbol{\nabla} \times \boldsymbol{A}_{i}$.  The total magnetic field is the superposition of the $N$ individual fields:
\beq
\boldsymbol{B} = \sum^{N} _{i=1} \boldsymbol{B}_{i} = \sum^{N} _{i=1} \boldsymbol{\nabla} \times \boldsymbol{A}_{i} \, .
\label{eq:magfield}
\eeq
In addition, we define another vector field, $\psi(\boldsymbol{r}) \boldsymbol{\mathbb{A}}$, at a position, $\boldsymbol{r}$, such that $\boldsymbol{\mathbb{A}}$ is the value of the field at $\boldsymbol{r} = 0$ and $\psi(\boldsymbol{r})$ is a scaling function with the properties that $\psi = 1$ at $\boldsymbol{r} = 0$, $\psi \rightarrow 0$ as $\boldsymbol{r} \rightarrow \infty$, and $\psi \geq 0~\forall~\boldsymbol{r}$.  Although several functions satisfy these properties, for convenience we chose $\psi(\boldsymbol{r}) \equiv \exp \left[ -\mathbb{S} r^{2} \right]$ where $r^{2}=x^{2}+y^{2}+z^{2}$ and $x$, $y$, and $z$ are the components of $\boldsymbol{r}$.  The scaling factor, $\mathbb{S}$, adjusts the influence of the data value at radius $r$.  We chose a value of the reciprocal mean of the squared radii to the data points:  $\mathbb{S}_{i} = 1/\left\langle r^{2}_{i} \right\rangle $.

We then define the vector potential, $\boldsymbol{A}_{i}$, at a position, $\boldsymbol{x}$, in terms of the new vector, $\psi_{i} \boldsymbol{\mathbb{A}}_{i}$: 
\beq
\boldsymbol{A}_{i}(\boldsymbol{x}) = \boldsymbol{\nabla} \times \left( \psi (\boldsymbol{x}-\boldsymbol{x}_{i}) \boldsymbol{\mathbb{A}}_{i} \right) \, ,
\label{eq:magpotential}
\eeq
where $\boldsymbol{\mathbb{A}}_{i}$ is the value of the vector potential at $\boldsymbol{x}_{i}$.  This means that the entire vector potential, $\boldsymbol{A}_{i}$, can be defined in terms of a single vector, $\boldsymbol{\mathbb{A}}_{i}$.

Combining Equation~\ref{eq:magfield} and Equation~\ref{eq:magpotential}, we can define the total magnetic field, $\boldsymbol{B}(\boldsymbol{x})$ at a position $\boldsymbol{x}$, as the superposition of $N$ vectors, $\boldsymbol{\mathbb{A}}_{i}$, scaled by a radial basis function, $\psi$:
\begin{align}
\boldsymbol{B}(\boldsymbol{x}) &= \sum ^{N}_{i=1} \boldsymbol{\nabla} \times \boldsymbol{A}_{i}(\boldsymbol{x}) = \sum ^{N}_{i=1} \boldsymbol{\nabla} \times \left( \boldsymbol{\nabla} \times \left( \psi (\boldsymbol{x}-\boldsymbol{x}_{i}) \boldsymbol{\mathbb{A}}_{i} \right) \right) \nonumber \\[6pt]
&= \sum ^{N}_{i=1} \left[ \boldsymbol{\nabla} \left( \boldsymbol{\nabla} \cdot \left( \psi (\boldsymbol{x}-\boldsymbol{x}_{i}) \boldsymbol{\mathbb{A}}_{i} \right) \right) \right. \nonumber \\[6pt]
& \quad \quad \quad \quad \left. - \boldsymbol{\nabla}^{2} \left( \psi (\boldsymbol{x}-\boldsymbol{x}_{i}) \boldsymbol{\mathbb{A}}_{i} \right) \vphantom{}\right] \, .
\label{eq:maginterp}
\end{align}
At this point, the $N$ values of $\boldsymbol{\mathbb{A}}_i$ are unknown, but since we are interpolating over $N$ data points, $\boldsymbol{B}_{i}$, we can set up a system of $N$ equations to solve for the unknowns:
\beq
\boldsymbol{B}_j(\boldsymbol{x}_{j}) = \sum ^{N}_{i=1} \boldsymbol{\nabla} \times \left( \boldsymbol{\nabla} \times \left( \psi (\boldsymbol{x}_{j}-\boldsymbol{x}_{i}) \boldsymbol{\mathbb{A}}_{i} \right) \right) \, .
\eeq
This ensures the second property of our desired scheme is met, namely $\boldsymbol{B}(\boldsymbol{x}_i) = \boldsymbol{B}_i(\boldsymbol{x}_i)$.

Lastly, because we originally defined the magnetic field as the curl of a vector potential, it will be divergence-free by construction, since the divergence of a curl is always zero $\boldsymbol{\nabla} \cdot \left( \boldsymbol{\nabla} \times \boldsymbol{A} \right) = 0$.  However, we can check our final scheme as well to verify that our introduction of an addition vector field, $\boldsymbol{\mathbb{A}}$ has not altered this property.  Using Equation~\ref{eq:maginterp}, we find:
\begin{align}
\boldsymbol{\nabla} \cdot \boldsymbol{B} &= \sum ^{N}_{i=1} \left[ \boldsymbol{\nabla} \cdot \boldsymbol{\nabla} \left( \boldsymbol{\nabla} \cdot \left( \psi (\boldsymbol{x}-\boldsymbol{x}_{i}) \boldsymbol{\mathbb{A}}_{i} \right) \right) \right. \nonumber \\[6pt]
& \quad \quad \quad \quad \left. - \boldsymbol{\nabla} \cdot \left( \boldsymbol{\nabla}^{2} \left( \psi (\boldsymbol{x}-\boldsymbol{x}_{i}) \boldsymbol{\mathbb{A}}_{i} \right) \right) \vphantom{}\right] \nonumber \\[6pt]
&= \sum ^{N}_{i=1} \left[ \boldsymbol{\nabla}^{2} \left( \boldsymbol{\nabla} \cdot \left( \psi (\boldsymbol{x}-\boldsymbol{x}_{i}) \boldsymbol{\mathbb{A}}_{i} \right) \right) \right. \nonumber \\[6pt]
& \quad \quad \quad \quad \left. - \boldsymbol{\nabla}^{2} \left( \boldsymbol{\nabla} \cdot \left( \psi (\boldsymbol{x}-\boldsymbol{x}_{i}) \boldsymbol{\mathbb{A}}_{i} \right) \right) \vphantom{}\right] = 0 \, .
\end{align}
Therefore, our interpolated magnetic field remains divergence-free.

\section{Flux-freezing with Spherical Symmetry}
\label{app:fluxfreeze}

\setcounter{figure}{0}
\makeatletter 
\renewcommand{\thefigure}{D\@arabic\c@figure}
\makeatother

\begin{figure*}[t]
	\centering
	\subfigure[]
	{\includegraphics[width=0.48\textwidth]{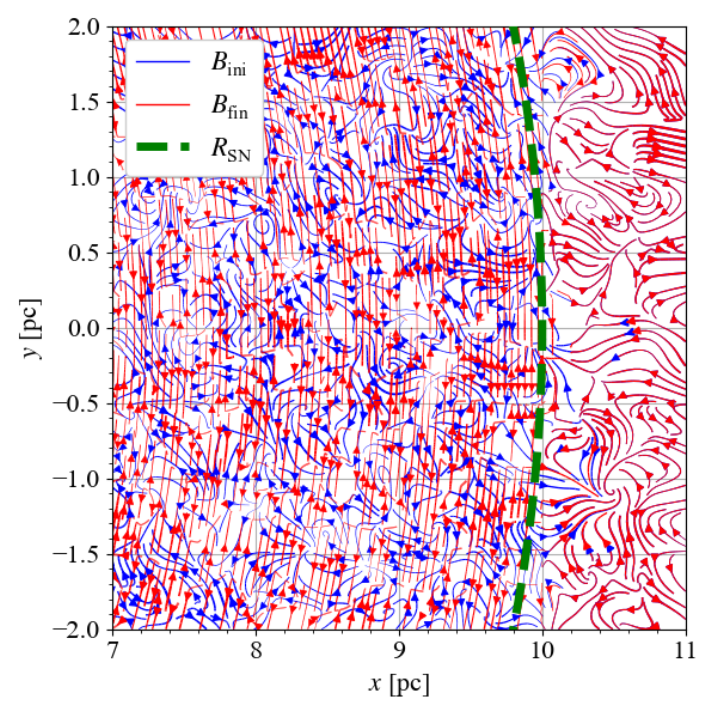}
		\label{fig:fluxfreezexy}} ~	
	\subfigure[]
	{\includegraphics[width=0.48\textwidth]{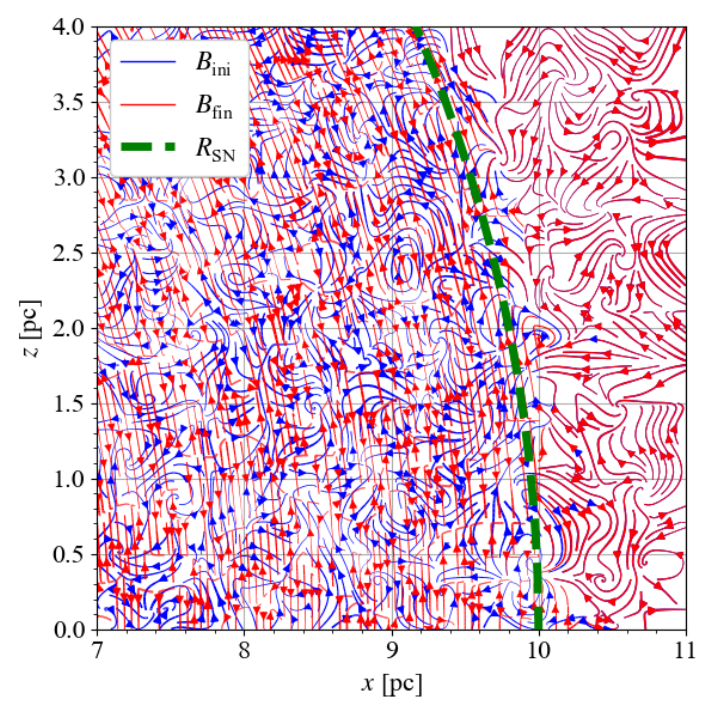}
		\label{fig:fluxfreezexz}} \\
	\caption{Comparison of initial and final magnetic field conditions after the passage of a shock wave with a generic Sedov profile.  The initial magnetic field (shown in blue) is stretched and compressed (red lines) as the shock wave passes through the medium.  Here the spherical shock is centered at the origin and has a radius of 10 pc (dashed green line).  [Note:  The slight variations between the initial and final magnetic fields just ahead of the shock wave are a product of the plotting algorithm not the calculation.]
		\label{fig:fluxfreeze}
	}
\end{figure*}

\begin{figure*}[t]
	\centering
	\includegraphics[height=7in]{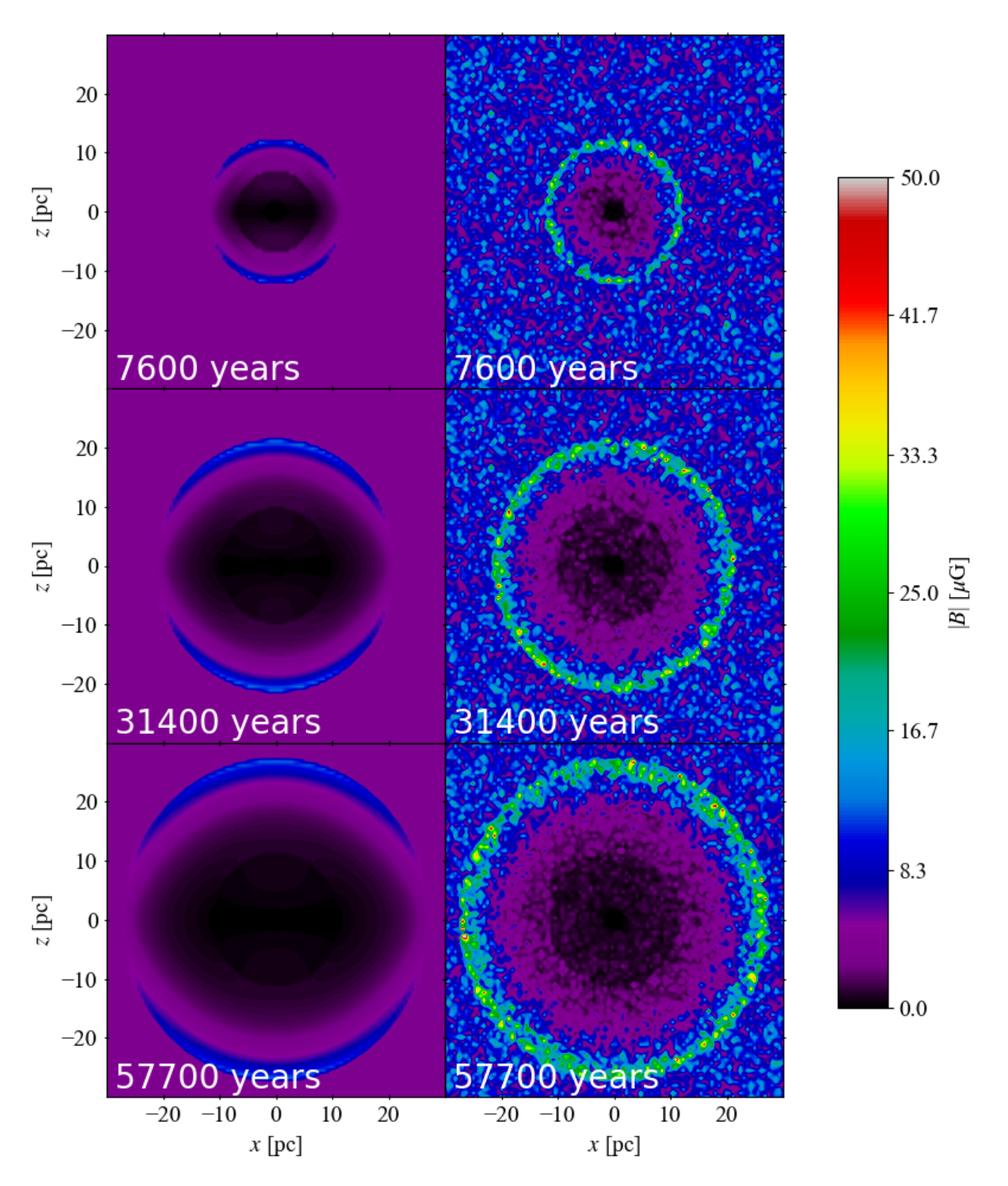}
	\caption[Magnetic Field Model Output.]{Outputs from magnetic field model for comparison to Figure 1, \cite{bals01}.  The left column is the uniform magnetic field case; the right column is the turbulent magnetic field case.  The output times are at 7600, 31,400, and 57,700 yr for the top, middle, and bottom rows, respectively.
		\label{fig:BalsaraComparison}
	}
\end{figure*}

In the case of a spherical expansion of a plasma, the magnetic fields will be ``frozen in'' the plasma as it expands.  If the expansion of the fluid can be determined entirely by the basic (i.e., non-MHD) fluid equations, then it is possible to solve for the magnetic field as the plasma expands.  Using the integral definition of magnetic flux:  $\mathbb{F}_{\rm mag} = \int \boldsymbol{B} \cdot d\boldsymbol{A}$, the initial and final magnetic fluxes through the surface containing a fluid element will be the same, i.e., $\mathbb{F}_{\rm mag, ini} = \mathbb{F}_{\rm mag, fin}$.  Since we are following a particular fluid element, the mass contained within will remain the same as well, i.e., $m_{\rm ini} = m_{\rm fin}$.

Using a spherical coordinate system with the origin at the center of the expansion, we define a fluid element with differential volume:
\beq
dV = r^{2} \sin \theta~dr~d\theta~d\phi \, ,
\eeq
and differential areas:
\begin{align}
dA_{\rm face} &= r^{2} \sin \theta~d\theta~d\phi = r^{2}~d\Omega \, , \\[6pt]
dA_{\rm top} &= r \sin \theta~dr~d\phi \, , \\[6pt]
dA_{\rm side} &= r~dr~d\theta \, ,
\end{align}
with $dA_{\rm face}$ the surface facing the direction of expansion, $dA_{\rm top}$ the upper surface, and $dA_{\rm side}$ one of the side surfaces of the fluid element.  The remaining three surfaces of the fluid element have the same areas, but because of Gauss' Law ($\oint \boldsymbol{B} \cdot d\boldsymbol{A} = 0$), we focus on three sides only.

As the fluid element moves away from the origin, by spherical symmetry, the angular properties of the fluid element will remain the same:
\beq
d\theta_{\rm ini} = d\theta_{\rm fin} \, , \quad d\phi_{\rm ini} = d\phi_{\rm fin} \, , \quad \sin \theta_{\rm ini} = \sin \theta_{\rm fin} \, , \nonumber
\eeq
\beq
\sin \theta_{\rm ini}~d\theta_{\rm ini}~d\phi_{\rm ini} = \sin \theta_{\rm fin}~d\theta_{\rm fin}~d\phi_{\rm fin} \nonumber
\eeq
\beq
\Rightarrow d\Omega_{\rm ini} = d\Omega_{\rm fin} \, .
\eeq
Additionally, the fluid element will compress and expand, but while its mass will remain constant, its density will change:
\beq
\rho_{\rm ini} = \frac{m_{\rm ini}}{r_{\rm ini}^{2}~dr_{\rm ini}~d\Omega_{\rm ini}} \, , \quad \quad \rho_{\rm fin} = \frac{m_{\rm fin}}{r_{\rm fin}^{2}~dr_{\rm fin}~d\Omega_{\rm fin}} \, ,
\eeq
\beq
\Rightarrow \rho_{\rm ini} r_{\rm ini}^{2}~dr_{\rm ini} = \rho_{\rm fin} r_{\rm fin}^{2}~dr_{\rm fin} \nonumber
\eeq
\beq
\Rightarrow \frac{dr_{\rm ini}}{dr_{\rm fin}} = \frac{\rho_{\rm fin} r_{\rm fin}^{2}}{\rho_{\rm ini} r_{\rm ini}^{2}} \, .
\label{eq:fluxdensity}
\eeq
For an infinitesimally small fluid element, the magnetic field will be uniform throughout the entire fluid element, and we can decompose the vector $\boldsymbol{B}$ into a component parallel to the direction of expansion, $\boldsymbol{B}_{\parallel}$, and a component orthogonal to the direction of expansion, $\boldsymbol{B}_{\perp}$:
\beq
\boldsymbol{B}_{\perp} \equiv \boldsymbol{B} \times \boldsymbol{\hat{r}} \, , \quad \boldsymbol{B}_{\perp} \equiv \boldsymbol{B} - \left( \boldsymbol{B} \cdot \boldsymbol{\hat{r}} \right) \boldsymbol{\hat{r}} \, ,
\eeq
\beq
\boldsymbol{B} = \boldsymbol{B}_{\perp} + \boldsymbol{B}_{\parallel} \, , \quad B^{2} = B_{\perp}^{2} + B_{\parallel}^{2} \, .
\eeq
With these definitions, we can calculate the flux through each surface:
\beq
d\mathbb{F}_{\rm mag} = \boldsymbol{B} \cdot d\boldsymbol{A}
\eeq
\beq
\Rightarrow d\mathbb{F}_{\rm mag, face} = B_{\parallel} r^{2}~d\Omega \nonumber
\eeq
\beq
\Rightarrow B_{\rm \parallel, ini} r_{\rm ini}^{2}~d\Omega_{\rm ini} = B_{\rm \parallel, fin} r_{\rm fin}^{2}~d\Omega_{\rm fin} \nonumber
\eeq
\beq
B_{\rm \parallel, fin} = B_{\rm \parallel, ini} \pfrac{r_{\rm ini}}{ r_{\rm fin}}^{2} \, .
\label{eq:paraflux}
\eeq
Defining $\alpha$ as the angle between the normal of the top surface and $\boldsymbol{B}_{\perp}$:
\beq
d\mathbb{F}_{\rm mag, top} = B_{\perp} \cos \alpha~r \sin \theta~dr~d\phi \nonumber
\eeq
\begin{align}
\Rightarrow &B_{\rm \perp, ini} \cos \alpha_{\rm ini}~r_{\rm ini} \sin \theta_{\rm ini}~dr_{\rm ini}~d\phi_{\rm ini} = \nonumber \\[6pt]
&= B_{\rm \perp, fin} \cos \alpha_{\rm fin}~r_{\rm fin} \sin \theta_{\rm fin}~dr_{\rm fin}~d\phi_{\rm fin} \nonumber
\end{align}
\beq
\Rightarrow B_{\rm \perp, fin} = B_{\rm \perp, ini} \pfrac{\cos \alpha_{\rm ini}}{\cos \alpha_{\rm fin}} \pfrac{r_{\rm ini}}{r_{\rm fin}} \pfrac{dr_{\rm ini}}{dr_{\rm fin}} \, . \nonumber
\eeq
Using Equation~\ref{eq:fluxdensity}, we have:
\beq
B_{\rm \perp, fin} = B_{\rm \perp, ini} \pfrac{\cos \alpha_{\rm ini}}{\cos \alpha_{\rm fin}} \pfrac{r_{\rm ini}}{r_{\rm fin}} \pfrac{\rho_{\rm fin} r_{\rm fin}^{2}}{\rho_{\rm ini} r_{\rm ini}^{2}}
\eeq
\beq
\Rightarrow B_{\rm \perp, fin} = B_{\rm \perp, ini} \pfrac{\cos \alpha_{\rm ini}}{\cos \alpha_{\rm fin}} \pfrac{\rho_{\rm fin} r_{\rm fin}}{\rho_{\rm ini} r_{\rm ini}}
\label{eq:topflux}
\eeq
\beq
d\mathbb{F}_{\rm mag, side} = B_{\perp} \sin \alpha~r \sin \theta~dr~d\phi 
\eeq
\begin{align}
\Rightarrow &B_{\rm \perp, ini} \sin \alpha_{\rm ini}~r_{\rm ini} \sin \theta_{\rm ini}~dr_{\rm ini}~d\phi_{\rm ini} = \nonumber \\[6pt]
&= B_{\rm \perp, fin} \sin \alpha_{\rm fin}~r_{\rm fin} \sin \theta_{\rm fin}~dr_{\rm fin}~d\phi_{\rm fin} \nonumber
\end{align}
\beq
\Rightarrow B_{\rm \perp, fin} = B_{\rm \perp, ini} \pfrac{\sin \alpha_{\rm ini}}{\sin \alpha_{\rm fin}} \pfrac{r_{\rm ini}}{r_{\rm fin}} \pfrac{dr_{\rm ini}}{dr_{\rm fin}} \, . \nonumber
\eeq
Using Equation~\ref{eq:fluxdensity}, we have:
\beq
B_{\rm \perp, fin} = B_{\rm \perp, ini} \pfrac{\sin \alpha_{\rm ini}}{\sin \alpha_{\rm fin}} \pfrac{r_{\rm ini}}{r_{\rm fin}} \pfrac{\rho_{\rm fin} r_{\rm fin}^{2}}{\rho_{\rm ini} r_{\rm ini}^{2}}
\eeq
\beq
\Rightarrow B_{\rm \perp, fin} = B_{\rm \perp, ini} \pfrac{\sin \alpha_{\rm ini}}{\sin \alpha_{\rm fin}} \pfrac{\rho_{\rm fin} r_{\rm fin}}{\rho_{\rm ini} r_{\rm ini}}
\label{eq:sideflux}
\eeq
In order for both Equation~\ref{eq:topflux} and Equation~\ref{eq:sideflux} to be true we must have:
\beq
\Rightarrow \frac{\cos \alpha_{\rm ini}}{\cos \alpha_{\rm fin}} = \frac{\sin \alpha_{\rm ini}}{\sin \alpha_{\rm fin}} = 1
\eeq
\beq
\Rightarrow B_{\rm \perp, fin} = B_{\rm \perp, ini} \pfrac{\rho_{\rm fin} r_{\rm fin}}{\rho_{\rm ini} r_{\rm ini}} \, .
\label{eq:perpflux}
\eeq
Combining Equation~\ref{eq:paraflux} and Equation~\ref{eq:perpflux} gives a means of relating initial conditions and final densities to the final magnetic field.  An example of the flux-freezing transformation is shown in Figure~\ref{fig:fluxfreeze}.

To check the reasonableness of our approach, we compared our method to a 3D, full-MHD simulation of the magnetic field.  \cite{bals01} carried out three-dimensional MHD simulations of SN occurring in a uniform and turbulent, magnetized ISM.  Using the same initial conditions as \cite{bals01} (namely $n_{\rm ISM} = 1~{\rm cm}^{-3}$, $T_{\rm ISM} = 8000$ K, $M_{\rm ej} = 5~\msol$, $E_{\rm SN} = 10^{51}$ erg, $\left\langle \left| \boldsymbol{B}_{\rm ISM} \right| \right\rangle = 3.57 \ \mu{\rm G}$ and $\sqrt{\left\langle B_{\rm ISM}^2 \right\rangle} = 8.94 \ \mu{\rm G}$ with a Kolmogorov spectrum) we compared our results to \cite{bals01}, see Figure~\ref{fig:BalsaraComparison} and compare to Figure 1 of \cite{bals01}.  At the final time step, $t = 57,700$ yr, our SN model results (in comparison with \citeauthor{bals01}) are:  $R_{\rm SN} = 27$ pc (25.5 pc), $T_{\rm SN} = 7.2 \times 10^5$ K ($4.2 \times 10^5$ K), $v_{\rm SN} = 133$ km s$^{-1}$ ($174$ km s$^{-1}$), $P_{\rm SN} = 3.7 \times 10^{-10}$ dyn cm$^{-2}$ ($5.2 \times 10^{-10}$ dyn cm$^{-2}$), $\rho_{\rm SN} = 3.3 \rho_{\rm ISM}$ ($3.3 \rho_{\rm ISM}$), $B_{\rm postshock} = 3 B_{\rm preshock}$ ($2.4 B_{\rm preshock}$, although \citeauthor{bals01} notes the expected value is $3.3 B_{\rm preshock}$).  \citeauthor{bals01} noted their resolution prevented them from getting the full magnetic compressional effect.  Given the qualitative and fairly close quantitative agreement between our hydrodynamic/magnetic field approach and the 3D, full MHD approach used by \cite{bals01}, we expect our approach to be a sufficiently good approximation of 3D MHD for the time period we examine.

\section{Grain Charging:  Processes}
\label{app:graincharging}

As grains move within the SNR, they will acquire/lose electrons and ions due to impacts with the plasma or photons.  Several processes can influence the total charge of the grain, so the total charging rate, $dq_{\rm gr}/dt$ is:
\beq
\frac{dq_{\rm gr}}{dt} = \sum_{i} {\cal I}_{i} \, ,
\eeq
which is summed over $i$ processes of currents, ${\cal I}_{j}$.  These currents are due to impinging plasma particles, ${\cal I}_{\rm imp}$, and the associated secondary electrons emitted, ${\cal I}_{\rm see}$, transmitted plasma particles, ${\cal I}_{\rm trans}$, and photoelectron emission, ${\cal I}_{\gamma}$.  The following derivations are the same as used by \citet{kim98}.

\subsection{Impinging Ions/Electrons}
\label{subsubsect:Jimp}

Charging by impinging plasma particles is caused by incident ions/electrons, $j$, impacting the surface of the grain, sticking, and altering the grain charge.  It is given by \citep{dwek92}:
\begin{align}
{\cal I}_{\rm imp} = &~2 \pi e \sum_{j} \left\lbrace Z_{j} \int_{v_{\rm A}}^{\infty} dv_{T} \right. \nonumber \\[6pt]
&\left. \times \int_{0}^{\pi} d\theta~C_{{\rm coll}, j}(v_{T}) f_{j}(v_{T}, \theta) v_{T}^{3} \sin \theta \vphantom{}\right\rbrace \, ,
\label{eq:Jimp}
\end{align}
where $v_{T}$ is the thermal velocity of the plasma.  The minimum impinging velocity, $v_{\rm A}$, is the given by:
\beq
v_{\rm A} = 
\begin{cases} 
	0 & \quad Z_{j}\Phi_{j} \leq 0 \, , \\[6pt]
	\left(2 Z_{j} e U_{\rm gr}/m_{j} \right)^{1/2} & \quad Z_{j}\Phi_{j} > 0 \, , \\[6pt]
\end{cases}
\eeq
with the collisional cross-section, $C_{{\rm coll}, j}(v_{T})$, given by:
\beq
C_{{\rm coll}, j}(v_{T}) = \pi a_{\rm gr}^{2} \left(1- \frac{2 Z_{j}e U_{\rm gr}}{m_{j} v_{T}^{2}} \right) \, .
\eeq
Because the dust grains will potentially have large relative velocities to the plasma, as well as large thermal velocities, we assume a Maxwellian velocity distribution, $f_{j}(v_{T}, \theta)$:
\begin{align}
f_{j}(v_{T}, \theta) =&~n_{j} \pfrac{m_{j}}{2 \pi k_{\rm B}T}^{3/2} \nonumber \\[6pt]
&\times \exp\left[ -\frac{m_{j}}{2 k_{\rm B}T} \left(v_{T}^{2} + v_{\rm rel}^{2} -2 v_{T} v_{\rm rel} \cos \theta \right) \right] \, ,
\end{align}
where $\theta$ is the angle between the thermal and relative velocities.

\subsection{Secondary Electron Emission}
\label{subsubsect:Jsee}

\setcounter{table}{0}
\makeatletter 
\renewcommand{\thetable}{E\@arabic\c@table}
\makeatother

{\renewcommand{\arraystretch}{1.75}
\begin{table}[t]
	\caption{Summary of Sputtering Yield and Escape Length ($\lambda_{\rm esc}$) Parameters}
	\label{tab:sputtering}
	\begin{tabularx}{0.48\textwidth}{C C C C C C c} \hline \hline
		Dust Species	& $E_{\rm bind}$ [eV]	& $Z_{\rm target}$	& $m_{\rm target}$ [$m_{\rm u}$] & $\kappa$		& $\pounds_{e}$	& $R_{m}(\pounds_{e})$ 	\\
		Fe 				& 4.31			& 26		& 56		& 0.23			& 1.5662	& 1.1891			\\ \hline
	\end{tabularx}
	{Values are from \citep[][and references therein]{noz06}.  Electron stopping ranges are based on outputs from the CASINO software \citep{drou07}. } \\
\end{table}}

If the impinging plasma particles have sufficient initial energy, $E_{\rm ini} = 2k_{\rm B}T + \frac{1}{2} m_{j} v_{\rm rel}^{2} + Z_{j}e U_{\rm gr}$ \citep{drain79,mckee87,kim98}, then after initially ejecting an electron, there is sufficient energy to eject additional electrons.  In this situation, the current of secondary electrons, ${\cal I}_{\rm see}$, is:
\begin{align}
{\cal I}_{\rm see} =&~2 \pi e \sum_{j} \left\lbrace \delta_{j}(E_{\rm ini}) \int_{E_{\rm min}}^{\infty} dE~\varrho_{j}(E) \right. \nonumber \\[6pt]
&\left. \times \int_{v_{\rm A}}^{\infty} dv_{T} \int_{0}^{\pi} d\theta~C_{{\rm coll}, j}(v_{T}) f_{j}(v_{T}, \theta) v_{T}^{3} \sin \theta \vphantom{}\right\rbrace \, ,
\label{eq:Jsee}
\end{align}
where the minimum required energy is $E_{\rm min} = \max[0, e U_{\rm gr}]$.  The type and energy of impacting plasma particles will determine the effectiveness of secondary emission, so the secondary electron yield, $\delta_{j}(E_{\rm ini})$, is given for electrons by \citep{drain79}:
\begin{align}
\delta_{e}(E_{\rm ini}) =&~\delta_{\rm max} \frac{8 (E_{\rm ini}/E_{\rm max})}{(1 + E_{\rm ini}/E_{\rm max})^{2}} \left( 1 - \exp\left[ -\frac{4a_{\rm gr}}{3\lambda_{\rm esc}} \right] \right) \nonumber \\[6pt] 
&\times f_{1}\pfrac{4a_{\rm gr}}{3R} f_{2}\pfrac{a_{\rm gr}}{\lambda_{\rm esc}} \, ,
\end{align}
where the fitting functions, $f_{1}$ and $f_{2}$, are given by:
\beq
f_{1}(\chi) = \frac{1.6 + 1.4 \chi^{2} + 0.54 \chi^{4}}{1 + 0.54 \chi^{4}} \, ,
\eeq
\beq
f_{2}(\chi) = \frac{1 + 2 \chi^{2} + \chi^{4}}{1 + \chi^{4}} \, ,
\eeq
and the escape length for electrons, $\lambda_{\rm esc}$ is:
\beq
\lambda_{\rm esc} = R_{e}(E_{\rm max})/R_{m}(\pounds_{e})^{\pounds_{e}} \, ,
\eeq
where the value of $R_{m}$ is given for various materials in Table \ref{tab:sputtering}.  The maximum yield from a bulk solid, $\delta_{\rm max}$, at energy, $E_{\rm max}$, is assumed to be 1.3 and 400 eV respectively \citep{CRC}.  The secondary yield for ions is given by the empirical formula in \citet{drain79}:
\begin{align}
\delta_{\rm ion}(E_{\rm ini}) = &~0.1 Z_{j}^{2} \frac{1 + (m_{H}/m_{j})(E_{\rm ini}/E_{1})}{\left[1 + (m_{H}/m_{j})(E_{\rm ini}/E_{2})\right]^{2}} \nonumber \\[6pt]
&\times 
\begin{cases} 
1 & \quad U_{\rm gr} \leq 0 \, , \\[6pt]
\displaystyle 1 + \pfrac{U_{\rm gr}}{\rm 1~V} & \quad U > 0 \, , \\[6pt]
\end{cases}
\end{align}
where $E_{1} = 500$ eV and $E_{2} = 35$ keV.

Lastly, the energy distributions $\varrho_{j}$ for secondary electrons emitted by electrons and ions are given by:
\begin{align}
\varrho_{e}(E) &= \frac{E}{2 E_{e}^{2}} \left[1 + \frac{1}{2} \pfrac{E}{E_{e}}^{2} \right]^{-3/2} \, , \\[6pt]
\varrho_{\rm ion}(E) &= \frac{1}{E_{\rm ion}} \left[1 + \frac{1}{2} \pfrac{E}{E_{\rm ion}}^{2} \right]^{-2} \, ,
\end{align}
where the most probable energies are $E_{e} = 2$ eV and $E_{\rm ion} = 1$ eV.

\subsection{Transmission of Ions/Electrons}
\label{subsubsect:Jtran}

The transmission \citep[also referred to as tunneling,][]{chow93} current of ions/electrons, ${\cal I}_{\rm tran}$, accounts for the plasma particles with sufficient velocity to penetrate completely through the grain without being captured:
\begin{align}
{\cal I}_{\rm tran} = &-2 \pi e \sum_{j} \left\lbrace Z_{j} \int_{v_{\rm B}}^{\infty} dv_{T} \right. \nonumber \\[6pt]
&\left. \times \int_{0}^{\pi} d\theta~C_{{\rm coll}, j}(v_{T}) f_{j}(v_{T}, \theta) v_{T}^{3} \sin \theta \vphantom{}\right\rbrace \, ,
\label{eq:Jtran}
\end{align}
where the minimum velocity, $v_{\rm B}$, required to pass through the grain is \citep{drain79,mckee87,kim98}:
\beq
v_{\rm B} = 
\begin{cases} 
	\displaystyle \sqrt{\frac{\left( 2k_{\rm B}T + \frac{1}{2} m_{j} v_{\rm rel}^{2} \right) \omega_{j}}{m_{j}}} & \quad Z_{j}\Phi_{j} \leq 0 \, , \\[12pt]
	\displaystyle \sqrt{\frac{\left( 2k_{\rm B}T + \frac{1}{2} m_{j} v_{\rm rel}^{2} \right) \left( \omega_{j} + Z_{j}\Phi_{j} \right)}{m_{j}}} & \quad Z_{j}\Phi_{j} > 0 \, . \\
\end{cases}
\eeq
From \citet{drain79}, the energy, $\left( k_{\rm B}T + \frac{1}{2} m_{j} v_{\rm rel}^{2} \right) \omega_{j}$, required to penetrate a grain is given by:
\beq
\left( k_{\rm B}T + \frac{1}{2} m_{j} v_{\rm rel}^{2} \right) \omega_{j} = 
\begin{cases} 
	\Delta_{j} & \quad Z_{j} < 0 \, , \\[6pt]
	\Delta_{j} + E_{\rm Bohr} & \quad Z_{j} > 0 \, . \\
\end{cases}
\eeq
Additionally, we assume that the ions emerge neutral because of recombination if their energy is below the Bohr Energy, $E_{\rm Bohr} = (m_{j}/m_{e}) E_{H}$ with $E_{H} = 13.6$ eV.

The penetration threshold energy, $\Delta_{j}$, is found using an energy-range relation and the size of the dust grain \citep{fitt74}:
\beq
R_{j}(\Delta_{j}) \equiv 4a_{\rm gr}/3 \, .
\eeq
The energy-range relation is based on measured stopping ranges for various particles into materials, and we used outputs from the SRIM software \citep{zieg85,zieg10} for ion and the CASINO software \citep{drou07} for electron stopping in materials and fit power-law profiles to the results in the form:
\beq
R_{j}(E) = \mathbb{R}_{j} E ^{\beta_{j}} \, .
\eeq
A compilation of the fit values are listed in Table \ref{tab:stopping}.

\begin{table}[t]
	\caption{Stopping Ranges for Various Particles onto Iron}
		\label{tab:stopping}
		\begin{tabularx}{0.48\textwidth}{ c C C C } \hline \hline
			\multicolumn{4}{ c }{\multirow{3}{*}{$\displaystyle R_{\rm incident \rightarrow target} =  \mathbb{R} \pfrac{E}{1 \ {\rm keV}}^{\beta}$}} \\ \\ \\ \hline
			Incident$^{a,b}$& Target	& $\mathbb{R}$ [nm]	& $\beta$	\\ \hline
			${\it e^{-}}$	& Fe	& 5.1457	& 1.5662	\\
			H	& Fe	& 7.8944	& 0.9667	\\
			He	& Fe	& 4.6270	& 0.9490	\\
			C	& Fe	& 2.2539	& 0.8049	\\
			N	& Fe	& 2.0788	& 0.7794	\\
			O	& Fe	& 1.9576	& 0.7627	\\
			Ne	& Fe	& 1.7949	& 0.7360	\\
			Mg	& Fe	& 1.6854	& 0.7061	\\
			Si	& Fe	& 1.5552	& 0.6753	\\
			S	& Fe	& 1.5055	& 0.6499	\\
			Ca	& Fe	& 1.4367	& 0.6139	\\
			Fe	& Fe	& 1.4116	& 0.5679	\\
			Ni	& Fe	& 1.4014	& 0.5609	\\
			Zn	& Fe	& 1.4128	& 0.5413	\\
			Kr	& Fe	& 1.4238	& 0.5076	\\ \hline
		\end{tabularx}
		{\raggedright
			{
			$^{a}$ - Electron stopping ranges are based on outputs from the CASINO software \citep{drou07}. \\
			$^{b}$ - Ion stopping ranges are based on outputs from the SRIM software \citep{zieg85,zieg10}.} \\
		}
\end{table}

\subsection{Photoelectron Emission}
\label{subsubsect:Jgamma}

\setcounter{figure}{0}
\makeatletter 
\renewcommand{\thefigure}{E\@arabic\c@figure}
\makeatother

The dust grains will be exposed to UV photons, and, depending on the grain material, electrons will be emitted from the surface of the grain.  The photoelectric current, ${\cal I}_{\gamma}$, then is given by:
\begin{align}
{\cal I}_{\gamma} =&~e \int_{W + E_{\rm min}}^{\infty} d(h\nu)~C_{\rm abs}(h\nu) \mathbb{F}_{\gamma, h\nu}(h\nu) Y_{\gamma}(h\nu) \nonumber \\[6pt]
&\times \int_{E_{\rm min}}^{E_{\rm max}} dE~\varrho_{\gamma}(E) \, ,
\label{eq:Jgamma}
\end{align}
with $E_{\rm max} = h\nu - W$, $h$ is Planck's constant, $\nu$ is the photon frequency, and $W$ is the work function required to emit an electron.  Following the example of \citet{kim98}, we set $E_{\rm work} = W$.  The photoelectric yield, $Y_{\gamma}(h\nu)$, is the number of electrons emitted per absorbed photon \citep{drain79}:
\begin{align}
Y_{\gamma}(h\nu) =&~\frac{\left( h\nu - E_{\rm work} + E_{\rm low} \right)^{2} - E_{\rm low}^{2}}{(h\nu)^{2} - E_{\rm low}^{2}} \nonumber \\[6pt]
& \times \left[ 1 - \left(1 - \frac{\lambda_{\rm esc}}{a_{\rm gr}} \right)^{3} \right] \, ,
\end{align}
where $E_{\rm work} = 8$ eV and $E_{\rm low} = 6$ eV.  The energy distribution, $\varrho_{\gamma}(E)$, of photoelectrons \citep{grard73}:
\beq
\varrho_{\gamma}(E) = \frac{E}{E_{\gamma}^{2}} \exp\left[- \frac{E}{E_{\gamma}}\right] \, ,
\eeq
where $E_{\gamma} = 1$ eV.

\begin{figure}[t]
	\begin{center}
		\includegraphics[width=0.48\textwidth]{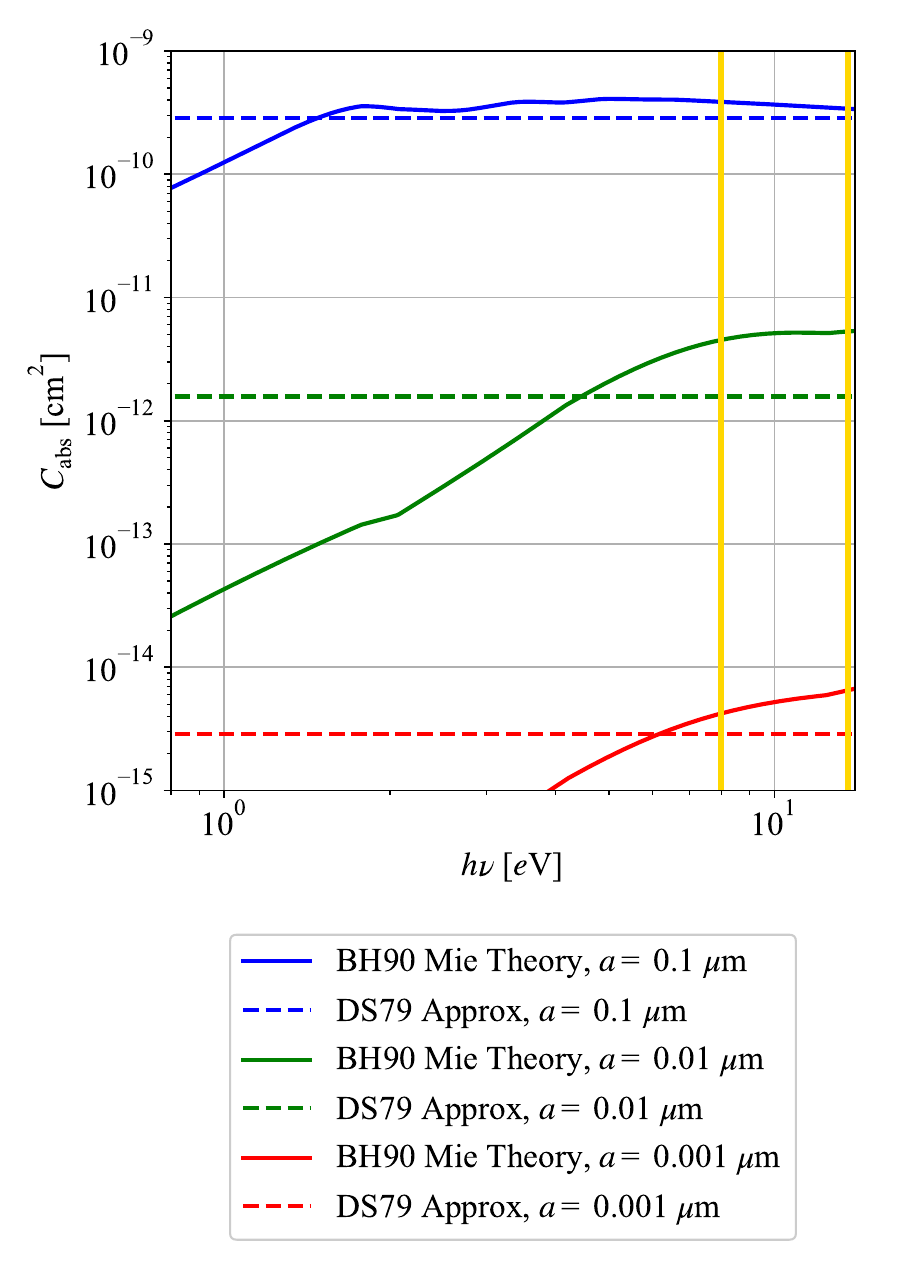}
		\caption[Comparison of Absorption Cross-Section, $C_{\rm abs}$, calculations.]{Comparison of Absorption Cross-Section, $C_{\rm abs}$, calculations.  The $C_{\rm abs}$ calculated using Mie theory is shown with solid lines, and the approximation used by \citet{drain79} is shown with dashed lines.  For the energy range appropriate for the photoelectron emission ($\sim 8-13.6$ eV, shown with yellow, vertical lines), the approximation provides a reasonable approximation with far less calculations.
			\label{fig:crosssect}
		}
	\end{center}
\end{figure}

Because the SNR is expected in our study to be non-radiative, we assume that the spectral photon flux, $\mathbb{F}_{\gamma, h\nu}(h\nu)$, is a blackbody at temperature, $T$, at the location of the grain inside the SNR \citep[with a dilution factor, $\mathbb{D} = 10^{-22}$,][]{drain79}:
\beq
\mathbb{F}_{\gamma, h\nu}(h\nu) = \frac{2 \pi \nu^{2}}{h c^{2}} \pfrac{1}{\exp\left[ \frac{h\nu}{k_{\rm B} T} \right]-1} \mathbb{D} \, ,
\label{eq:photonflux}
\eeq
or the average interstellar background \citep[see e.g.,][]{drain11} outside the SNR.  Additionally, the absorption cross-section of the grain, $C_{\rm abs}(h\nu)$, is dependent on the photon energy, grain size, and complex index of refraction.  The complex indices of refraction for iron from \citet{poll94} were used and $C_{\rm abs}(h\nu)$ was calculated using Mie theory and the procedure from \citet{bh83}.  However, this method is extremely calculation-intensive, and in order to simplify calculations, the $C_{\rm abs}(h\nu)$ approximation given by \citet{drain79} was used:
\beq
C_{\rm abs}(h\nu) = \frac{\pi a_{\rm gr}^{3}}{a_{\rm gr} + 0.01 \ \mu{\rm m}} \, .
\eeq
This approximation shows good agreement with calculation using Mie theory for iron within the region in which we are interested.  Comparisons for various grain sizes are shown in Figure~\ref{fig:crosssect}.

\section{Grain Charging:  Analytic Description}
\label{app:chargingapprox}

\setcounter{figure}{0}
\makeatletter 
\renewcommand{\thefigure}{F\@arabic\c@figure}
\makeatother

\begin{figure*}[t]
	\centering
	\subfigure[]
	{\includegraphics[width=0.48\textwidth]{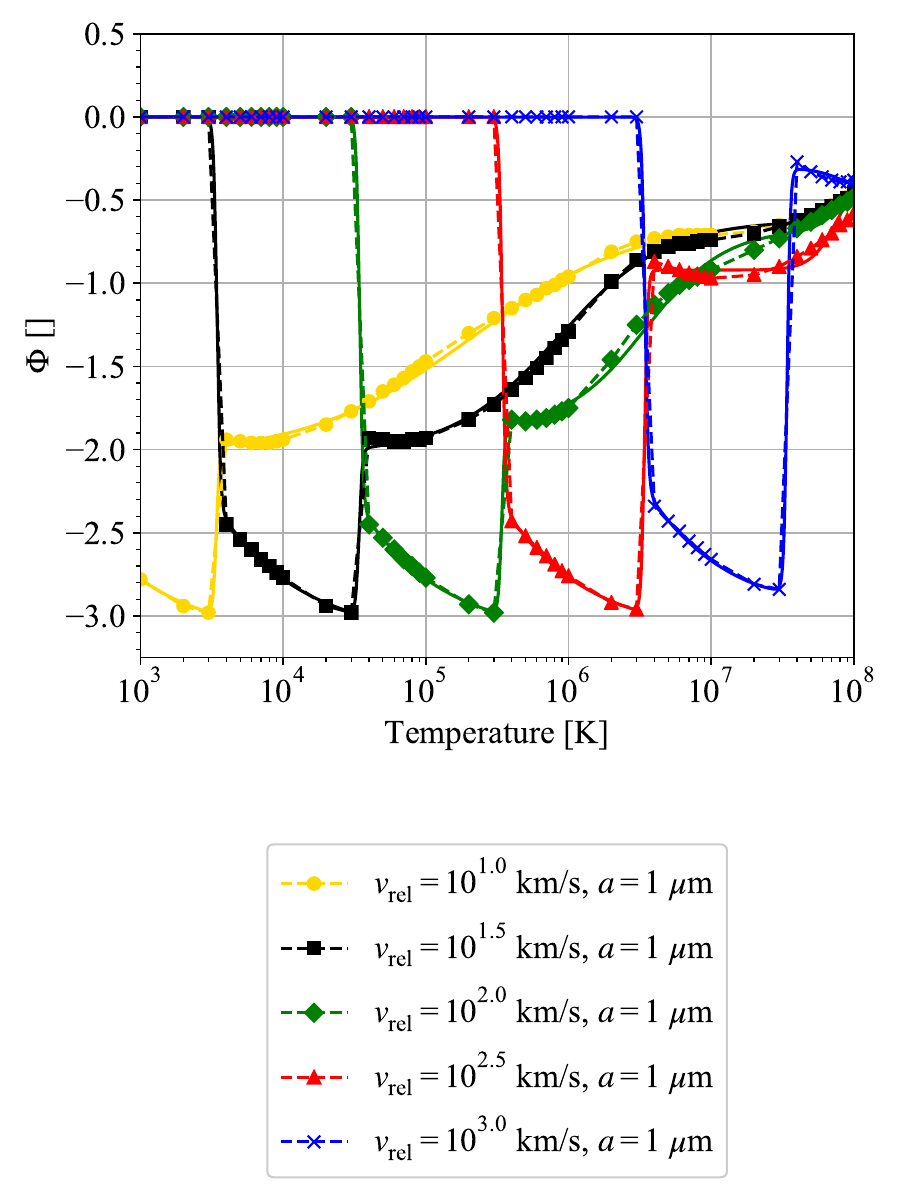}
		\label{fig:chargevel}} ~	
	\subfigure[]
	{\includegraphics[width=0.48\textwidth]{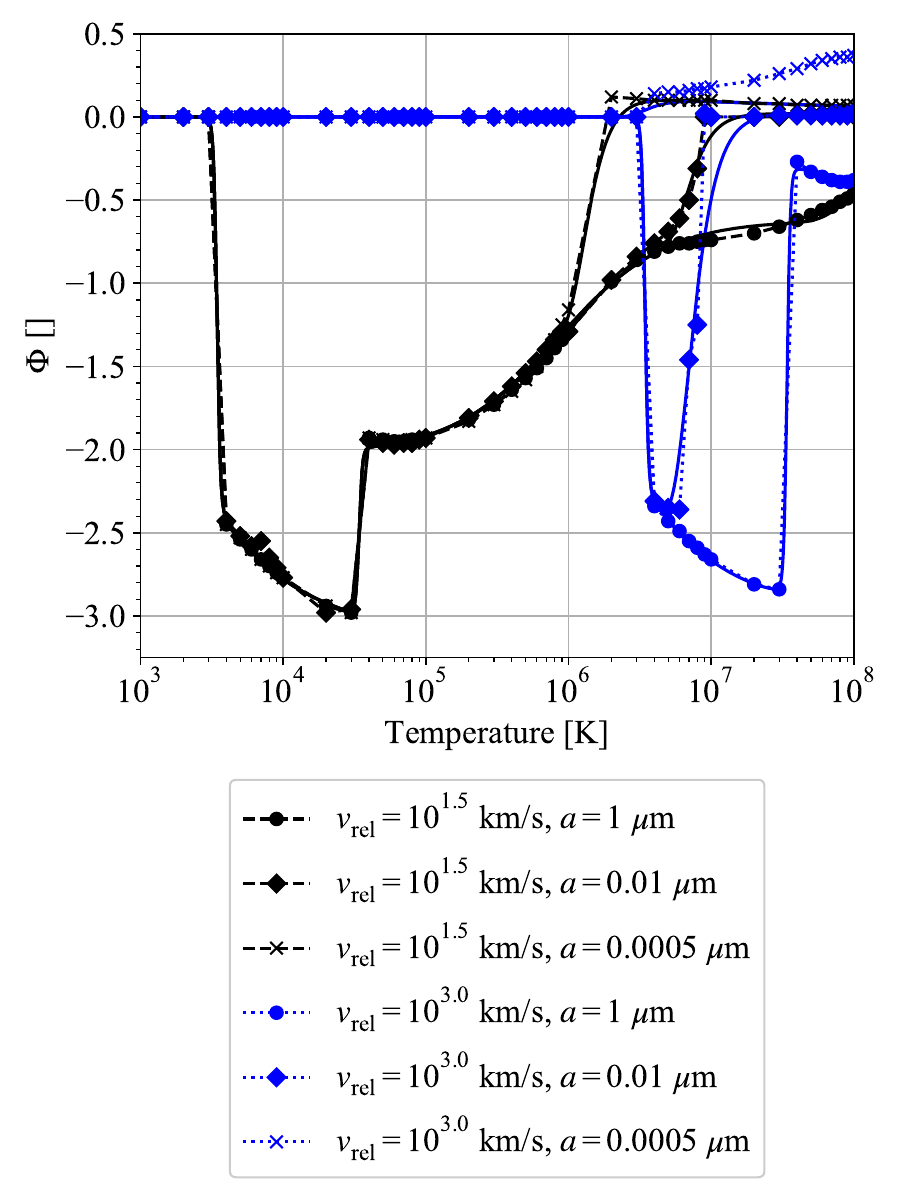}
		\label{fig:chargerad}} \\
	\caption{Sample potential parameters, $\Phi$, within the SNR for Fe grains.  The left panel (a) shows the potential parameter for a 1-$\mu$m grain at various relative velocities, and the right panel (b) shows the potential parameter for various grain sizes.  In both panels, the numerically-solved values (not including field emission) are shown with data marks and dash-dotted lines, while the analytical fit is shown with solid lines. \\[18pt]
		\label{fig:chargeanalytic}
	}
\end{figure*}

Within the SNR, the grain charge transitions through a range of values depending on the dominant charging process.  Figure~\ref{fig:chargeanalytic} shows the results for selected values and the analytic fitting function, which takes the form:
\begin{align}
\Phi = &\left[ \Phi_{\rm imp} (1-\Psi_{\rm B}) + \Phi_{\rm stationary} + \Phi_{\rm see1} \Psi_{\rm B} \right. \nonumber \\[6pt]
&\left. + \Phi_{\rm see2} \Psi_{\rm B} \Psi_{\rm C} \vphantom{}\right] \Psi_{\rm A} \Psi_{\rm F}  (1-\Psi_{\rm D}) \nonumber \\[6pt]
& + \Phi_{\rm tran} \Psi_{\rm D} \Psi_{\rm E} + \Phi_{\rm thermal} \Psi_{\rm B} (1-\Psi_{\rm D}) (1-\Psi_{\rm F}) \, ,
\label{eq:chargeanalytic}
\end{align}
where the ``$\Phi$'' quantities represent the corresponding charging regime and the ``$\Psi$'' quantities are scaling functions that scale and transition between the different charging regimes.

At temperatures where the thermal velocity is much smaller than the relative velocity ($v_{T} \ll v_{\rm rel}$), the side of the grain that is opposite from the relative motion is effectively shielded from impacting electrons/ions.  Additionally, because the relative velocity dominates, the electrons and ions will impact the grain at similar frequency; this results in a nearly neutral grain charge (e.g., for $v_{\rm rel} = 10^{1.5}$ km s$^{-1}$, this occurs for $T<3 \times 10^{3}$ K, see Figure~\ref{fig:chargeanalytic}).

As the thermal velocity approaches the relative velocity ($v_{T} \lesssim v_{\rm rel}$), the frequency of impinging electrons will dominate over impinging ions, driving the grain charge negative:
\begin{align}
\Phi_{\rm imp} &= -0.084+1.112 \times 10^{-3}~v_{7}^{2} + \pfrac{T_{\rm rel}}{T_{5}}^{0.75} \, , \\[6pt]
\Psi_{\rm A} &= \frac{T_{5}^{\Theta_{\rm A}}}{T_{\rm imp}^{\Theta_{\rm A}}+T_{5}^{\Theta_{\rm A}}} \times \max\left[ 0, \frac{T_{\rm tran}-T_{\rm imp}}{ \left| T_{\rm tran}-T_{\rm imp} \right| } \right] \, ,
\end{align}
with 
\begin{align}
v_{7} &= \pfrac{v_{\rm rel}}{10^{7} \ {\rm cm~s^{-1}}} \, , \nonumber \\[6pt]
T_{5} &= \pfrac{T}{10^{5} \ {\rm K}} \, , \nonumber \\[6pt]
T_{\rm rel} &= 0.2506~v_{7}^{2} \, , \nonumber \\[6pt]
T_{\rm imp} &= 0.3433~v_{7}^{2} \, , \nonumber \\[6pt]
T_{\rm tran} &= \frac{10.57}{1-\exp\left[ -\pfrac{\lambda_{\rm esc}}{a_{\rm gr}}^{0.75} \right]} \, , \nonumber \\[6pt]
\Theta_{\rm A} &= 36.99 \, , \nonumber
\end{align}
(e.g., for $v_{\rm rel} = 10^{1.5}$ km s$^{-1}$ $\hat{=}$ $T \approx 3 \times 10^{3}$ K).

When the thermal velocity is about the same value as the relative velocity ($v_{T} \approx v_{\rm rel}$), the grain charge will approach (but not reach in this case because of the inclusion of additional charging processes) its stationary value, $\Phi_{\rm stationary}$ \citep{spitz41}:
\beq
\exp \left[ \Phi_{\rm stationary} \right] = \sqrt{\frac{m_{e}}{m_{\rm ion}}} \left( 1-\Phi_{\rm stationary} \right) \, .
\eeq
In solving for the steady-state values, we chose an approximate composition of the ECSN ejecta:  $n_{\rm H} \approx 0.01 \ {\rm cm}^{-3}$, $n_{\rm He} \approx 1 \ {\rm cm}^{-3}$, $n_{\rm Fe} \approx 0.6 \ {\rm cm}^{-3}$, and $n_{\rm Ni} \approx 0.4 \ {\rm cm}^{-3}$.  For this composition, since He is much more numerous than H, and He is much more mobile compared to the Fe and Ni atoms, the plasma will behave similar to a pure-He environment, so $m_{\rm ion} \approx 4~m_{\rm u}$, at least with respect to grain charging, and  $ \Phi_{\rm stationary} \approx -3.049$ (e.g., for $v_{\rm rel} = 10^{1.5}$ km s$^{-1}$ $\hat{=}$ $T \in \left( 3 \times 10^{3}, 3 \times 10^{4} \right)$ K).

At higher thermal velocities, secondary electron emission will begin to dominate ($v_{T} \gtrsim v_{\rm rel}$), increasing the grain charge:
\begin{align}
\Phi_{\rm see1} &= 1.740 \left( 1-\exp\left[ -0.1037~v_{7}^{2} \right] \right)+1.005 \, , \\[6pt]
\Phi_{\rm see2} &= \max\left[ 0, -0.2267~v_{7}^{2}+1.430 \right] \, , \\[6pt]
\Psi_{\rm B} &= \frac{T_{5}^{\Theta_{\rm B}}}{T_{\rm see1}^{\Theta_{\rm B}}+T_{5}^{\Theta_{\rm B}}} \, , \\[6pt]
\Psi_{\rm C} &= \frac{T_{5}^{\Theta_{\rm C}}}{T_{\rm see2}^{\Theta_{\rm C}}+T_{5}^{\Theta_{\rm C}}} \, ,
\end{align}
with
\begin{align}
T_{\rm see1} &= 3.404~v_{7}^{2} \, , \nonumber \\[6pt]
T_{\rm see2} &= 34.82~v_{7}^{1.223} \, , \nonumber \\[6pt]
\Theta_{\rm B} &= 38.48 \, , \nonumber \\[6pt]
\Theta_{\rm C} &= 0.3545 \ln \left[ v_{7} \right]+1.563 \, , \nonumber
\end{align}
(e.g., for $v_{\rm rel} = 10^{1.5}$ km s$^{-1}$ $\hat{=}$ $T \in \left( 3 \times 10^{4}, 10^{6} \right)$ K).

Near $T = 10^{6}$ K, the transmission/tunneling current will become important, further increasing the grain charge:
\begin{align}
\Phi_{\rm tran} &= 0.1953~T_{5}^{-0.162} \, , \\[6pt]
\Psi_{\rm D} &= \exp\left[ -\pfrac{T_{\rm cr}}{T_{5}}^{4} \right] \, , \\[6pt]
\Psi_{\rm E} &= \exp\left[ -10^{-4} \pfrac{a_{\rm gr}}{\lambda_{\rm esc}}^{4} \right] \, ,
\end{align}
with 
\beq
T_{\rm cr} = \max\left[ T_{\rm tran}, T_{\rm imp} \right] \, , \nonumber
\eeq
(e.g., for $v_{\rm rel} = 10^{1.5}$ km s$^{-1}$ $\hat{=}$ $T \in \left( 10^{6}, 3 \times 10^{7} \right)$ K).

At higher temperatures, the electrons and ions will again impact the grain at relatively similar frequencies, resulting in a relatively neutral grain charge:
\begin{align}
\Phi_{\rm thermal} &= 0.1862 \ln\left[ T_{5} \right]-1.756 \, , \\[6pt]
\Psi_{\rm F} &= \exp\left[ -\pfrac{T_{5}}{T_{\rm thermal}}^{4} \right] \, , 
\end{align}
with
\beq
T_{\rm thermal} = \max\left[ 703.8, 9.964~v_{7}^{2} \right] \, ,
\eeq
(e.g., for $v_{\rm rel} = 10^{1.5}$ km s$^{-1}$ $\hat{=}$ $T > 3 \times 10^{7}$ K).

Within the SNR, because of the extremely small dilution factor ($\mathbb{D} = 10^{-22}$, Equation~\ref{eq:photonflux}), photoelectron emission plays almost no role in grain charging.  However, outside the SNR, it becomes the dominant process, driving the grain potential to $U_{\rm gr} = 5.6$ V, since it will be in interstellar space and subject to the ISRF \citep{drain79,drain11}.

Lastly, we also establish potential limits to account for field emission \citep{mckee87}:
\beq
\Phi_{\rm min} \leq \Phi \leq \Phi_{\rm max} \, , \nonumber
\eeq
with
\begin{align}
\Phi_{\rm min} = -&11.6 \pfrac{a_{\rm gr}}{0.1 \ \mu{\rm m}}\pfrac{10^{5} \ {\rm K}}{T} \, , \nonumber \\[6pt]
\Phi_{\rm max} =~~&348 \pfrac{a_{\rm gr}}{0.1 \ \mu{\rm m}}\pfrac{10^{5} \ {\rm K}}{T} \, .
\end{align}
This completes our description of the grain charge so that:
\beq
U_{\rm gr}(r) = 
\begin{cases}
	\Phi k_{\rm B}T/e & \quad r \leq R_{\rm SN} \, , \\[6pt]
	5.6 \ {\rm V} & \quad r > R_{\rm SN} \, . \\[6pt]
\end{cases}
\eeq

\section{Grain Dynamics}
\label{app:graindynamics}

The combined drag of both collisional and Coulomb sources is given by \citep{drain79,drain11}:
\begin{align}
F_{\rm drag} =&~2 \pi a_{\rm gr}^{2} k_{\rm B}T \nonumber \\[6pt] 
&\times \left\lbrace \sum_{j} n_{j} \left[ G_{0}\left( s_{j} \right) + Z_{j}^{2} \Phi^{2} \ln \left[ \Lambda / Z_{j} \right] G_{2}\left( s_{j} \right) \right] \right\rbrace \, ,
\end{align}
where:
\begin{align}
G_{0}\left( s \right) \equiv &\left( s^{2} + 1 - \frac{1}{4 s^{2}}\right) \erf [s] \nonumber \\[6pt]
&+ \frac{1}{\sqrt{\pi}} \left( s + \frac{1}{2 s} \right) \exp\left[ -s^{2} \right]  \, , \\[6pt]
G_{2}\left( s \right) \equiv &\frac{\erf [s]}{s^{2}} - \frac{2}{s \sqrt{\pi}} \exp\left[ -s^{2} \right] \, , \\[6pt]
\Phi \equiv & \frac{e U_{\rm gr}}{k_{\rm B}T} \, , \\[6pt]
s_{j} \equiv &\pfrac{m_{j} v_{\rm rel}^{2}}{2k_{\rm B}T}^{1/2} \, , \\[6pt]
\Lambda \equiv &\frac{3}{2 a_{\rm gr} e\left| \Phi \right| } \pfrac{k_{\rm B}T}{\pi n_{e}}^{1/2} \, , \\[6pt]
\erf [\chi] \equiv &\frac{2}{\sqrt{\pi}} \int_{0}^{\chi} \exp \left[ - \varUpsilon^{2 }\right] d\varUpsilon \, , 
\end{align}
where we use cgs/esu units.  The $G_{0}(s)$ term accounts for collisional drag, and the $G_{2}(s)$ term accounts for Coulomb drag.  Approximations exist for both, but we used the exact forms given here for completeness.  The drag force is summed over all plasma species, $j$, within the plasma (e.g., $p^{+}$, $e^{-}$, $\alpha$, C, etc.), each with number density $n_{j}$.  The velocity parameter, $s$, depends on the relative velocity between the grain and plasma, $v_{\rm rel}$, mass of the impacting plasma particle, $m_{j}$, and the temperature of the plasma, $T$ (we assume all constituents are at the same temperature, i.e., $T_{j} = T \ \forall \ j$).  Similarly, the potential parameter, $\Phi$ depends on the electric potential of the grain, $U_{\rm gr} = q_{\rm gr}/a_{\rm gr}$, where $q_{\rm gr}$ is the charge of the grain.  The charge number of the plasma particle is $Z_{j}$, $k_{\rm B}$ is the Boltzmann constant, and $e$ is the elementary charge.

\section{Grain Sputtering}
\label{app:grainsputtering}

The erosion rate due to sputtering (both kinetic and thermal) is given by \citep{dwek92}, and we use the approach by \citet{noz06,bisc16}:
\begin{align}
\frac{da_{\rm gr}}{dt} =&~-\frac{m_{\rm sp}}{4 \rho_{\rm gr}} \sum_{j} \frac{n_{j}}{s_{j}} \pfrac{8k_{\rm B}T}{\pi m_{j}}^{1/2} \exp\left[ -s_{j}^{2}\right] \nonumber \\[6pt]
&\times \int d\epsilon_{j}~\sqrt{\epsilon_{j}} \exp\left[-\epsilon_{j}\right] \sinh \left( 2 s_{j} \sqrt{\epsilon_{j}} \right) Y_{j}^{0}(\epsilon_{j}) \, ,
\end{align}
where $m_{\rm sp}$ is the mass of the sputtered atom (i.e., the average atomic mass for the dust composition, $m_{\rm sp, {\rm Fe}} = 56 m_{\rm u}$), and $\rho_{\rm gr}$ is the density of the dust grain.  Additionally, the angle-averaged sputtering yield given by: $\left\langle Y_{j}(E_{j}) \right\rangle_{\theta} = 2 Y_{j}^{0}(E_{j})$ \citep{drain79}, and the backward sputtering yield at normal incidence, $Y_{j}^{0}(E)$, is given by \citep{bohd84}:
\begin{align}
Y_{j}^{0} \left( \epsilon_{j} \right) =&~160 \ \frac{\rm atoms}{\rm ion} \pfrac{S_{j}(E)}{\rm 1~erg~cm^{2}} \pfrac{\rm 4.31~eV}{E_{\rm bind}} \nonumber \\[6pt]
&\times \pfrac{\xi_{j}(\zeta_{j})}{\kappa \zeta_{j} + 1} \left( 1 - \pfrac{E_{\rm th}}{E}^{2/3} \right) \nonumber \\[6pt]
&\times \left( 1 - \pfrac{E_{\rm th}}{E} \right)^{2} \, ,
\end{align}
where $\kappa$ is a free parameter that is adjusted to fit experimental data, $E_{\rm bind}$ is the surface binding energy (see Table \ref{tab:sputtering}), and $\zeta_{j} = m_{\rm target}/m_{j}$ is the ratio of the target atom mass, $m_{\rm target}$, to the incident atom mass, $m_{j}$.  The threshold energy, $E_{\rm th}$, to induce sputtering is given by \citep{and81,bohd84}:
\beq
E_{\rm th} = 
\begin{cases} 
	\displaystyle \frac{E_{\rm bind}}{g_{j}(1-g_{j})} & \quad \text{for } m_{j}/m_{\rm target} \leq 0.3 \, , \\[12pt]
	\displaystyle 8 E_{\rm bind} \pfrac{m_{j}}{m_{\rm target}}^{1/3} & \quad \text{for } m_{j}/m_{\rm target} > 0.3 \, , \\
\end{cases}
\eeq
where $g_{j} = 4 m_{j} m_{\rm target} \left( m_{j} + m_{\rm target} \right)^{-2}  $ is the maximum fraction energy transfer in a head-on elastic collision.  The stopping cross-section, $S_{j}(E)$, is given by \citep{sigm81}:
\beq
S_{j}(E) = 4 \pi a_{\rm sc} Z_{j} Z_{\rm target} e^{2} \frac{m_{j}}{m_{j} + m_{\rm target}} \varsigma_{j}(\epsilon_{j}) \, ,
\eeq
and the screening length, $a_{\rm sc}$, for interaction between nuclei is:
\beq
a_{\rm sc} = 0.885 a_{\rm Bohr} \left( Z_{j}^{2/3} + Z_{\rm target}^{2/3} \right)^{-1/2} \, ,
\eeq
where $a_{\rm Bohr} = 0.529 \ {\rm \AA}$ is the Bohr radius.  An approximation of the function, $\varsigma_{j}(\epsilon_{j})$ is given by \citet{mats84}:
\beq
\varsigma_{j}(\epsilon_{j}) = \frac{3.441 \sqrt{\epsilon_{j}} \ln \left[ \epsilon_{j} + 2.718 \right]}{1 + 6.35 \sqrt{\epsilon_{j}} + \epsilon_{j}\left( 6.882 \sqrt{\epsilon_{j}} - 1.708 \right)} \, ,
\eeq
where the reduced energy, $\epsilon_{j}$, is:
\beq
\epsilon_{j} = \pfrac{m_{\rm target}}{m_{j} + m_{\rm target}} \pfrac{a_{\rm sc}}{Z_{j} Z_{\rm target} e^2} E \, .
\eeq
The function $\xi_{j}(\zeta_{j})$, depends on the energy distribution deposited into the target, and we used the derivation by \citet{noz06} for $\zeta_{j} \in [0.3, 56]$:
\beq
\xi_{j}(\zeta_{j}) = 
\begin{cases}
	0.2 & \quad \zeta_{j} \leq 0.5 \, , \\[6pt]
	0.1 \zeta_{j}^{-1} + 0.25 \left( \zeta_{j} - 0.5 \right)^{2} & \quad 0.5 < \zeta_{j} \leq 1 \, , \\[6pt]
	0.3 \left( \zeta_{j} - 0.6 \right)^{2} & \quad 1 < \zeta_{j} \, . \\
\end{cases}
\eeq

{}

\end{document}